\documentclass[preprint2]{aastex}

\usepackage{epstopdf}
\usepackage{graphics}
\usepackage{lscape}
\usepackage{psfig}

\def\ntot{118}
\def\ninde{83}

\def\gloz{$\ell(z)_{\rm GRB}$}
\def\mgloz{\ell(z)_{\rm GRB}}
\def\qloz{$\ell(z)_{\rm QSO}$}

\def\loz{\ell(z)}
\def\gloz{$\ell(z)$}
\def\intl{\int\limits}
\def\samI{Sample {\rm I}}
\def\samF{Sample {\rm F}}
\def\samH{Sample {\rm H}}
\def\samL{Sample {\rm L}}

\def\lya{Ly$\alpha$}

\def\CIVdblt{{\rm C~}\kern 0.1em{\sc iv}~$\lambda\lambda 1548, 1550$}
\def\MgIIdblt{{\rm Mg~}\kern 0.1em{\sc ii}~$\lambda\lambda 2796, 2803$}
\def\NVdblt{{\rm N}\kern 0.1em{\sc v}~$\lambda\lambda 1238, 1242$}  
\def\OVIdblt{{\rm O}\kern 0.1em{\sc vi}~$\lambda\lambda 1031, 1037$}
\def\SiIVdblt{{\rm Si~}\kern 0.1em{\sc iv}~$\lambda\lambda1394, 1403$}
\def\AlIIIdblt{{\rm Al~}\kern 0.1em{\sc iii}~$\lambda\lambda1855,1863$}
\def\FeIIdblt{{\rm Fe~}\kern 0.1em{\sc ii}~$\lambda\lambda 2383, 2600$}

\def\AlII{\hbox{{\rm Al~}\kern 0.1em{\sc ii}}}
\def\AlIII{\hbox{{\rm Al~}\kern 0.1em{\sc iii}}}
\def\CaI{\hbox{{\rm Ca}\kern 0.1em{\sc i}}}
\def\CaII{\hbox{{\rm Ca}\kern 0.1em{\sc ii}}}

\def\CrII{\hbox{{\rm Cr~}\kern 0.1em{\sc ii}}}
\def\CI{\hbox{{\rm C~}\kern 0.1em{\sc i}}}
\def\CII{\hbox{{\rm C~}\kern 0.1em{\sc ii}}}
\def\CIII{\hbox{{\rm C~}\kern 0.1em{\sc iii}}}
\def\CIV{\hbox{{\rm C~}\kern 0.1em{\sc iv}}}
\def\CV{\hbox{{\rm C}\kern 0.1em{\sc v}}}

\def\HI{\hbox{{\rm H~}\kern 0.1em{\sc i}}}
\def\HII{\hbox{{\rm H~}\kern 0.1em{\sc ii}}}

\def\Lya{\hbox{{\rm Ly}\kern 0.1em$\alpha$ }}
\def\Lyb{\hbox{{\rm Ly}\kern 0.1em$\beta$}}
\def\Lyg{\hbox{{\rm Ly}\kern 0.1em$\gamma$}}
\def\Lyfive{\hbox{{\rm Ly}\kern 0.1em$5$}}
\def\Lysix{\hbox{{\rm Ly}\kern 0.1em$6$}}
\def\Lyseven{\hbox{{\rm Ly}\kern 0.1em$7$}}
\def\Lyeight{\hbox{{\rm Ly}\kern 0.1em$8$}}
\def\Lynine{\hbox{{\rm Ly}\kern 0.1em$9$}}
\def\Lyten{\hbox{{\rm Ly}\kern 0.1em$10$}}
\def\HeI{\hbox{{\rm He}\kern 0.1em{\sc i}}}
\def\HeII{\hbox{{\rm He}\kern 0.1em{\sc ii}}}
\def\FeI{\hbox{{\rm Fe~}\kern 0.1em{\sc i}}}
\def\FeII{\hbox{{\rm Fe~}\kern 0.1em{\sc ii}}}
\def\FeIII{\hbox{{\rm Fe~}\kern 0.1em{\sc iii}}}
\def\MnII{\hbox{{\rm Mn}\kern 0.1em{\sc ii}}}
\def\MgI{\hbox{{\rm Mg~}\kern 0.1em{\sc i}}}
\def\MgII{\hbox{{\rm Mg~}\kern 0.1em{\sc ii}}}
\def\MgIII{\hbox{{\rm Mg~}\kern 0.1em{\sc iii}}}
\def\MgIV{\hbox{{\rm Mg~}\kern 0.1em{\sc iv}}}
\def\NaI{\hbox{{\rm Na}\kern 0.1em{\sc i}}}
\def\NV{\hbox{{\rm N}\kern 0.1em{\sc v}}}
\def\NII{\hbox{{\rm N}\kern 0.1em{\sc ii}}}
\def\NIII{\hbox{{\rm N}\kern 0.1em{\sc iii}}}
\def\NiI{\hbox{{\rm Ni~}\kern 0.1em{\sc i}}}
\def\NiII{\hbox{{\rm Ni~}\kern 0.1em{\sc ii}}}
\def\OVI{\hbox{{\rm O}\kern 0.1em{\sc vi}}}
\def\OI{\hbox{{\rm O}\kern 0.1em{\sc i}}}
\def\OII{\hbox{[{\rm O}\kern 0.1em{\sc ii}]}}
\def\SiII{\hbox{{\rm Si~}\kern 0.1em{\sc ii}}}
\def\SiIII{\hbox{{\rm Si~}\kern 0.1em{\sc iii}}}
\def\SiIV{\hbox{{\rm Si~}\kern 0.1em{\sc iv}}}
\def\SII{\hbox{{\rm S}\kern 0.1em{\sc ii}}}
\def\SIII{\hbox{{\rm S}\kern 0.1em{\sc iii}}}
\def\SIV{\hbox{{\rm S}\kern 0.1em{\sc iv}}}
\def\TiII{\hbox{{\rm Ti}\kern 0.1em{\sc ii}}}
\def\ZnII{\hbox{{\rm Zn~}\kern 0.1em{\sc ii}}}

\def\swift{\emph{Swift \,}}

\def\xray{X-ray}
\def\swift{\emph{Swift}}
\def\gmos{\emph{GMOS}}

\def\dz{$\Delta z$}

\def\simlt{\mathrel{\hbox{\rlap{\hbox{\lower4pt\hbox{$\sim$}}}\hbox{$<$}}}}
\def\simgt{\mathrel{\hbox{\rlap{\hbox{\lower4pt\hbox{$\sim$}}}\hbox{$>$}}}}
\newcommand{\hiz}{high-$z$}

\begin{document}

\title{An Independent Measurement of the Incidence of MgII Absorbers
  along Gamma-Ray Burst Sightlines: the End of the Mystery?}


\author{
	A.~Cucchiara\altaffilmark{1}, 
	J.~X.Prochaska\altaffilmark{1},
	G.~Zhu\altaffilmark{2},
	B.~M\'enard\altaffilmark{2,3,4},
	J.~P.~U. Fynbo\altaffilmark{5},
	D.~B. Fox\altaffilmark{6}, 
	H.-W.~Chen\altaffilmark{7},
	K.~L. Cooksey\altaffilmark{8}
	S.~B. Cenko\altaffilmark{9},
	D.~Perley\altaffilmark{10},
	J.~S. Bloom\altaffilmark{9},
	E.~Berger\altaffilmark{11},
	N.~R. Tanvir\altaffilmark{12},
	V.~D'Elia\altaffilmark{13},
	S.~Vergani\altaffilmark{14},
	S.~Lopez\altaffilmark{15}
	R. Chornock\altaffilmark{11},
	Thomas deJaeger\altaffilmark{15}
	}
	\email{acucchia@ucolick.org}

\altaffiltext{1}{Department of Astronomy and Astrophysics, UCO/Lick Observatory, University of California, 1156 High Street, Santa Cruz, CA 95064, USA}
\altaffiltext{2}{Department of Physics \& Astronomy, Johns Hopkins University, Baltimore, MD 21218 U.S.A.} 
\altaffiltext{3}{Kavli Institute for the Physics and Mathematics of the Universe,
Tokyo University, Kashiwa 277-8583, Japan} 
\altaffiltext{4}{Alfred P. Sloan fellow} 
\altaffiltext{5}{Dark Cosmology Centre, Niels Bohr Institute, University of Copenhagen, Juliane Maries Vej 30, 2100 Copenhagen, Denmark} 
\altaffiltext{6}{Department of Astronomy \& Astrophysics, Pennsylvania State University, University Park, PA 16802, USA} 
\altaffiltext{7}{Department of Astronomy \& Astrophysics, Kavli Institute for Cosmological Physics, University of Chicago, Chicago, IL 60637, USA}
\altaffiltext{8}{MIT Kavli Institute for Astrophysics \& Space Research, 77 Massachusetts Avenue, 37-685, Cambridge, MA 02139, USA}
 \altaffiltext{9}{Department of Astronomy, University of California, Berkeley, CA 94720-3411, USA}
\altaffiltext{10}{Caltech}
\altaffiltext{11}{Harvard-Smithsonian Center for Astrophysics, 60 Garden Street, Cambridge, MA 02138, USA} 
\altaffiltext{12}{Department of Physics and Astronomy, University of Leicester, University Road, Leicester, LE1 7RH, UK} 
\altaffiltext{13}{Istituto Nazionale di Astrofisica - Osservatorio Astronomico di Roma, Via di Frascati 33, I-00040 Monte Porzio Catone (RM), Italy}
\altaffiltext{14}{GEPI, Observatoire de Paris, CNRS, Univ. Paris Diderot, 5 Place Jules Jannsen, F-92195, Meudon, France}
\altaffiltext{15}{Departamento de Astronom\'{i}a, Universidad de Chile, Casilla 36-D, Santiago, Chile}

\begin{abstract}
In 2006, Prochter et al.\ reported a statistically significant
enhancement of very strong \ion{Mg}{2} absorption systems
intervening the sightlines to gamma-ray bursts (GRBs) relative to the incidence of
such absorption along quasar sightlines.  This counterintuitive result, 
has inspired a diverse set of 
astrophysical explanations (e.g.\ dust, gravitational lensing) but
none of these has obviously resolved the puzzle.  Using the largest
set of GRB afterglow spectra available, we reexamine the purported
enhancement.  In an independent sample of GRB spectra with a survey
path 3 times larger than Prochter et al., we measure the incidence per
unit redshift of $\geq1$\AA\ rest-frame equivalent width \ion{Mg}{2} absorbers at $z
\approx 1$ to be \gloz$= 0.18 \pm 0.06$.  This is fully consistent with
current estimates for the incidence of such absorbers along quasar
sightlines.  Therefore, we do not confirm the original enhancement and
suggest those results suffered from a statistical fluke.
Signatures of the original result do remain in our full sample
(\gloz\ shows an $\approx 1.5$ enhancement over \qloz), but the
statistical significance now lies at  $\approx90\%$\,c.l.
Restricting our analysis to the subset of high-resolution spectra of
GRB afterglows (which overlaps substantially with Prochter et al.), we
still reproduce a statistically significant enhancement of \ion{Mg}{2}
absorption.  
The reason for this excess, if real, is still unclear since there is no
connection between the rapid afterglow follow-up process with
echelle (or echellette) spectrographs and the detectability of strong
\ion{Mg}{2} doublets. Only a larger sample of such high-resolution data
will shed some light on this matter.

\end{abstract}



\section{Introduction }
\label{sec:intro}

In the last decade, the study of the inter-galactic (IGM) and circum-galactic 
medium (CGM) has received
a great boost thanks to large spectroscopic surveys of distant
quasars, in particular the dataset provided by the 
Sloan Digital Sky Surveys \citep{York:2000uq}. 
These objects randomly sample thousands of lines of sight and, 
being bright background sources of light, probe gas and matter 
located in foreground objects.  

One of the most commonly surveyed set of transitions in quasar spectra
is the \ion{Mg}{2} doublet at $\lambda\lambda 2796,2803$\AA.  Its
common detection stems from the large rest wavelength (which makes them
easily detectable by most optical spectrographs when the absorber is located at redshift $z=0.5-2.2$), the relatively
high abundance of Mg, and the strength of this resonance-line doublet.  
The \ion{Mg}{2} systems are frequently classified in terms of 
the rest-frame equivalent width, $W_r$, of the bluer component 
as ``weak'' ($W_{2796} <0.3$ \AA), 
``strong'' ($W_{2796} >0.3$ \AA),
and ``very strong''  \citep[$W_{2796} >1.0$ \AA, like in][]{Rodriguez-Hidalgo:2012oq}.
For simplicity, throughout the paper we will refer to this 
last category as ``strong'', since it is the only one pertinent to 
this work.
 \ion{Mg}{2} doublet lines have been surveyed extensively 
from $z \approx 0.1-2.5$ in the optical passband and
now to $z = 5.2$ with near-IR spectroscopy 
\citep[e.g.][]{Steidel:1992kx,Nestor:2005fk,Prochter:2006kx,
Quider:2011fk,Simcoe:2011uq,zhu:2012aa}.
The results indicate that while the weak and strong absorbers
incidence show small if any evolution with redshift, the 
very strong \MgII\ absorbers present an increasing trend
up to  $z\sim3$ before declining at higher redshift
\citep{Prochter:2006kx,Matejek:2012bh}.
This evolution rather closely tracks the cosmic star formation history
\citep{Prochter:2006kx,zhu:2012aa}, suggesting that some systems may be causally connected to
on-going star formation \citep{Menard:2011vn,Matejek:2012bh}, although, accurate analysis
of the SDSS survey needs to be carefully taken into account in order to
avoid technical biases \citep{Lopez:2012cr}.

For several decades now, strong \ion{Mg}{2} absorption has been
associated with gas in and around galaxies.  Early work identified a
small sample of $L \approx L^*$ galaxies at modest impact parameters
($\rho \approx 10-50$\,kpc) to quasars exhibiting strong \ion{Mg}{2}
absorption \citep{Bergeron:1986fk,Lanzetta:1987vn,Steidel:1993fk} , although no significant 
trend has been found for a population of Luminous Red Galaxies \cite[e.g. ][and references therein]{Bowen:2011nx}.  

These observations motivate
the association of \ion{Mg}{2} gas with the outer disk and/or CGM of
these galaxies.
Several QSOs line of sights presenting 
\MgII\ absorbers have been explored in order to probe the extent and the 
baryon content around low-$z$ galaxies 
\citep[][and reference therein]{Kacprzak:2012kx,Chen:2008dq}, 
as well as a
diagnostic of the inner part of these galaxies' interstellar medium \citep{Bowen:1995bh}.
Also, a stack analysis was performed by \cite{Zibetti:2007ys} using light profiles
(from associated galaxies) of quasars exhibiting strong \ion{Mg}{2}
absorption in the SDSS. With their image-stacking technique they
studied the cross-correlation between the \MgII\ gas 
and the galaxy light from 10 to 200 kpc, finding that strong \MgII\
absorbers may be explained by models that include metal-enriched 
outflows from star-forming/bursting galaxies.   
Most recently, several attempts to trace the
covering fraction and nature of \ion{Mg}{2} absorption by targeting
known galaxies with coincident background quasars have been performed
\citep{Barton:2009zr,Chen:2010uq,Werk:2012vn}.
Their results indicate the mean covering fraction increases from $\sim70\%$ for 
$W_r\geq0.3$ \AA\ to $\sim80\%$ for $\geq0.1$ \AA, confirming that
extended \MgII\ absorbing haloes are a common feature around
normal galaxies.
Finally, it 
has been found that ``strong'' 
absorbers are often associated with nearby (within 75 kpc) 
$\sim0.1-5 L^*$ galaxies along the line of sight
\citep{Kacprzak:2008qf,Nestor:2011fk,Chen:2010uq}.

The survey and analysis of \ion{Mg}{2} gas is no longer limited to
quasar spectroscopy. For example, researchers have now used distant galaxies to
probe foreground galaxies enabling searches at very small impact
parameter \citep{Rubin:2011ly} 
and statistical `maps' of the absorption 
correlated with the foreground galaxy orientation \citep{Bordoloi:2011ve}.
Similarly, Gamma-ray Bursts (GRBs), with their extraordinarily bright optical afterglows
provide not only direct information on their host galaxies, but also
trace matter intercepting their lines of sight \citep{Metzger:1997lr}. 

The advantage of using GRBs as background sources is twofold: 
first, they can be observed up to very high redshifts  
\citep{Kawai:2006fk,tfl09,sdc09,Cucchiara:2011lr},
which allows one to explore a larger redshift path length, and second, their discovery 
is largely unbiased with respect to intrinsic properties of their hosts 
(extinction, luminosity or mass). 
When a GRB fades away, they leave the line of sights clear for future deep observations in
order to search for the \ion{Mg}{2} counterparts \citep{Vreeswijk:2003uq,Jakobsson:2004vn,Schulze:2012ys,Chen:2012kx}.
One of the first attempt to identify the nature of three absorbers along GRB 060418 was performed by 
\cite{Pollack:2009zr}, which identified the absorbers to be $L\sim 0.1-1 L^*$ galaxies
at very small impact parameter from the GRB location ($\rho \lesssim 10\ h^{-1}$ kpc).
Deep imaging of several other fields have confirmed these early findings  \citep{Chen:2009ly}.  
On the other hand, the number of GRBs discovered and spectroscopically observed 
is several orders of magnitude less then the number of quasars available 
in large optical surveys (e.g. SDSS DR8).
This difference has been reduced, however, with the success of the {\it
  Swift} satellite providing the discovery and follow-up of several hundred
GRBs \citep{Gehrels:2004fj,Gehrels:2009fk}.

Shortly after the launch of {\it Swift} in November 2004, a survey of \ion{Mg}{2}
absorption in GRB afterglow spectroscopy was performed for an early
sample of {\it Swift} bursts and a heterogeneous sample of pre-existing GRB
spectra \citep[][P06 hereafter]{Prochter:2006fr}.
The authors revealed an extremely puzzling result: the incidence of
strong ($W_{2796} \ge 1$\AA)
intervening \ion{Mg}{2} absorbers was about 4 times
higher along GRB sightlines than quasar sightlines.  Despite the small
sample size, the statistical
significance of their dataset was high: the null
hypothesis that GRBs and quasar spectra would show identical
incidences of strong, foreground \ion{Mg}{2} absorption 
was ruled out at $\gtrsim 99.99\%$\ confidence.

The authors proposed several hypotheses that might explain the
difference, which have since been studied in greater detail: 
1) a possible intrinsic origin of these absorbers associated near the GRBs
themselves \citep{Cucchiara:2009mgii,Bergeron:2011lr}; 
2) a significant dust bias along QSO lines of sight
\citep{mnt08,pvs07,Budzynski:2011uq}; 
3) a geometric effect difference due to the sizes of the emitting regions 
between GRBs and QSOs \citep{fbs07,pvs07, Lawther:2012fk} ; 
4) a gravitational lensing effect
\citep{Vergani:2009fj,Porciani:2001vn,Rapoport:2011kx}. 
Subsequent work has ruled out several of these possibilities 
and none appears to be sufficient on its own to
explain the observations. 
After seven years of the \swift\ mission and more than two hundred
GRBs with spectroscopic confirmations, this mystery remains.

It is important to emphasize that the original P06 work,
and even the studies that have followed, have relied on
a small sample of GRB afterglow spectra.   Even the largest analysis
to date 
analyzed only 26 line of sights (finding 22 absorbers), 
for a total redshift path of $\Delta z =31.55$ \citep{Vergani:2009fj}.
Furthermore, no study has analyzed a completely
independent set of GRB sightlines from the P06 analysis.

In this paper, we use 
data obtained primarily during the \swift\ era by several facilities,
to obtain the most complete sample of GRB afterglow spectra and the
largest  redshift pathlength available to date. 
From this parent sample, we are able to construct sub-samples which are
entirely independent from the original work of P06.  
Similarly, we can study possible instrumental biases (e.g. spectral resolution) 
which may affect the final results.

The paper is structured as follow: in  \S\ref{sec:data} we describe our dataset and the 
data analysis procedure, while in  \S\ref{sec:redpath}  we present our 
procedure for defining the redshift path density per GRB sightline.
\S\ref{sec:search}  describes the search 
methodology to identify possible \MgII\ systems along every line of
sight,
with distinction between different datasets (e.g. high-resolution vs. low-resolution, strong vs weak \MgII\ equivalent width). 
Finally, in Section \S\ref{sec:result} and \S\ref{sec:discussion} we
 present our findings, including interesting sub-samples results,
and we summarize them in light of possible steps forward into
understanding this puzzling phenomenon. All the quoted errors, 
unless otherwise stated, are considered at 1-$\sigma$ confidence level.


\section{Data Selection}
\label{sec:data}

The acquisition of an optical spectrum from a given GRB afterglow is a complex
and unrepeatable process. During the \swift\ era, the timelapse
between the discovery of the gamma-ray emission (by the \swift/BAT instrument) and
the afterglow localization (by XRT and/or UVOT on-board the spacecraft)
is generally less than a few minutes, with some exceptions due to observability 
constraints which delay the satellite to slew towards the BAT
position (e.g., due to the small angular separation between the GRB and the Moon or the Sun). 

The on-board localization has an accuracy between several arcminutes
 (BAT only) to subarcseconds (UVOT). The immediate transmission to the ground
of all the \swift-acquired data via the Gamma-Ray Burst Network 
\citep{Barthelmy:1995lr} allows rapid (seconds to hours) follow-up 
with ground-based optical/IR telescopes.

In most of the cases presented in this paper, rapid follow-up
spectroscopic observations were triggered as soon as an \xray\ counterpart
position was delivered (XRT identified more than 98\% of the BAT GRBs).
Before the actual trigger is sent to the opportune telescope, we check the 
current telescope/camera set-up available, the visibility window at the telescope site, 
and weather conditions. A finding chart is usually provided to the telescope 
operator, using archival images (usually SDSS or USNO catalogues), 
which also helps to identify the afterglow. 
Once the trigger is sent to the telescope, usually via 
Target of Opportunity (ToO) programs, 
a long acquisition image of the field is acquired. 
In the eventuality that a ``new'' source has been found inside the XRT error circle
(by comparison with the finding chart), a spectroscopic sequence is executed.

In other cases, especially for \hiz\ bursts, robotic, real-time follow-up
by different facilities have provided similarly accurate identification
within the first hour, using redder filters than the ones available on
\swift/UVOT \citep[e.g. the GROND and RAPTOR instruments,][]{Greiner:2008fk,Vestrand:2002uq}. 
Thanks to the prompt responses, different groups have been 
able to obtain spectroscopic observations of the optical afterglow
when it was 
bright enough to detect absorption lines, which generally yields a
definitive estimate of the GRB redshift. 
For this purpose, the identification of fine-structure lines 
represents a secure determination of the GRBs host galaxies 
and the GRB circumstellar environment \citep{Prochaska:2006vn}.
Other secure identification is the presence of a Damped Lyman-$\alpha$
system, which also has been signature of a typical \hiz\ GRB host galaxy environment.
Whenever these features are not present, we assume that the 
higher redshift system of absorption features
from several ionized transitions corresponds to the GRB redshift, but 
obviously does not guarantee that these features do not rise, instead,
in foreground objects along the GRB line of sight  \citep[see for example
GRB 071003,][]{Perley:2008zr}.

For GRBs that occurred before the launch of the \swift\ satellite
a similar procedure was followed: the gamma-ray identification was 
made by high-energy facilities in orbit (BATSE, HETE-2 spacecrafts), while \xray\ 
observations were due to slower response missions (like Beppo-SAX), 
which required several hours to repoint. 
This inevitable delay propagated often to a late follow-up
observation, which, in few cases, led to a spectroscopic
sequence to executed.

For our parent sample, we set out to obtain the optical afterglow
spectra for all GRBs with reported redshifts, restricted as follows. 
Because of our interest in detecting \MgII\ lines, and because most 
spectrographs have wavelength coverage beginning at $\sim4000$\AA\ (or
poor UV sensitivity), we require 
the \MgII\ doublet rest-frame wavelength to be redshifted beyond 
this limit.
This leads us to include all the publicly available spectra obtained 
from GRBs with  redshift higher than $z_{\rm GRB}=0.5$.

The spectra analyzed in this paper were obtained with facilities
across the world, including the Gemini Observatory, Keck Observatory,
and Very Large Telescope.
Many of these data were obtained by our respective research teams,
although several tens were taken from public data archives or were
kindly contributed by members of the community.
Table~\ref{tab:sample} lists all of the sightlines with reported GRB
redshifts where we were able to retrieve a spectrum.
The last column lists the literature references for GRB afterglow
spectra that have been previously published.

A small sample of seven GRB spectra, as mentioned in \S\ref{sec:intro}, was 
obtained during the pre-\swift\ era: these GRBs were discovered by
non-GRB 
dedicated missions, like the Interplanetary Network (GRB 000926),
Beppo-SAX (GRB 010222) and
HETE-2 (GRB 020813,GRB 021004,GRB 030226,GRB 030323,GRB 030429), and 
followed from the ground several hours (if not days) after the events were discovered.
Neverthless, these data have sufficiently high quality to be included
in our work.

Our large dataset consists of a total of \ntot\ GRB afterglows observed by different facilities and 
instruments, including the Gemini
telescopes with the GMOS instruments (46 spectra), the Very Large Telescope equipped with 
FORS1, FORS2, UVES, and the X-Shooter spectrographs (55), the North Optical Telescope
(NOT) with the ALFOSC camera (6), the Keck telescopes with the HIRES,LRIS and
ESI instruments (13), and the Magellan Clay telescope with the MagE spectrograph (1). 
Finally, we also include two spectra obtained by the KAST spectrograph 
mounted on the 3-m Shane Telescope at Lick Observatory and several 
from the Magellan and the Telescopio Nazionale Galileo (TNG).
The spectral resolution of these data ranges from 450 km/sec (or $\sim13$\AA\ , NOT/ALFOSC) to 7 km/sec ($\sim0.13$\AA, HIRES/UVES). 
This large variety of data give us the opportunity to test different subsamples 
drawn from the overall \ntot\ GRBs. 
All the data presented are part of a public repository of GRB spectra\footnote{\tt http://grbspecdb.ucolick.org/}.


The VLT sample has been obtained with a specific set of observational criteria,
making the best use of the different instruments available and the 
technical improvement for ToO observations, like 
the \emph{Rapid Response Mode (RRM)}, which allows
the observer to execute the required observation remotely and without
a
major intervention by the telescope operators \citep[see][]{Fynbo:2009lr}.

Most of the FORS1/2 data are part of the catalog presented in \cite{Fynbo:2009lr} and \cite{de-Ugarte-Postigo:2012kx}, 
while most of the high-resolution ones (HIRES, ESI, UVES) were already 
published in single-GRB papers or as part of \cite{Vergani:2009fj}. 
A large fraction of the Gemini spectra are presented here for the first time, and are the 
result of our group's follow-up efforts over the last 7 years 
\citep[see also ][]{Cucchiara:2010fk}. We encourage the readers to refer to the reference 
in the last column in Table \ref{tab:sample} for the data reduction procedures and the original published papers.
In the following sections we will briefly review the reduction procedure for the Gemini and the X-Shooter
data.

\subsection{Gemini sample} 
These datasets are part of several follow-up programs for which Target of Opportunity time was awarded between 2005 and 2011. 
All the data included were obtained with the Gemini Multi-object spectrographs \citep[GMOS;][]{Hook:2004fj}. The typical observation sequence consists of two spectra in two dithered positions along
the slit (usually 1\arcsec\ wide) in order to facilitate sky-line subtraction. 
Immediately before or after the science 
frames, a ThAr lamp is observed and a flat field is obtained in order to allow data reduction ``on-the fly''.

We used the {\tt GEMINI/GMOS} data analysis packages under the IRAF\footnote{IRAF is distributed by the National Optical 
Astronomy Observatory, which is operated by the Association for Research 
in Astronomy, Inc., under cooperative agreement with the National Science 
Foundation.} environment in order to perform
the basic reduction, flat fielding and wavelength calibration. Cosmic
rays were identified and replaced by a median of the surrounding 
pixels which were not flagged as bad pixels. For this purpose we
used the {\tt lacos\_spec} tool \citep{van-Dokkum:2001fk}. 
Finally, one frame was subtracted from the other to 
remove the strongest skylines. 
This procedure provides good results at $\lambda < 8000$\AA, but
leaves significant residuals 
in the reddest portion of the spectra where the \gmos\
spectrographs suffer substantial CCD-fringing. 
Therefore, the extracted error arrays associated with the Gemini-GMOS
data reflect these higher-noise patterns at longer wavelengths.

One dimensional spectra were then extracted using the IRAF {\tt APALL}
tool and coadded weighting each spectrum by the inverse of its variance spectrum 
in order to increase the $S/N$ of the final result.  
The {\tt APALL} package also produces a 1-d array with the poissonian
statistical error and the 1-d sky background (estimated in regions selected 
far from the object trace, so to avoid any spurious contamination). 
These last two arrays have been summed in quadrature to obtain the final 
error array per pixel. In some cases we assess the quality of the extracted
error array with the estimated RMS of the data-array and modified
the latter in order to fully account for the poissonian fluctuations in the 
actual data.
Finally, using the {\tt splot} routine we estimated
signal-to-noise over the whole wavelength range 
(also reported in Table \ref{tab:sample}).


\subsection{X-Shooter data}
Data for GRB 090926A and GRB 100418A, were obtained via the 
ESO Archive\footnote{Based on observations made with ESO Telescopes at
  the La Silla Paranal Observatory under programs ID 60.A-9427(A) and
  ID 085.A-0009(B)} and reduced with version 1.3.7 
of the X-Shooter pipeline \citep{Goldoni:2006fk} in physical mode.
The spectra in the UVB and VIS arms were used for the redshift
pathlength estimate as well as the \MgII\ search. We do not use NIR
arm due to the high level of contamination from 
skylines in the infrared. Furthermore, the infrared sample of QSO spectra, largely obtained
by \cite{Matejek:2012bh}, are still small compared to the large
compilation from SDSS.

\subsection{Subsamples}
\label{sec:subs}
The set of spectra listed in Table~\ref{tab:sample} comprises our
full sample for analysis which we refer to as \samF.  This sample
maximizes the survey path for \ion{Mg}{2} absorption along GRB
sightlines.  From this parent sample, we consider several subsamples
for the same analysis.  Most important is the independent subsample
(\samI) which ignores all of the data analyzed in the original
paper of P06.  We focus first and foremost on this subsample to
perform a complementary study.  
In addition, we consider two other subsamples which cut the data
according to spectral resolution: we combined all the high-resolution 
spectra, obtained with echelle or echellette
spectrographs (ESI, HIRES, MagE and UVES) 
in \samH\ and, all other data in \samL.
These are summarized in Table~\ref{tab:subsam}.

\section{Survey Path}
\label{sec:redpath}

The starting point of a survey for intervening absorption-line systems
is to estimate the redshift path density $g(z)$.  This function
expresses, as a function of redshift, the number of unique sightlines for which an absorption line
could be detected in the survey to a limiting equivalent width. 
In practice, one determines for each spectrum those
regions that have sufficient S/N and are free of strong blending by
terrestial or intrinsic gas.  These specific windows define redshift
intervals $j$,  $[z_1^{i}, z_2^{i}]_j$, for the $i$th quasar (or GRB) where $g(z)=1$
within each window and zero otherwise.  By integrating $g(z)$ across the full
spectrum, one recovers the redshift path $\Delta z_i$ covered by the source.

To properly determine $g(z)$
for each sightline, several issues must be considered 
to minimize systematic effects that could bias the search. 
First, we exclude from the search the wavelengths (or the redshift ranges) 
that fall in the atmospheric telluric 
bands, which heavily absorb the afterglow flux, rendering it very
difficult to identify
any features (intrinsic to the GRB, or QSO, host or intervening). Since some of our spectra extend
towards the near-IR regime, we also consider atmospheric absorption at these wavelengths.
A complete list of the excluded regions is presented in Table \ref{tab:exclreg}.

\begin{figure*}[t!]
\epsscale{1.0}
\includegraphics[scale=0.60,angle=0]{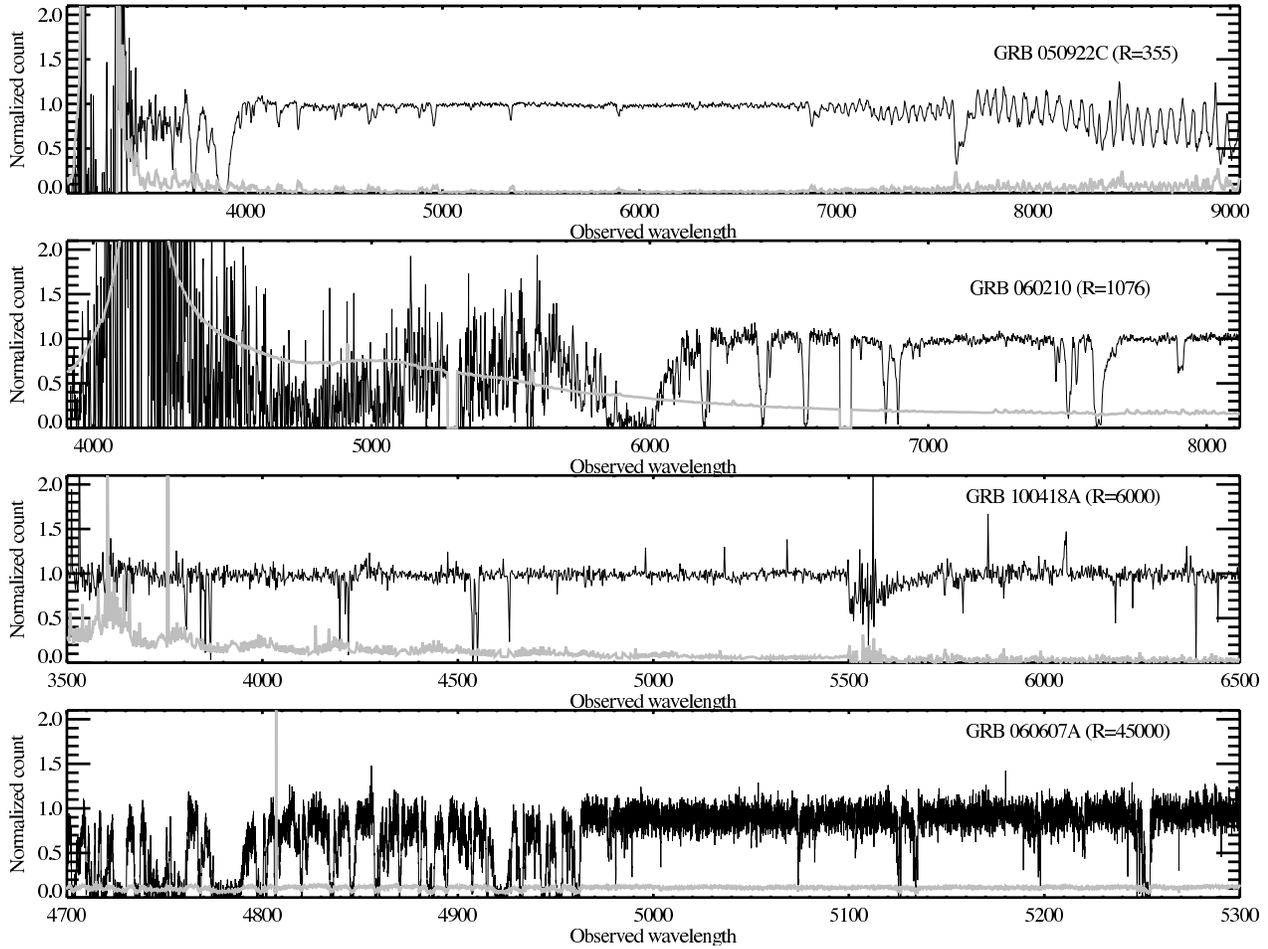}
\caption{Comparison between 4 different GRB spectra obtained with different spectrographs and different
resolving power. From top to bottom: GRB 050922C observed with
ALFOSC; GRB 060210 observed with Gemini/GMOS;
GRB 100418A observed with the UV arm of VLT/X-Shooter; section of GRB 060607A
observed with VLT/UVES. In
all the panes the grey curve represent the associated $1\sigma$ error spectrum.} 
\label{fig:stack}
\end{figure*}

Second, GRB afterglow spectra exhibit strong absorption lines belonging to ionic species 
located in the progenitor environment and up to tens kpc along the line of sight. 
The number of detected host features varies depending on the
brightness of the afterglow, the properties of the host galaxy, 
the signal-to-noise
and the resolution of the spectrograph (see Figure \ref{fig:stack}).
\cite{Christensen:2011fk} created a high-$S/N$ composite spectrum using 66 afterglow
spectra obtained with low and mid-resolution spectrographs.
Strong absorption lines were identified as well as weak ones previously 
undetected in the individual spectra. Since most of these 
lines are common in GRB host galaxies
we compile a sublist 
of these absorption features to be excluded in our 
redshift pathlength calculation. We included also some of the most common
fine-structure transitions.
To be more conservative, 
a region equivalent to one half-resolution element both blueward and redward of
the observed central wavelength of the considered transition (as
set by the redshift of the host galaxy) has had $g(z)$ set to zero.
In the cases of high-resolution spectra, the minimum size of the masked region 
is 200~km/s.
A complete list of these features is also presented in Table \ref{tab:exclreg}.

Finally, for some of the spectrographs (e.g.\ UVES, GMOS), the
spectral coverage is non-contiguous due to 
 gaps between detector chips and/or the use of multiple cameras. 
Regions without data were simply masked in the $g(z)$ evaluation.
In several cases, regions beyond $\sim8000$\AA\ were heavily effected 
by fringing, even after correcting for it in the data processing.
We opted for a visual inspection of the data and decided on a case-by-case 
which regions needed to be excluded for the search.

\begin{figure*}[t!]
\epsscale{1.0}
\includegraphics[scale=0.60,angle=0]{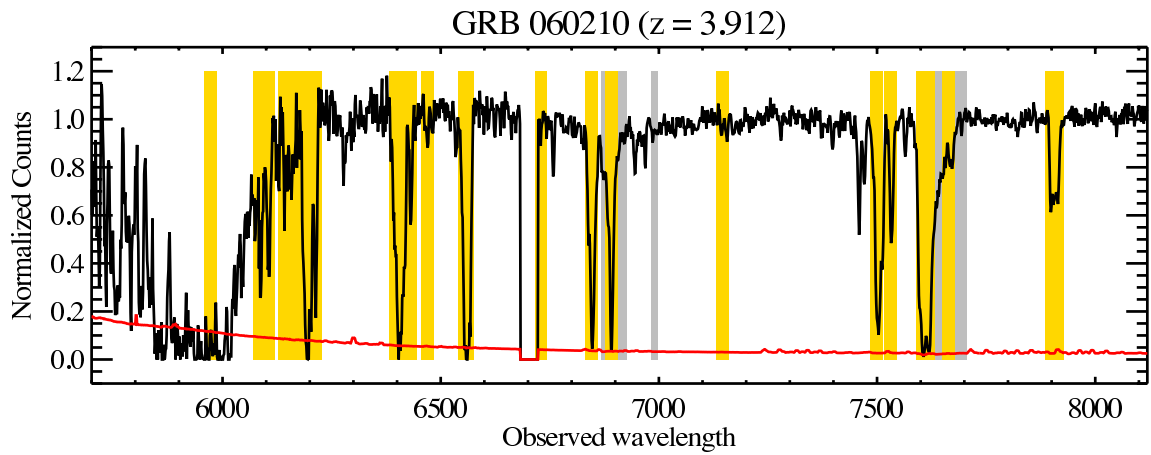}
\caption{GRB 060210 Gemini spectrum. This example shows our excluded
  regions for the purpose of estimating the survey path for
  intervening \ion{Mg}{2} absorption
  taking into account the host absorption lines (in gold, from the tabulation of
\cite{Christensen:2011fk})  as well as telluric lines (in gray). The red curve is the 1$\sigma$ spectrum associated 
with the data.
} 

\label{fig:regions}
\end{figure*}

As an example, in Figure~\ref{fig:regions} we present the 
Gemini/GMOS spectrum of GRB 060210, where 
all of the masked regions are indicated. 
It is clear from Figure~\ref{fig:regions} that the maximum
redshift, $z_{max}$, 
allowed for our 
intervening \MgII\ search is dictated by the host galaxy redshift, in particular, we
 begun our search starting  1500 km s$^{-1}$ blueward the 
corresponding \MgII\ feature (or $z_{max}=z_{GRB}-0.015$).
Also, the minimum, $z_{min}$,  is indicated either by the bluest wavelength 
covered by the spectrograph or, as in the case of GRB 060210, by the presence 
of the \Lya\ feature (at rest-frame  $\lambda_{\rm Ly\alpha}^{\rm rest}=1215.67$ \AA) such that 
$z_{min}= (\frac{\lambda_{\rm Ly\alpha}^{\rm obs}}{2796}-1) +0.05$, 
where $\lambda_{\rm Ly\alpha}^{\rm obs} = \lambda_{\rm Ly\alpha}^{\rm rest}(1+z_{\rm
  GRB})$
(or equivalently $\sim5000$ km s$^{-1}$ redward the \Lya\ feature). 
We do not extend the search for intervening \ion{Mg}{2} into the \Lya\
forest and we avoid the (typically) very strong damped \lya\
absorption profile of the GRB host galaxy.

Once these regions have been excluded we determined the $5\sigma$ equivalent width 
limit per pixel, using the variance spectrum associated to each
object (shown as red in Figure~\ref{fig:regions}), 
considering a simple gaussian profile of $FWHM= S/2.35$, where $S$ is the 
resolution element. 
At each unmasked pixel in the spectrum, we query whether the $5\sigma$
equivalent width limit exceeds a given rest-frame survey limit for
\ion{Mg}{2}~2796 (e.g.\ 1\AA).  If the limit is satisfied, we query
whether the corresponding \ion{Mg}{2}~2803 line lies in an unmasked
region.  If both of these criteria are satisfied, $g(z)=1$ for the
redshift interval covered by that pixel otherwise we set $g(z) = 0$.
This generally leads to a series of discontinuous redshift intervals
for the \ion{Mg}{2} survey, as listed in Table~\ref{tab:path}.

\begin{figure*}[t!]
\epsscale{1.0}
\includegraphics[scale=0.60,angle=0]{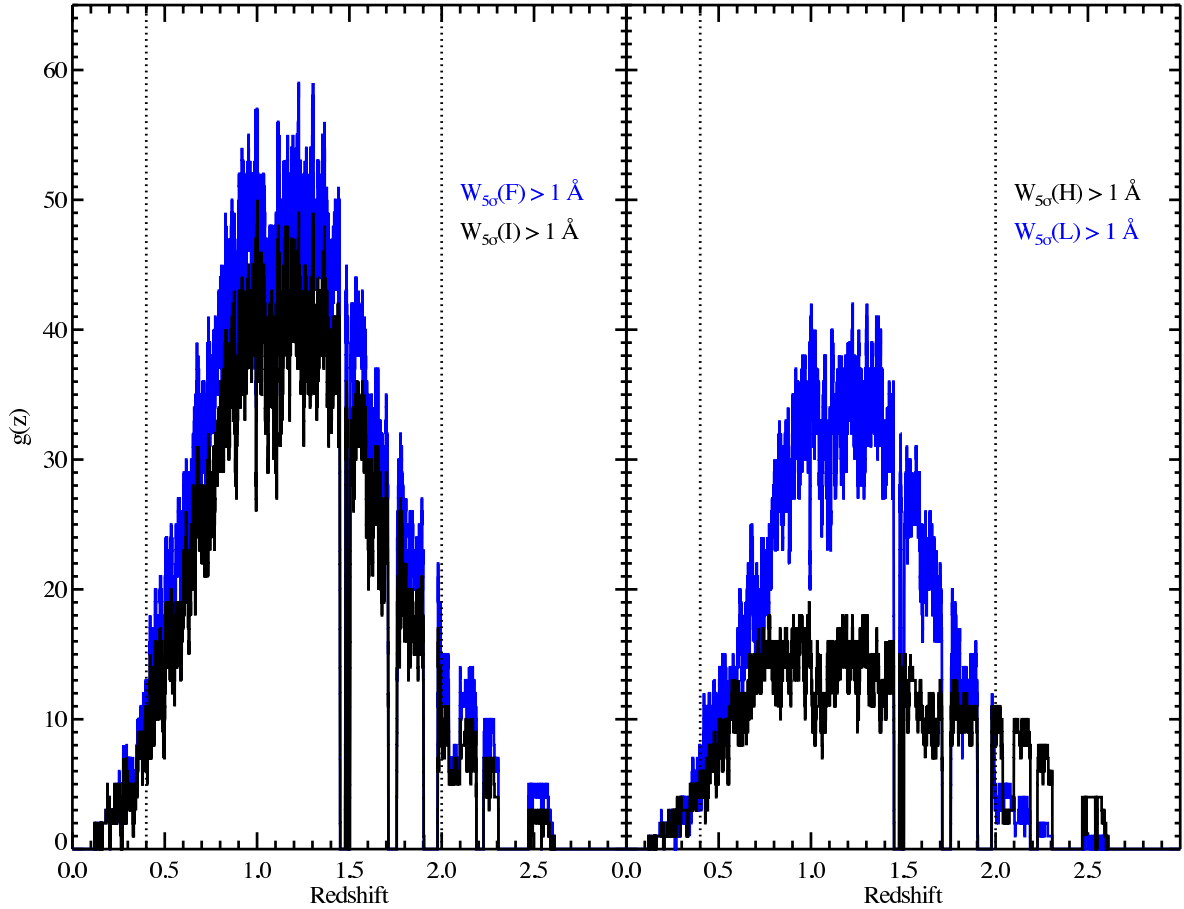}
\caption{\emph{Left:}Redshift path density for 1 \AA\  rest-frame equivalent widths at 5$\sigma$ detection limit for the \samF\ (blue) and \samI\ (black). Dotted vertical lines represent the quasar selection regions, where the \MgII\ doublet is detectable in the SDSS spectral coverage. \emph{Right:} similar plot for \samH\ and \samL.
  }
\label{fig:gz}
\end{figure*}

Figure \ref{fig:gz} presents the total redshift path density for \samF\ and
\samI, 
which represents the number of GRB sight-lines available for our \MgII\
search as a function of redshift.   These are shown for a limiting
rest-frame equivalent width of 1\AA\ at 5$\sigma$ confidence. 
It is immediately clear from this figure that we accurately excluded 
from our analysis the telluric lines regions (i.e.\ at $z_{\rm MgII} \sim 1.5, 1.7, 1.9,
$ and $2.4$). Also, the analysis is mainly performed where we have the majority of the
searchable path, in the $0.4 \lesssim z \lesssim 2.2$ interval range, due to the larger statistical sample from the GRB and the QSO samples.

The total $\Delta z$ of the survey crudely expresses its statistical
power.   This may be calculated by simply summing the 
$\Delta z_i$ values for each source.
For the full sample (F), a redshift pathlength of $\Delta z =55.5$ 
for the 1\AA\ equivalent width limit.
For the independent sample (I), we find \dz\ $ =44.9$.
The latter represents a $\sim$4 times  larger survey sample 
than P06.

\section{Identifying and measuring intervening \MgII\ absorbers}
\label{sec:search}

We described in the previous section the construction of the redshift path density
which defines, for each spectrum, the regions
where an intervening \MgII\ doublet may be detected at $5\sigma$
significance. 
Independent of this calculation, we have searched each sightline for the
presence of \MgII\ absorbers.

Using the algorithm for 
optimal extraction \citep{Horne:1986dq}, 
we constructed an equivalent width spectrum by convolving the
normalized data with a Gaussian profile with width set by  
the resolution of the spectrograph and weighting 
the flux at each wavelength by the associated variance. 
 From this array, we determined every feature satisfying a
$5\sigma$ detection threshold via an automatic procedure
similar to the \CIV\ doublet search performed in \cite{Cooksey:2010uq}.
We considered each line as a
possible \MgII\ absorber, and confirmed this association through the
presence of a proper \MgII\ doublet (both in velocity separation and
relative $W_r$). Finally, we inspected every candidate identified 
by this procedure visually, confirming the presence of a genuine doublet
with the additional identification of other common features (e.g. \MgI\, \FeII).
We also accurately measured the equivalent widths of the doublet components
via line profile fitting (see Table~\ref{tab:interv}). 
As sanity check, each GRB sightline was also manually
inspected by the lead authors in search of \MgII\ doublets that might be missed by the
automatic screening process. We found two doublets in addition to the 
candidates automatically identified which may be \MgII\ features (see \S \ref{sec:com}
for our completeness analysis).
Also,  we found some doublets which were 
misidentified as \MgII\ doublets: in reality these features were 
host galaxy fine-structure transitions or other metal lines belonging to other intervening systems
(e.g. GRB 061121).
We estimated that our total redshift pathlength  would be decreased of a factor  
of $\lesssim6 \%$ if we would have masked also these features, therefore 
we prefer not to exclude these spectral regions to preserve a maximum searchable path.
Again, our visual inspection prevent these features to be accounted in our
\MgII\ search.

\begin{figure*}[t!]
\figurenum{10a}
\epsscale{2.5}
\plotone{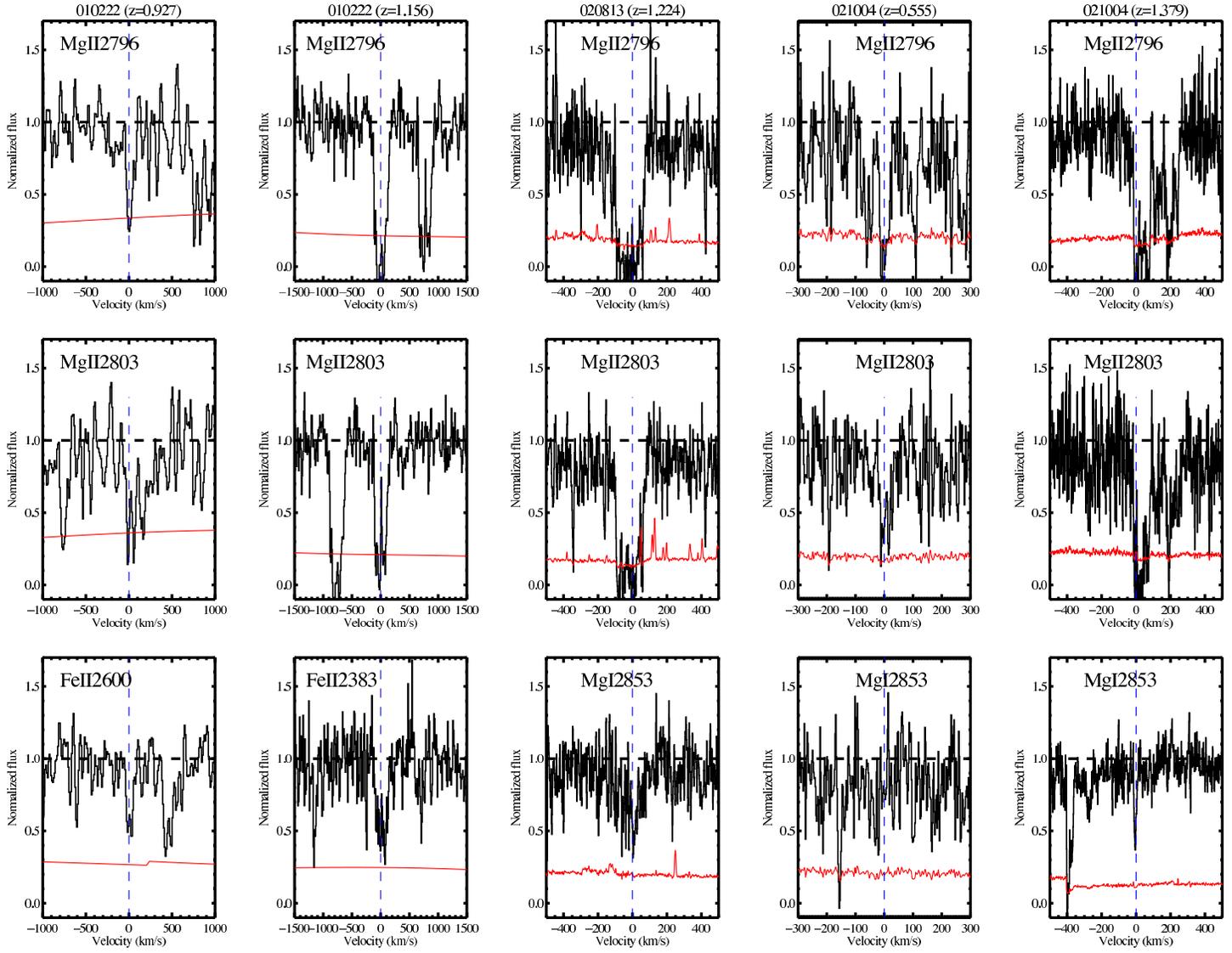}
\caption{Line profiles of strong \MgII\ transitions along GRB
  sightlines.
} 
\label{fig:lines1}
\end{figure*}

\begin{figure*}[t!]
\epsscale{2.5}
\figurenum{10b}
\plotone{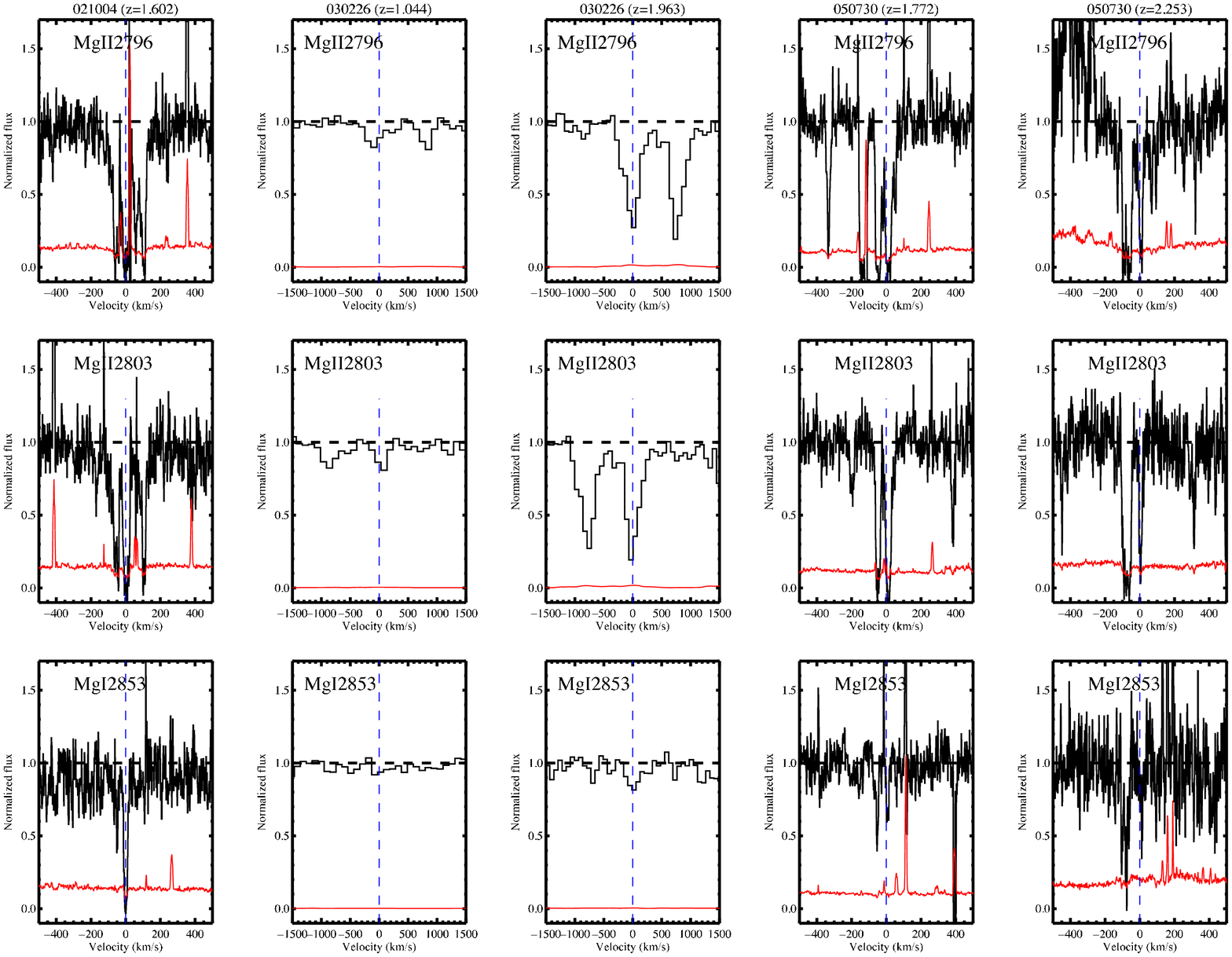}
\caption{}
\figurenum{9b}
\label{fig:lines2}
\end{figure*}

\begin{figure*}[t!]
\figurenum{10c}
\epsscale{2.5}
\plotone{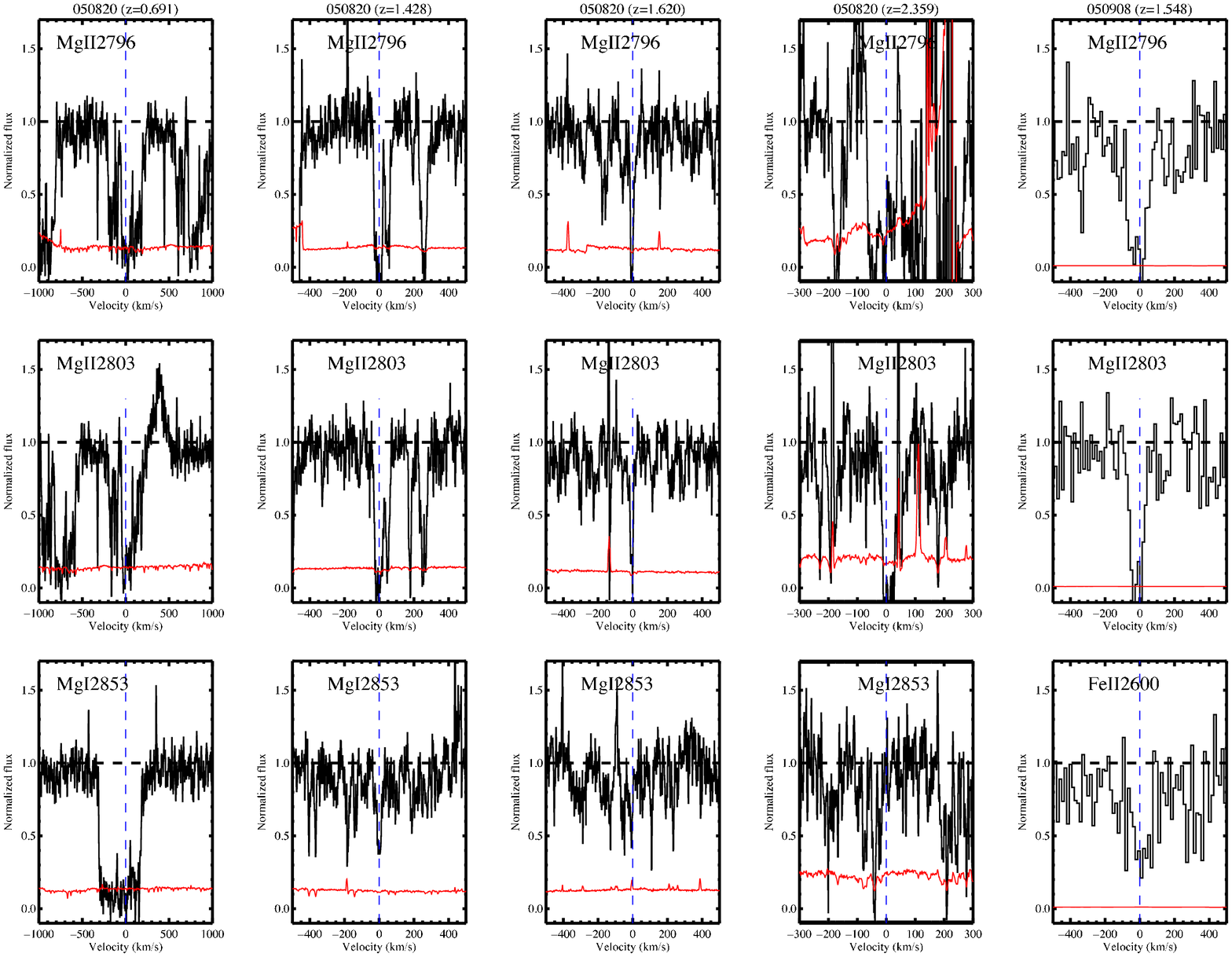}
\figurenum{9c}
\label{fig:lines3}
\caption{}
\end{figure*}

\begin{figure*}[t!]
\figurenum{10d}
\epsscale{2.5}
\plotone{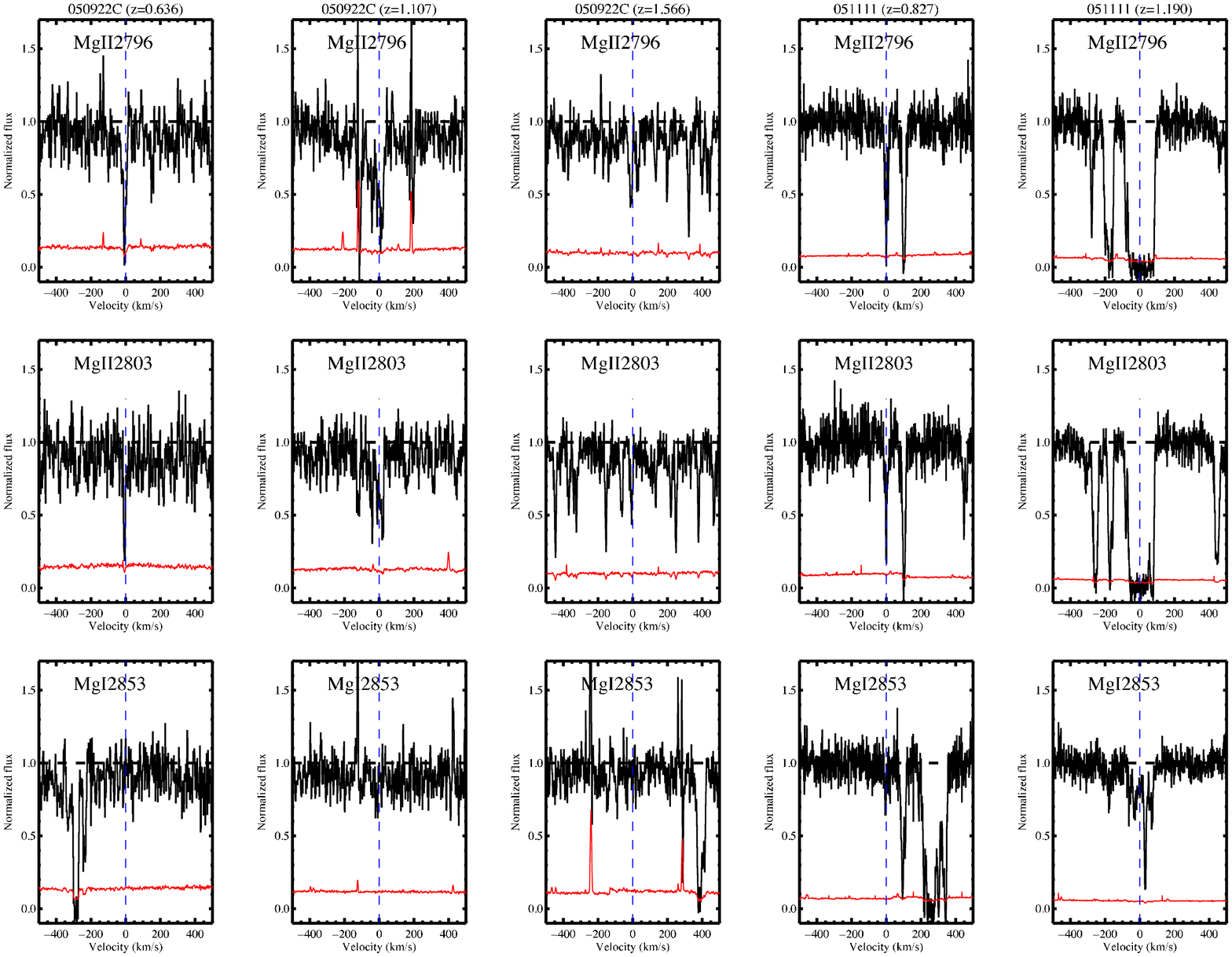}
\label{fig:lines4}
\caption{}
\end{figure*}

\begin{figure*}[t!]
\figurenum{10e}
\epsscale{2.5}
\plotone{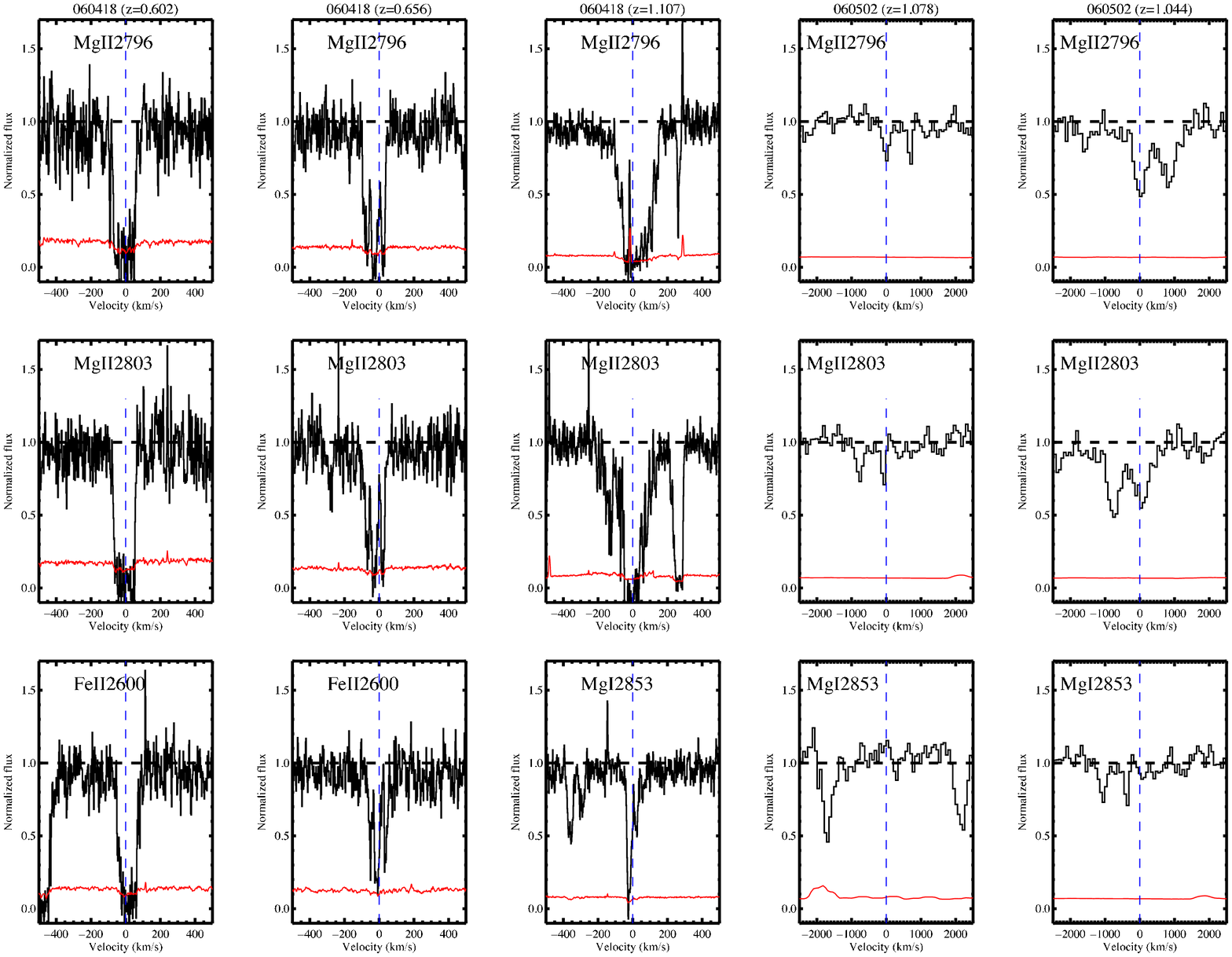}
\label{fig:lines5}
\caption{}
\end{figure*}

\begin{figure*}[t!]
\figurenum{10f}
\epsscale{2.2}
\plotone{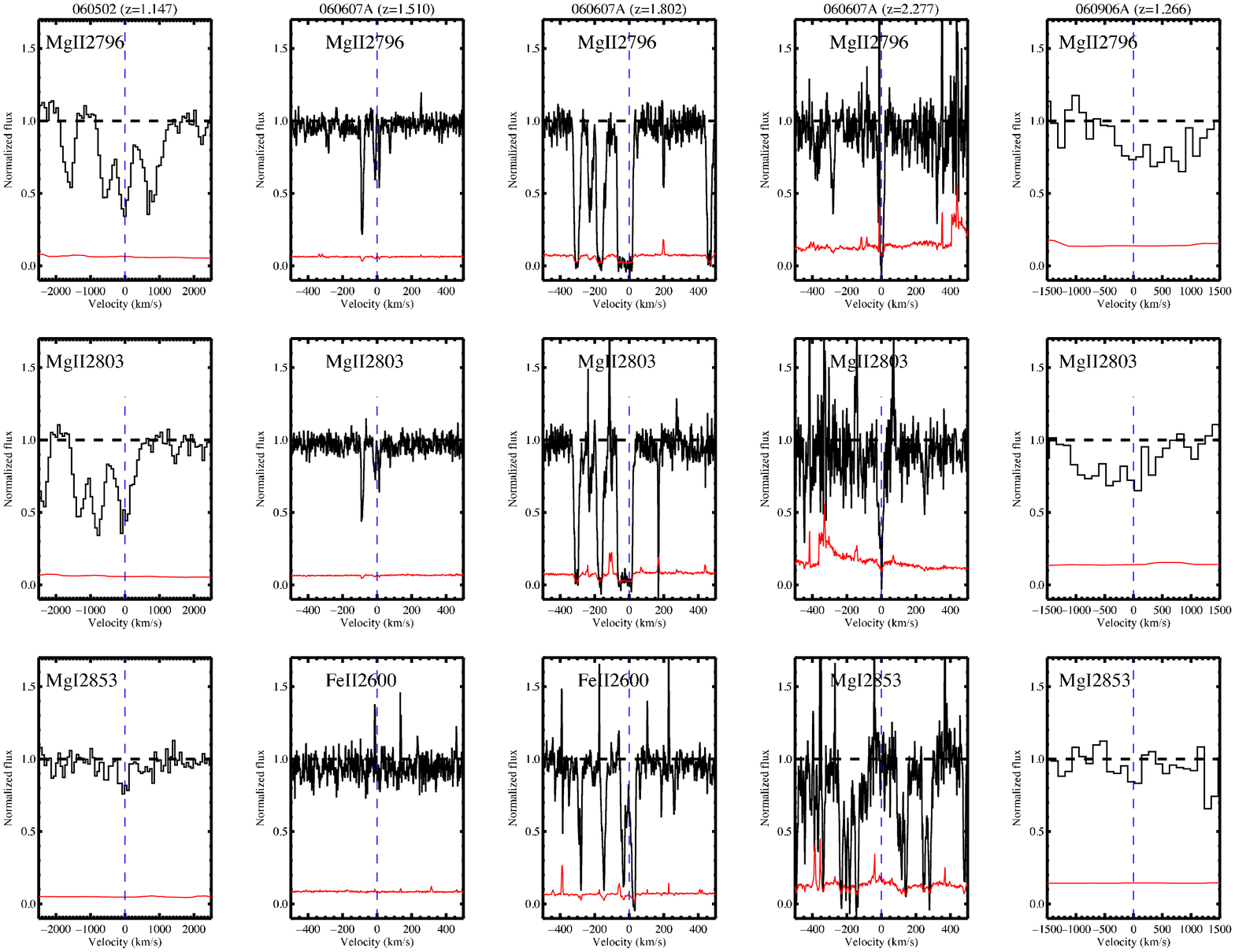}
\label{fig:lines6}
\caption{}
\end{figure*}

\begin{figure*}[t!]
\figurenum{10g}
\epsscale{2.5}
\plotone{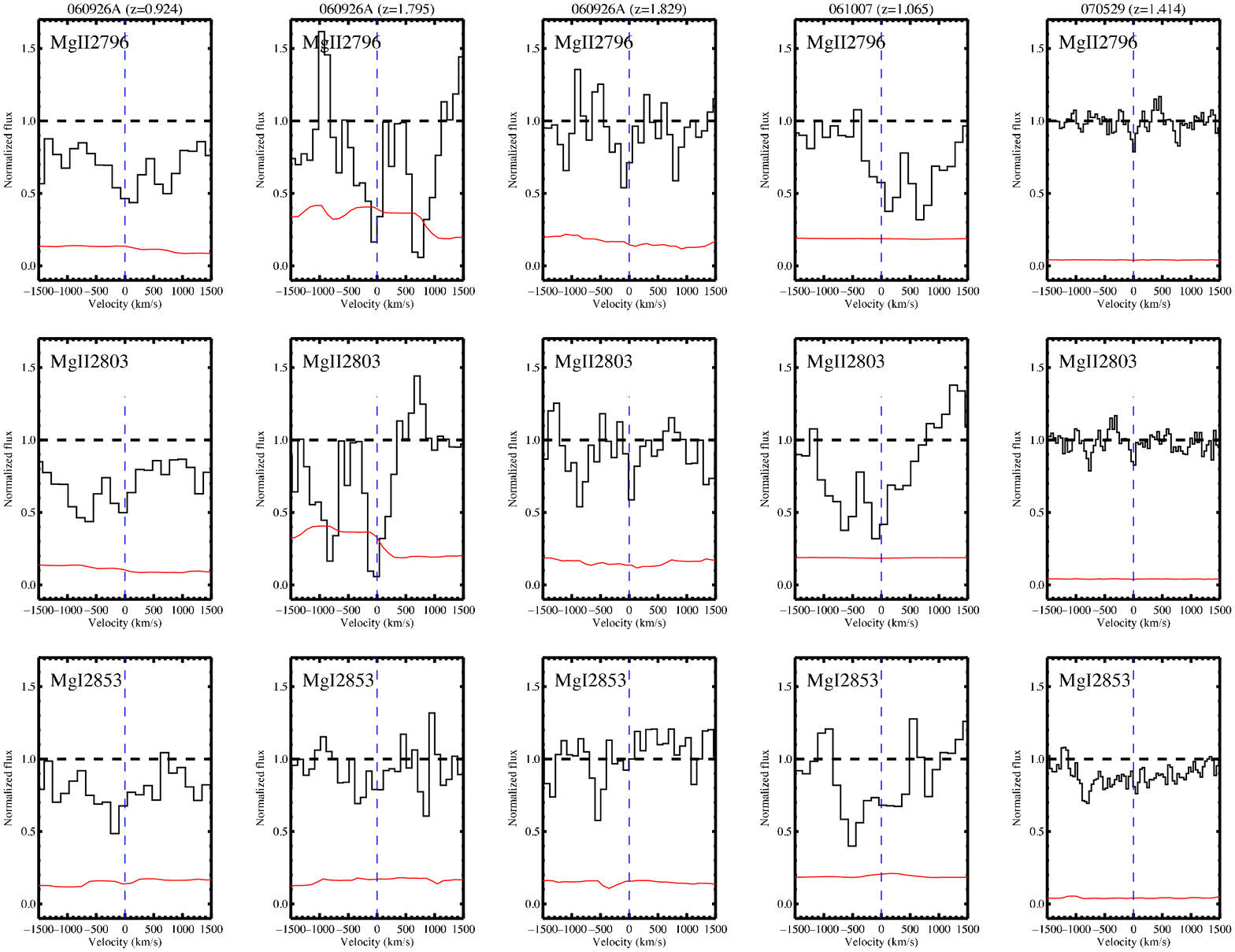}
\label{fig:lines7}
\caption{}
\end{figure*}

\begin{figure*}[t!]
\figurenum{10h}
\epsscale{2.5}
\plotone{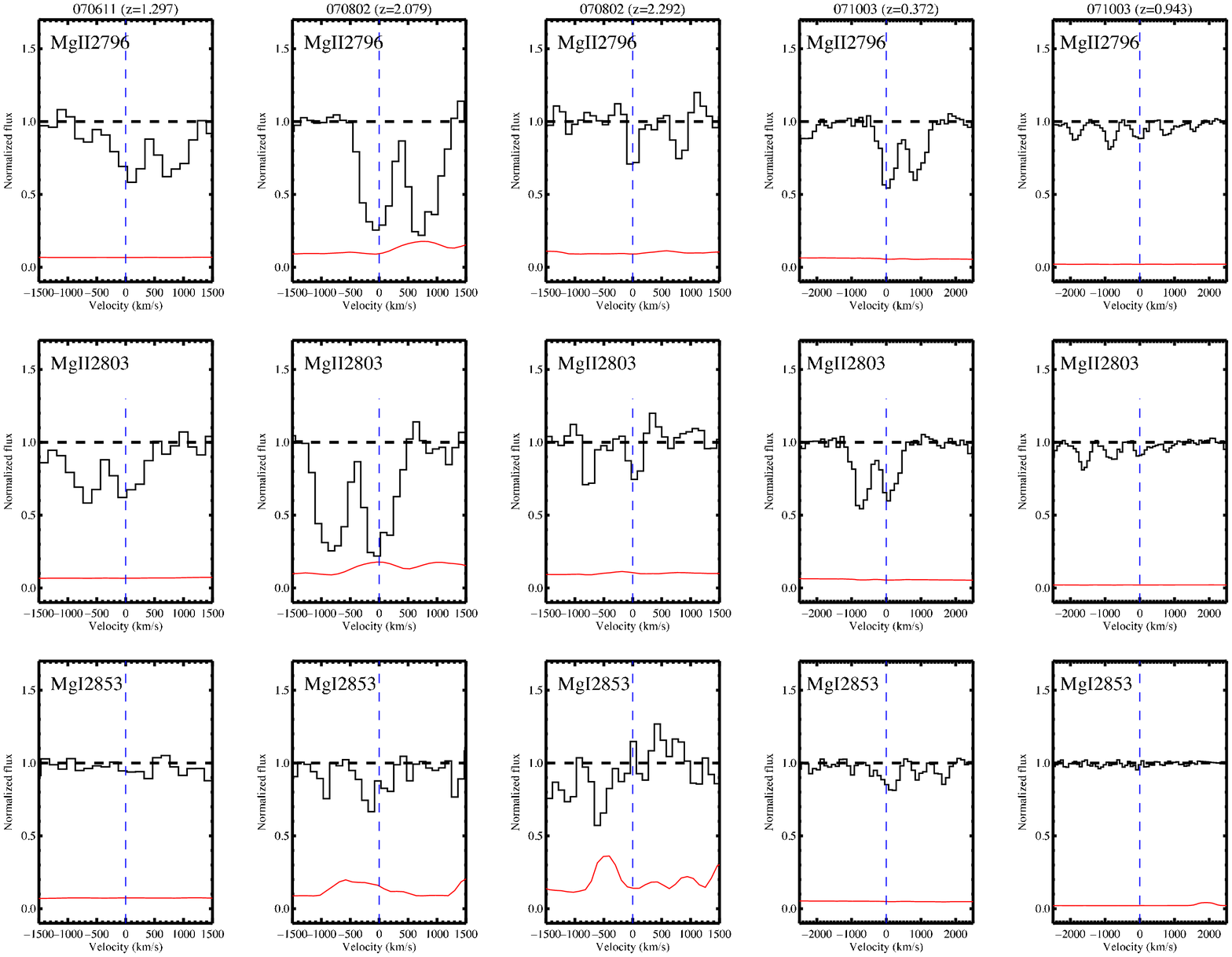}
\label{fig:lines8}
\caption{}
\end{figure*}

\begin{figure*}[t!]
\figurenum{10i}
\epsscale{2.5}
\plotone{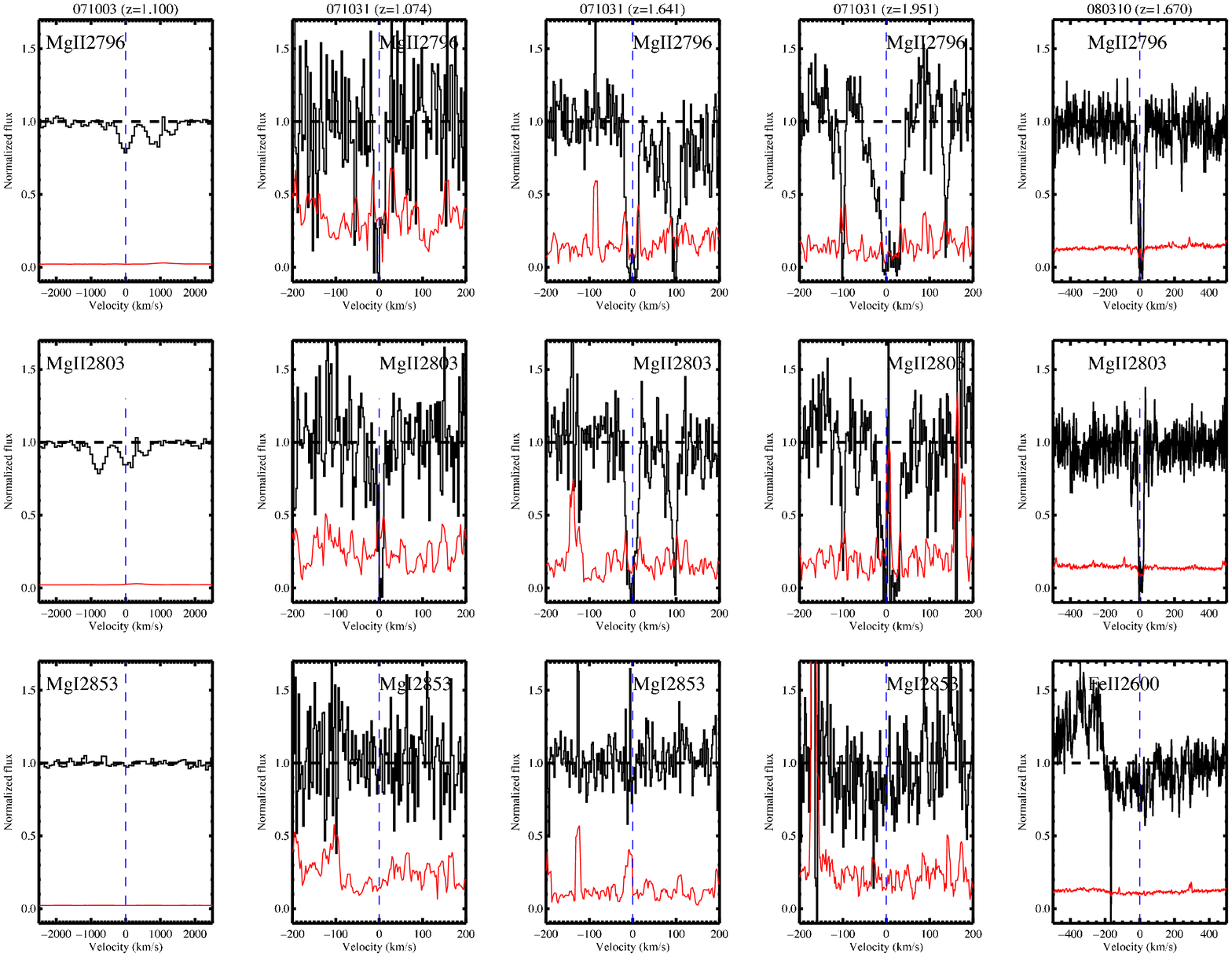}
\label{fig:lines9}
\caption{}
\end{figure*}

\begin{figure}[t]
\figurenum{10j}
\epsscale{2.5}
\plotone{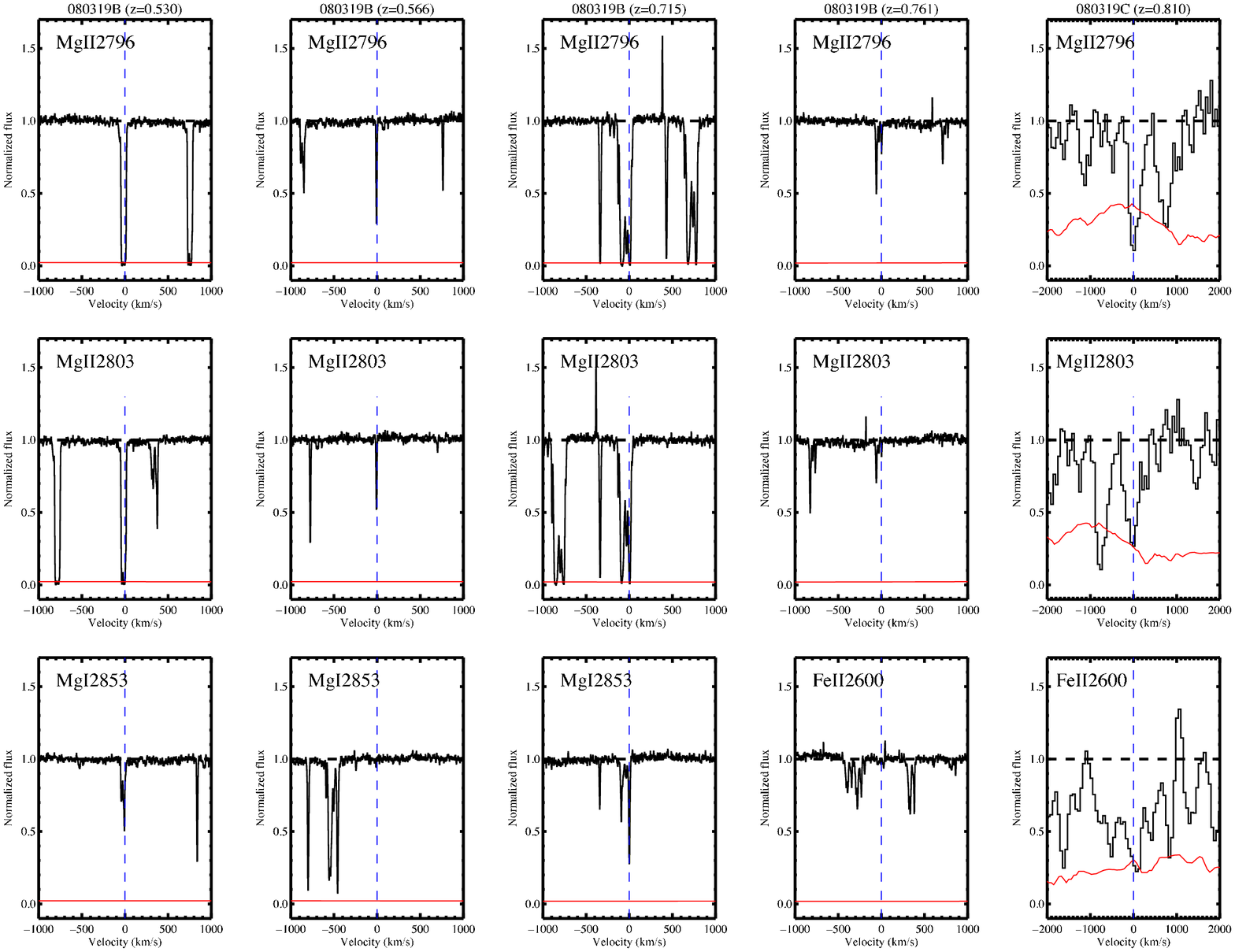}
\label{fig:lines10}
\caption{}
\end{figure}

\begin{figure*}[t]
\figurenum{10k}
\epsscale{2.5}
\plotone{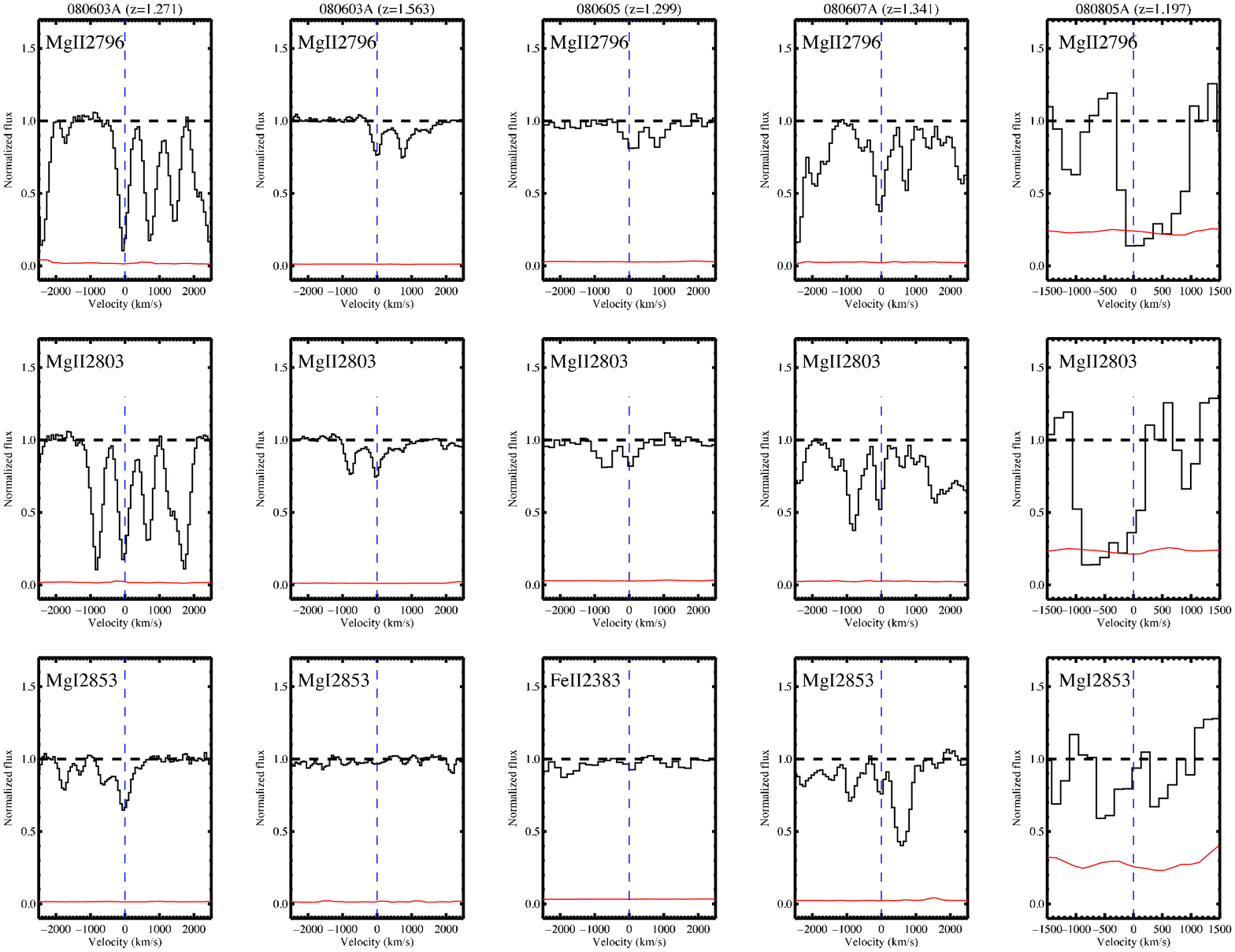}
\label{fig:lines11}
\caption{}
\end{figure*}

\begin{figure*}[t]
\figurenum{10l}
\epsscale{2.5}
\plotone{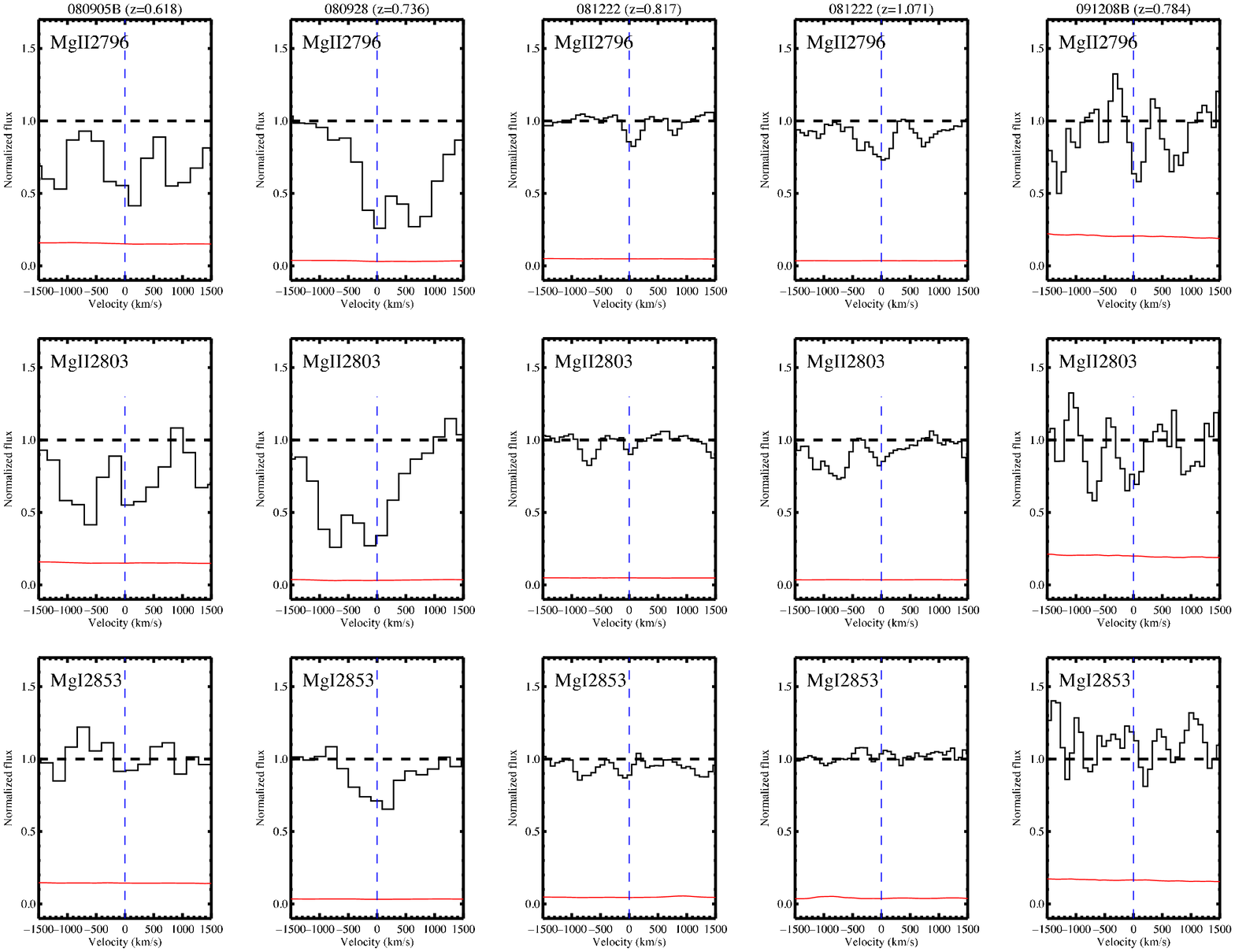}
\label{fig:lines12}
\caption{}
\end{figure*}

\begin{figure*}[t]
\figurenum{10m}
\epsscale{2.5}
\plotone{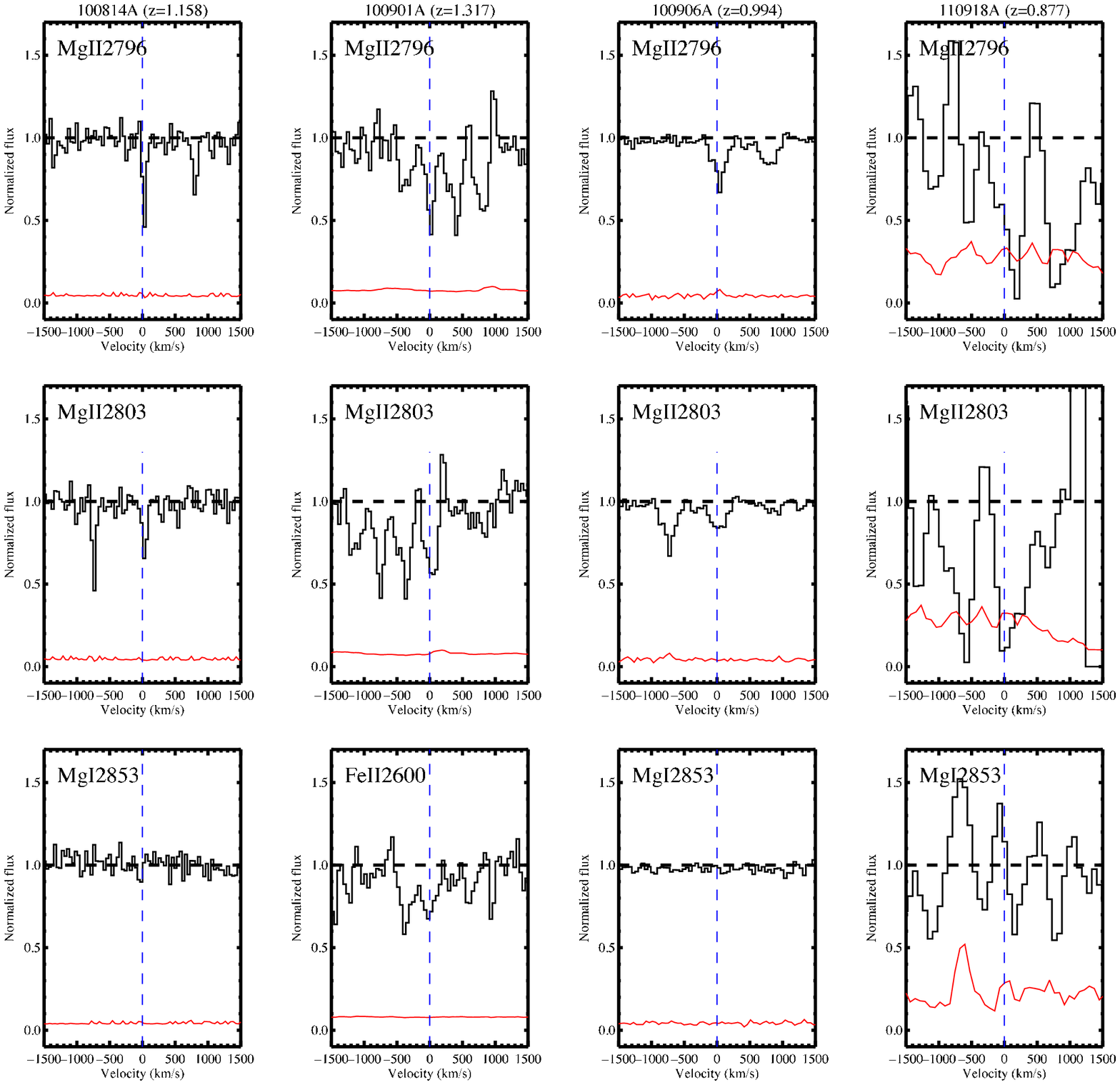}
\label{fig:lines13}
\caption{}
\end{figure*}

Table \ref{tab:interv} lists the \MgII\ systems that have been
discovered, and Figures \ref{fig:lines1}--\ref{fig:lines13} show the 
line-profiles of all the strong \MgII\ systems in combination, 
when available, with other metal features.
The $W_r$ values for the \MgII\ doublet were estimated by fitting the 
line-profiles with Gaussian profiles or in the case of line-black 
saturated transitions (e.g.\ high resolution
data) by pixel summation. 
We estimated the uncertainty in 
our $W_r$ values by summing the pixel-by-pixel 
variance in quadrature.

\subsection{Completeness estimate}
\label{sec:com}
\begin{figure*}[t!]
\epsscale{1.0}
\includegraphics[scale=0.60,angle=0]{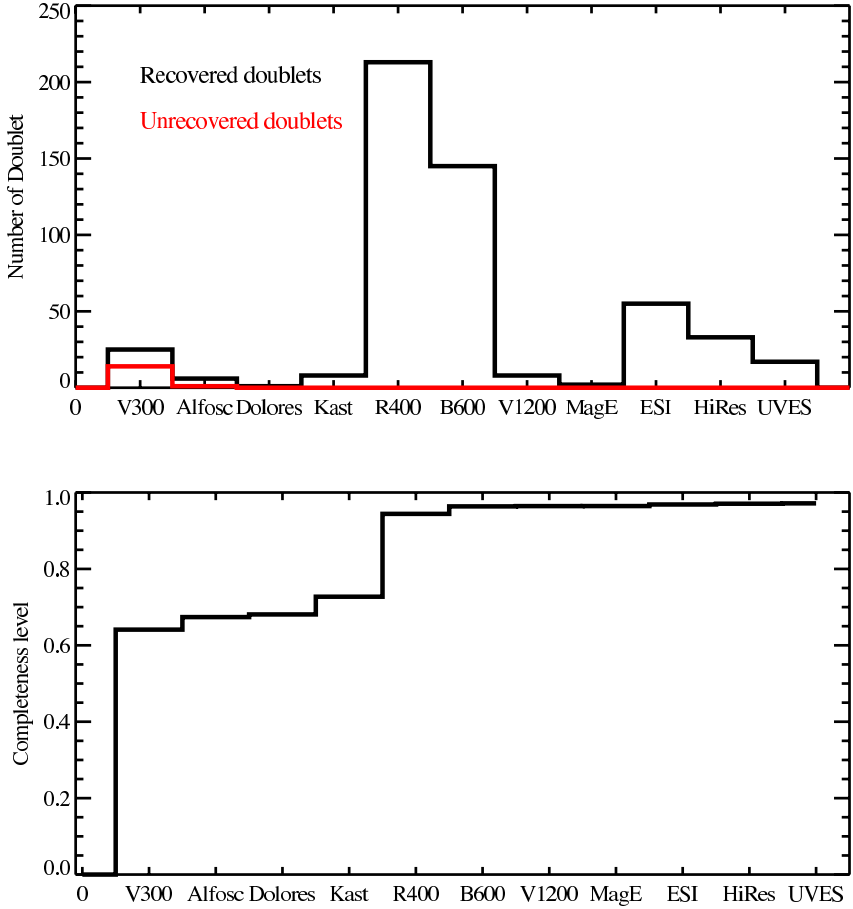}
\caption{\emph{Top:} Number of recovered (black) and unrecovered (red) strong doublets ($W_r > 1$\AA)
in our mock sample ordered by instrument resolution.  
  Most of the missed doublets, besides being in searchable regions of the spectra ($g(z)=1$),
   are missed or misidentified due to self-blending with other
  features (like other intervening systems metal lines, or wings of
  GRB host features). Therefore the automatic procedure usually fails to identify both members of the doublet
  due to the low-resolution of the instrument. 
   \emph{Bottom:} Completeness level ordered by
  spectral resolution. From our mock sample we derived a final
  completeness level of $\sim 98\%$.}  

\label{fig:complet}
\end{figure*}

It is important to note that the sample under consideration has been obtained 
by a large variety of  facilities and, also, GRBs have been observed at different epochs 
(meaning at different afterglow brightness) as well as 
with different atmospheric conditions. Therefore, it is worthwhile evaluating 
our completeness in finding very strong \MgII\ absorbers at the considered
$5\sigma$ confidence level.
To assess our completeness we inserted mock \MgII\ features 
into our spectra (taking into account the
S/N and the resolution of the original spectra)  and then we re-process these new datasets
via our automatic procedure. 
The injected features, in a number which is drawn by a poisson distribution centered on the 
expected number of absorbers (\gloz $\times \Delta z$) for the GRB sample,
 have random equivalent widths between 0.05 to 5\AA\ and
a maximum number of  seven sub-components, each with a range of 
doppler parameters $b=5-20\ $km/s. These features were inserted between
$z_{min}$ and $z_{max}$ as defined in Sec.\ref{sec:redpath} per each GRB.
We repeated this process 50 times per sightline
for a total of 5250 iterations.
We compared the number of injected strong features ($W_r \geq 1$\AA) that should 
be automatically identified because they were located in regions of the spectra were $g(z)= 1$, 
(accordingly to Sec. \ref{sec:redpath})
with the actual recovered list: we conclude that $\approx 98\%$ of the systems were 
correctly identified and detected as genuine strong \MgII\ doublets.
Figure \ref{fig:complet} shows the result of our completeness test: on the top panel 
we show the total number of absorbers correctly identified (black histogram) and not
(in red) depending on the instrument resolution. It is clear that the
lowest-resolution spectra (e.g.\ the V300 grating with the FORS1
spectrograph) have a low completeness level using our automated search
but that the other spectra give excellent results. 
For the missed absorbers in the lowest-resolution data, we find
that these doublets
are usually self-blended (resembling a single broad line) or they are blended with the profile wings  
of other lines (intervening systems metal lines or host galaxy features) preventing the automatic
identification of both doublet components. 

We examined again our original sample and we confidently retrieved only
two such cases: GRB 090812 and a possible absorber at $z=1.055$ and GRB 070110
with a possible doublet at  $z=1.5875$.
Nevertheless, including such features, which were 
not automatically recovered, does not effect our conclusions.

In the bottom panel of Figure \ref{fig:complet} we present our cumulative completeness 
level with increasing resolution. Again, as noticed previously, we reach $\sim97\%$ level
around the resolution of the Gemini-GMOS instrument (R400 grating,
$R\sim 1200$), whose spectra 
provide the best combination of Signal-to-Noise ratio and resolving power to properly identify
the population of strong \MgII\ absorbers characterizing the GRB intervening system 
population.
 
\section{Results}
\label{sec:result}

\subsection{Incidence \gloz}

\begin{figure*}[t!]
\epsscale{1.0}
\includegraphics[scale=0.60,angle=0]{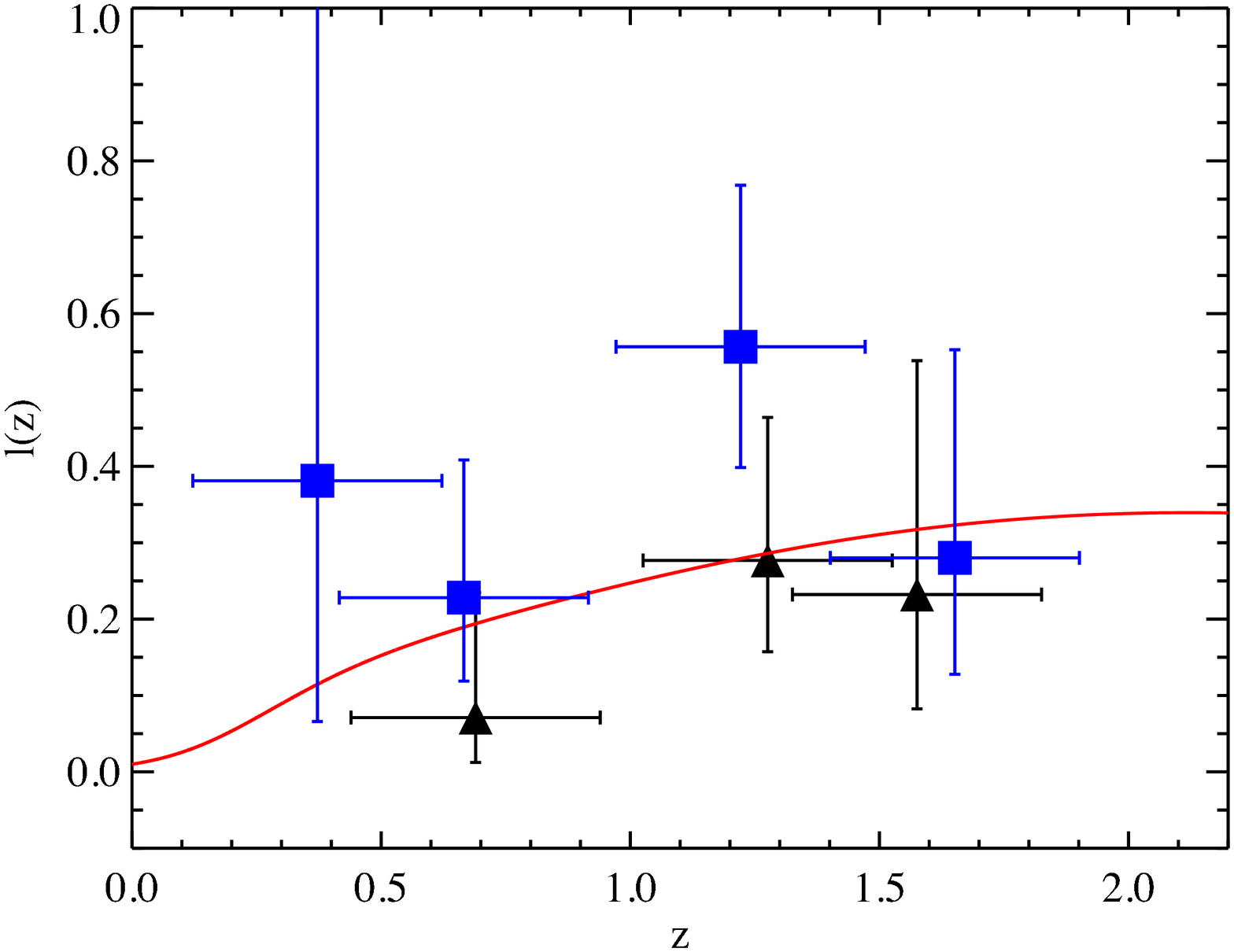}
\caption{$\loz$\ evolution of intervening \ion{Mg}{2} absorbers
  ($W_{2796} \ge 1$\AA) for our sample of GRB sightlines: triangles
  and square symbols refer to the \samI\ and  \samF\
  respectively. The red curve shows the
evolution of the \MgII\ incidence along quasar sightlines as recently computed by 
\cite{zhu:2012aa}.   We derive an average $\loz=0.20$ for \samI, in agreement
with the prediction, while $\loz=0.36$ for \samF, indicating a slight overabundance of 
absorbers compared to the QSOs.} 
\label{fig:nzevol}
\end{figure*}

Combining the results from the previous two sections, we may estimate
the incidence of \MgII\ absorption per unit redshift \gloz\ (also
referred to as $dN/dz$ or $dn/dz$).  The standard estimator for \gloz\
to a limiting $W_r$ is the observed ratio of the number of absorbers
discovered, $N$, having $W \ge W_r$ in a given redshift interval $[z_1, z_2]$ to the total
redshift pathlength searched, $\Delta z$, in that redshift interval

\begin{equation}
\loz = \frac{N}{\Delta z}
\end{equation}
with

\begin{equation}
\Delta z = \intl_{z_1}^{z_2} g(z) \, dz \;\;\; .
\end{equation}
Figure~\ref{fig:nzevol} presents our \gloz\ estimates for $W_r > 1$\AA\
\MgII\ absorbers, for  \samI\ and \samF\ restricted between $z=0.4$ and 2. 
The error estimates assume Poisson statistics for $N$ and correspond to 68\%\
confidence. 
The values for the GRB sightlines have roughly constant value
with redshift at $\mgloz \approx 0.18$ and $\mgloz \approx 0.36$ for \samI\ and \samF\ 
respectively.  

For comparison, we display a fit to the measured
\qloz\ values for $W_r \ge 1$\AA\ \MgII\ absorbers discovered along the
thousands of quasar sightlines drawn from the SDSS \citep{zhu:2012aa}.
These spectra have been chosen to have $S/N\gtrsim 15$, 
so to assure a high confident statistical sample of \MgII.
  
This quasar sample was searched for \MgII\ absorbers
at wavelengths redward of the strong \CIV\ quasar 
line ($\lambda_{rest}=1550$\AA) and blueward 
of the reliable response of the Sloan fibers up to the quasar's
\ion{Mg}{2} emission line.  


From Figure~\ref{fig:nzevol} it is evident that while the \samI\ 
follows the expected distribution derived from the 
QSO analysis, \samF\ still presents a modest excess of absorbers.
In the case of \samF, for instance,  $\Delta z=55.5$ and
the number of absorbers identified is $N_{obs}=20$ ($N_{exp}=13$).
Overall the $\loz_{\rm GRB,F}= 0.36\pm 0.09$, a factor $\sim1.5$ greater than 
the expected quasar  density of absorbers ($\loz_{\rm QSO,F} = 0.24$).
Considering the independent sample, which as mentioned in \S\ref{sec:subs} 
excludes all the lines of sight in PO6, we obtain $\loz_{\rm GRB,I}= 0.18 \pm 0.06 $. 
Following the same analysis, similar results are evident using the high-resolution and the 
low-resolution samples (\samH\ and \samL): in these cases we identify an 
overabundance of strong \MgII\ absorbers in the high-resolution sample, leading to a 
$\ell(z)_{H}=0.64$, a factor 2.6 larger then the expected
($\loz_{\rm QSO,H}=0.25$).
We summarize our analysis in Table \ref{tab:subsam}.

\begin{figure*}[t!]
\epsscale{1.0}
\includegraphics[scale=0.90,angle=0]{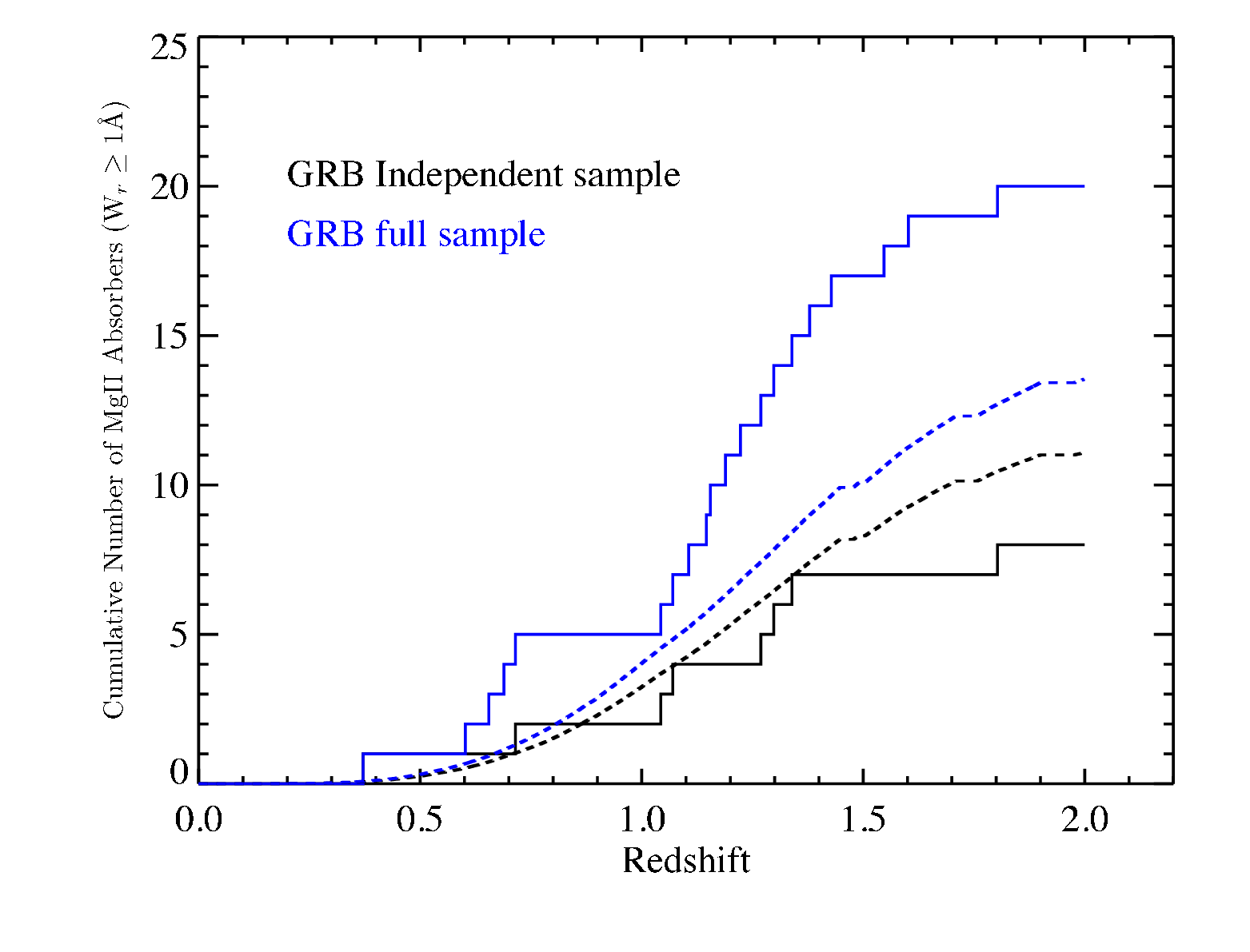}
\caption{Cumulative distribution of strong \MgII\ absorbers along
  GRB sightlines for \samI\  and
\samF\ (black and blue solid curves, respectively).  These are
compared to the predicted incidence based on measurement along QSO
lines of sight (dashed curves).  The independent \samI\ actually shows
fewer absorbers than expected while a modest excess remains in \samF.
Neither result corresponds to a statistically significant difference
from the QSO results.
}
\label{fig:nzcum}
\end{figure*}

Figure \ref{fig:nzcum} shows the cumulative distribution of \MgII\
absorbers detected from GRB \samF\  and \samI, together with the quasar
estimates.   We may compare these results against the predicted
cumulative distribution functions for a QSO survey with identical
search path to the GRB analysis by simply convolving
the GRB $g(z)$ with \qloz:

\begin{equation}
N^{\rm QSO}_{\rm cumul}(z>z') = \intl_{0.4}^{z'} \ell(z)_{\rm QSO}\
g(z) \, dz \;\; .
\end{equation}
It is evident that the full \samF\ exhibits a modest excess of 
$\sim 30\%$, but that the independent 
\samI\ shows no excess.
The new results for \samI\ do not confirm earlier
works which reported an excess of strong \ion{Mg}{2} absorption along
GRB sightlines.

\subsection{Monte Carlo Analysis}
\label{sec:Monte}
To assess the significance of these results,
in particular the observed excess for \samF, 
we perform a Monte Carlo analysis as follows.
First, we selected a set of 12700 SDSS quasars from  \cite{zhu:2012aa} 
that have a continuous $g(z) = 1$ redshift path density
from $z_{\rm min}$ to $z_{\rm max}$, where $z_{\rm min}$ is the
greater of 0.4 and $(1+z_{\rm QSO})\times
\lambda_{\rm CIV}^{\rm rest}/\lambda_{\rm MgII}^{\rm rest}$
and $z_{\rm max}=min[z_{\rm QSO}-0.04,2.2]$.
This is the brighter subset of quasars in the SDSS with
correspondingly higher S/N spectra.  Restricting our Monte Carlo
analysis to this QSO sample facilitates the generation of random
samples with a survey path identical to the GRB analysis.

For GRBs with $z<1.5$, an SDSS QSO matched in redshift will
cover the survey path of the GRB analysis.
For a given GRB,
we selected all quasars
close in redshift space to $z_{GRB}$ (usually in the range
$z_{GRB} \leq z_{QSO} \leq z_{GRB}+0.04$   there were always at least 50 such
quasars). In each Monte Carlo realization we 
randomly picked one and  
by construction adopted the $g(z)$ from
the reference GRB spectrum. 
We then identified the total number of absorbers discovered by
\cite{zhu:2012aa}
along the lines of sight of these quasars and recorded those that
satisfy the $W_r  > 1$\AA\ limit and have $g(z)=1$. 

For $z_{\rm GRB} > 1.5$, the \ion{Mg}{2} survey performed by
\cite{zhu:2012aa} using the SDSS quasars does not extend as low in
redshift as our GRB analysis because those authors truncated the
search bluer then the \ion{C}{4} emission peak.  As a result, we considered
two approaches to handling this difference.  The cleanest approach is
to artificially truncate the GRB analysis at the same starting
redshift as the quasars, i.e., 

\begin{equation}
z_{\rm min, GRB} = \frac{(z_{\rm QSO}+1)\lambda_{\rm
    CIV}}{\lambda_{\rm MgII}} - 1.
\end{equation}
The other `hybrid' approach, which maximizes the survey path of this
Monte Carlo comparison, is to introduce a second random quasar (with $z
< 1.5$) to cover the redshift path at $z<z_{\rm min,GRB}$ in
the GRB spectrum.  
In these cases, the minimum quasar redshift is $z_{min,QSO}= (1+z_{GRB})\times
\lambda_{\rm CIV}^{\rm rest}/\lambda_{\rm MgII}^{\rm rest} - 1$.  
Finally, for very high-redshift GRBs ($z_{\rm GRB} > 2.2$) the second 
quasar has to be chosen such that $z_{min,QSO}=min[ 
(1+z_{GRB})\times \lambda_{\rm CIV}^{\rm rest}/\lambda_{\rm MgII}^{\rm rest} - 1,2 ]$ and $z_{max,QSO}= 
(1+z_{GRB})\times \lambda_{\rm Ly\alpha}^{\rm rest}/\lambda_{\rm MgII}^{\rm rest} - 1$,
which allows us to select at least 50 QSOs 
covering the desired redshift path coverage.

\begin{figure*}[t!]
\epsscale{1.0}
\includegraphics[scale=0.60,angle=0]{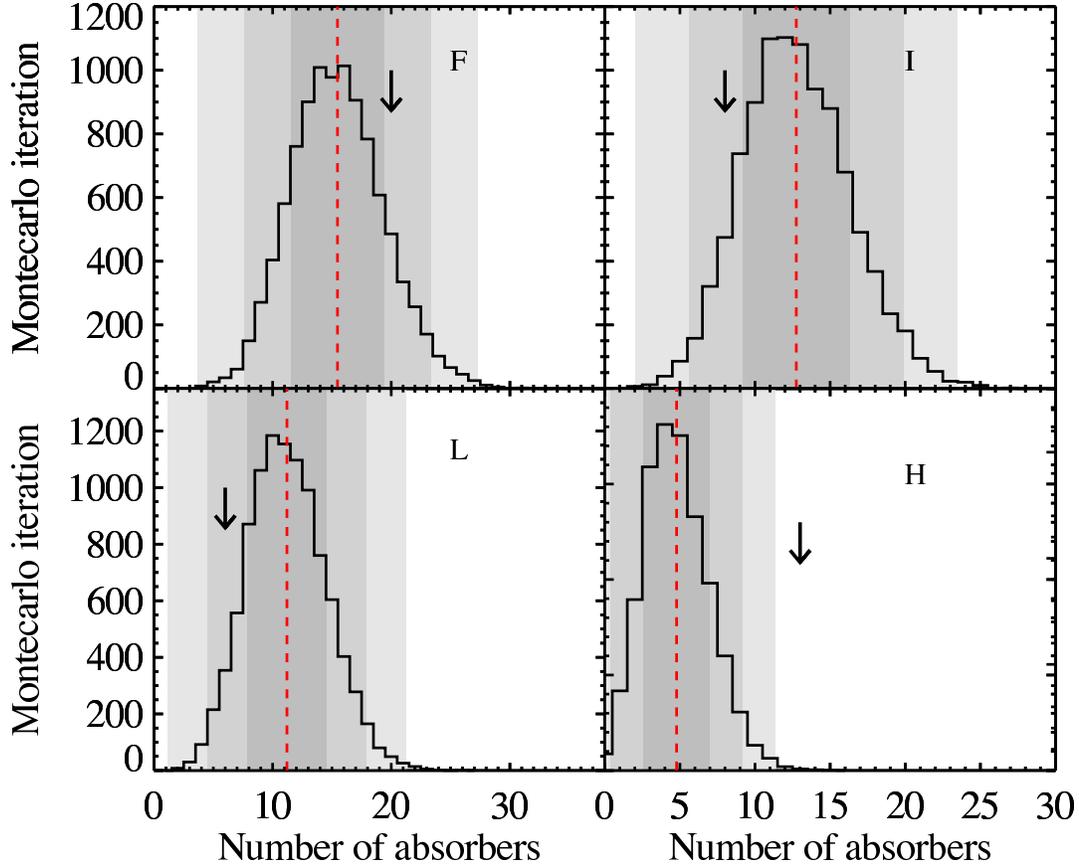}
\caption{The black curves show the distribution of recovered \MgII\
  absorbers along Monte Carlo realizations of quasar sightlines
  designed to match the $g(z)$ survey path of the GRB samples (from top left clockwise, \samF, 
  \samI, \samL, and \samH).
  The red-dashed lines trace the mean number of absorbers for each
  distribution and the shaded regions represent the 1,2, and 3$\sigma$
  confidence interval assuming Poisson statistics.
  The solid arrow in each panel denotes the number of \ion{Mg}{2}
  absorbers detected for each subsample of GRBs.  Only the high-resolution
  Sample~H exhibits a statistically significant excess, but we caution
  that this sample has substantial overlap with the original P06
  work. 
} 
\label{fig:monte}
\end{figure*}

We ran ten thousand Monte Carlo iterations using both approaches and we
recorded for each iteration the number of \ion{Mg}{2} absorbers
recovered.  We performed this analysis for each of the GRB samples.
Figure ~\ref{fig:monte} presents our outcomes using the hybrid
approach, though no relevant
differences are present using the truncated redshift path.

The results indicate that the incidence of \ion{Mg}{2}
absorbers detected in our independent \samI\ are consistent with the
results along quasar sightlines.  In fact, we recovered a slightly greater
number of absorbers on average along the quasar sightlines.
Furthermore,
the analysis shows that  there is no statistically significant discrepancy 
between the expected total number of absorbers along the QSOs and the
full parent GRB sample
(\samF).  In $6\%$ of our simulated quasar lines of sight, 
we observed a number of absorbers equal to or larger  than the \samF\
(corresponding to a $1.6\sigma$ significance). 
Only in the case of the high-resolution sub-sample, \samH\, is there a
statistically significant excess.  This sample, however, is dominated
by the sightlines analyzed in previous works (e.g.\ P06). 
We discuss this result further in the following section.
A summary of our Monte Carlo analysis is given 
in Table~\ref{tab:subsam}.

\begin{figure*}[t!]
\includegraphics[scale=0.90,angle=0]{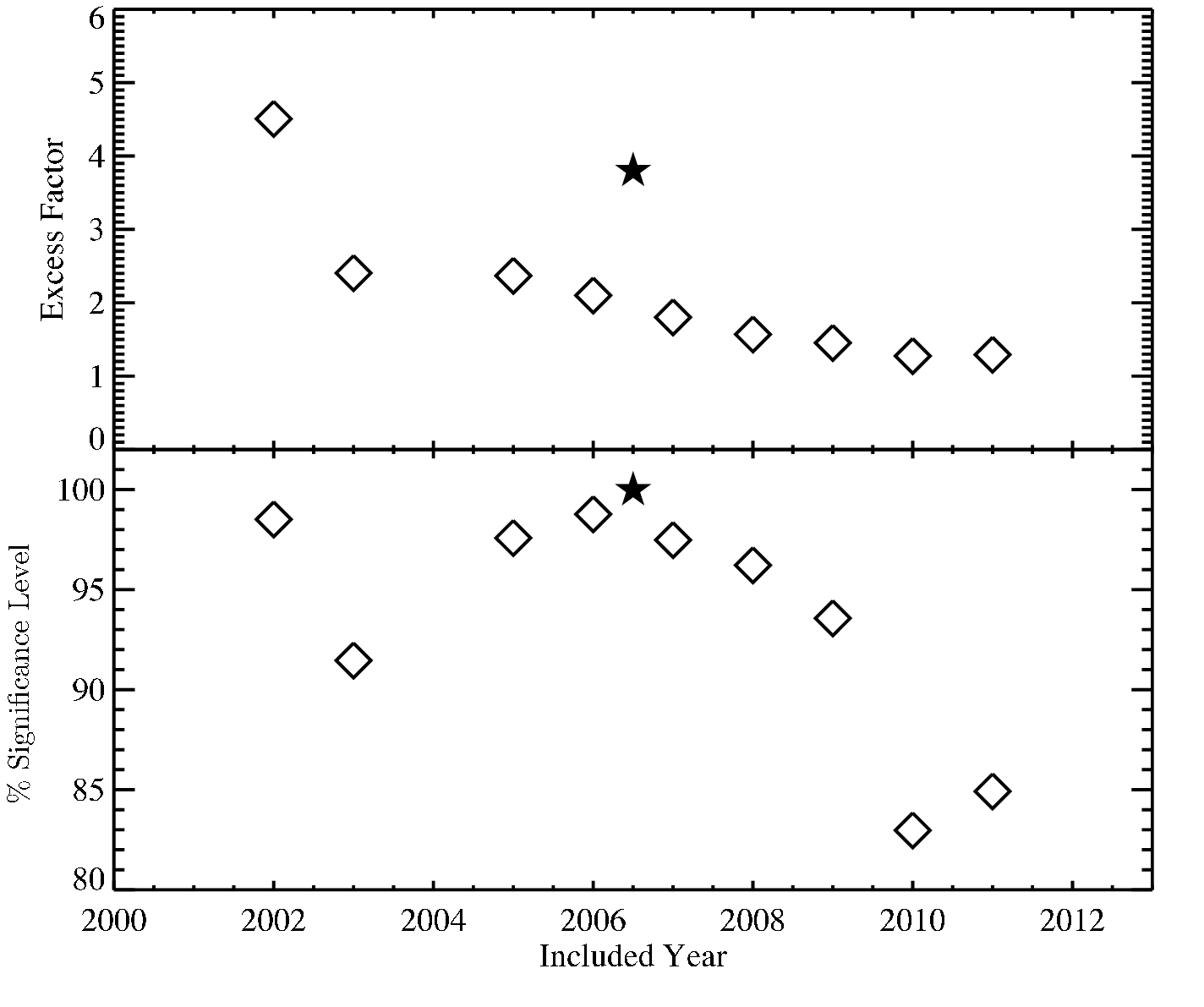}
\caption{$Top:$ Observed excess in the incidence of strong
  \ion{Mg}{2} absorption along GRB sightlines relative to that
  predicted from observations along quasar sightlines.   This is shown
  as a function of historical time where 
  each bin includes all the GRB lines of sights until December
  31$^{st}$ of the specified year, as extracted from \samF. 
  The filled star marks the results published by PO6.
  $Bottom:$  Confidence level at which the excess factor has been 
detected based on  the Monte Carlo analysis described in $\S$~\ref{sec:Monte}.
} 
\label{fig:yearplot}
\end{figure*}

It is further illuminating to estimate the statistical significance of
the \ion{Mg}{2} enhancement along GRB sightlines as a function of
historical time.  Figure~\ref{fig:yearplot} shows the results of a
Monte Carlo analysis for each year, where we include all GRBs from that
year and any previous.
Until the end of 2006 a significant ($\gtrsim3\sigma$)
excess was present.  Since that time, the statistical significance has
steadily declined and the current full sample (which has several times the
survey path of P06) has only a modest statistical significance.  At
present, we do not find a statistically significant difference in the
incidence of strong \ion{Mg}{2} absorbers between GRB and quasar
sightlines.  

\subsection{Other Characteristics of the \ion{Mg}{2} GRB Sample}
\begin{figure*}[t!]
\epsscale{1.0}
\includegraphics[scale=0.90,angle=0]{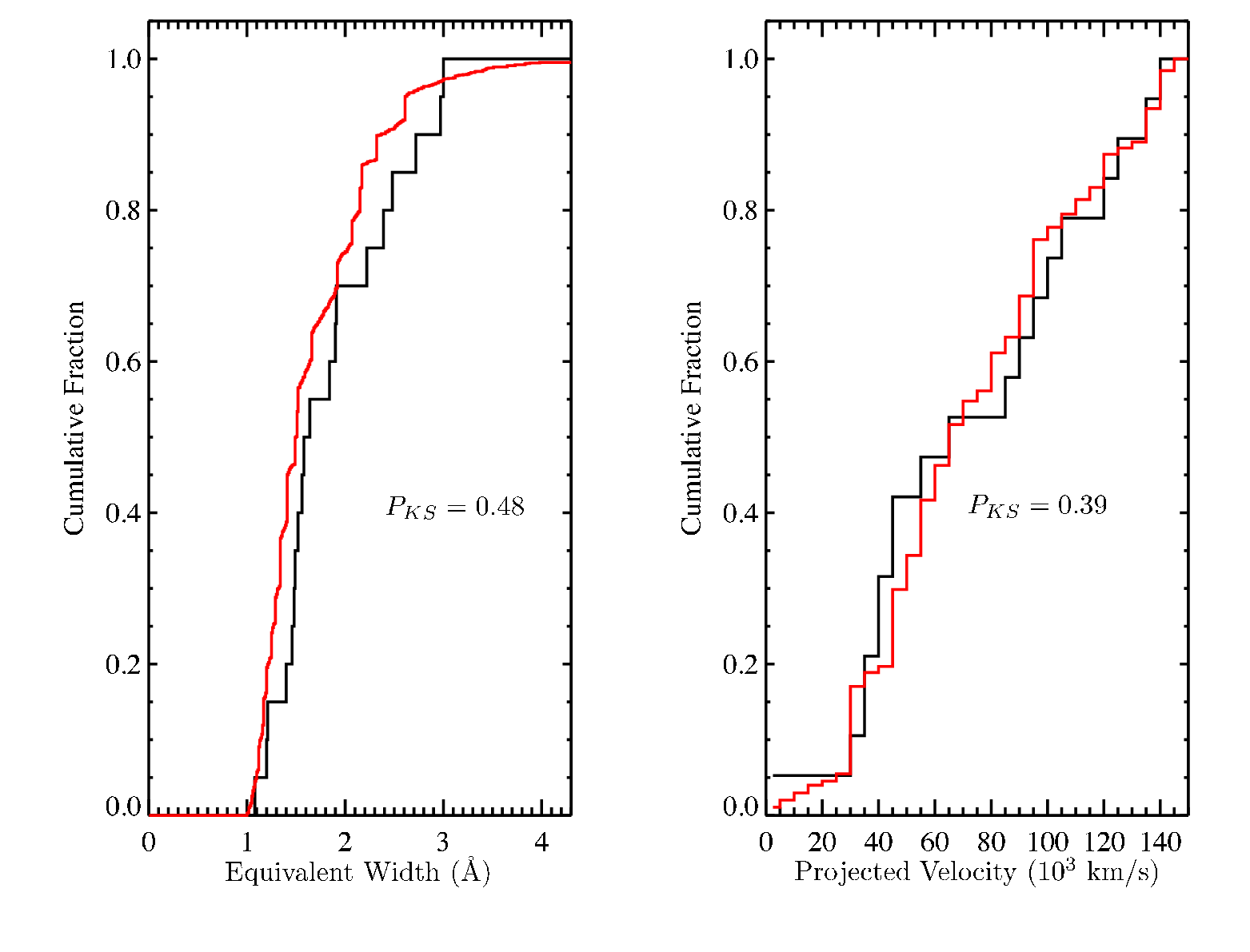}
\caption{$Left:$ Rest-frame cumulative equivalent width distribution of the \MgII\ absorbers in \samF\
(black) and the QSOs absorbers from our Monte Carlo analysis (red).
$Right:$ Cumulative distribution of the relative velocity  (black for the GRB and red for the QSO absorbers), 
assuming every absorber is local to the QSO or the GRB host galaxy and
is moving towards the observer mimicking a foreground intervening
system at lower redshift).
Kolmogorov-Smirnov tests reveal that the properties of the \ion{Mg}{2}
absorbers along GRB and QSO sightlines are consistent with having
been drawn from the same parent population.}
\label{fig:ewdistr}
\end{figure*}

In Fig. \ref{fig:ewdistr} we present the cumulative distributions of
the equivalent widths and relative velocities for the strong
\MgII\ full sample.  The latter is calculated 
assuming that these intervening systems are local to the GRB 
environments and are moving at such velocity towards the
observer to mimic a lower redshift system  \citep[see also][for the
intrinsic properties of a small sample of such
systems]{Cucchiara:2009mgii}. 
As previously observed, more than $50\%$ of the intervening systems 
would require ejection velocities larger than $50,000$~km s$^{-1}$, making very unlikely
an intrinsic origin of these absorbers. 
Recently, \cite{Bergeron:2011lr}, based on 
similar distribution of strong \MgII\ absorbers along blazars,
have suggested a possible theoretical model for producing such high
relative velocities.

The red curves represent similar quantities from our Monte Carlo
analysis of the QSO sightlines.  
A KS-test analysis shows for both metrics that the GRB and QSO
absorbers 
are consistent with having been drawn from the same parent population 
($P_{KS}=0.48$ and 
$P_{KS}=0.39$, for the $W_r$ and the projected velocity respectively).

\section{Discussion and Conclusion}
\label{sec:discussion}
We have presented the largest compilation to date of GRB 
spectroscopic data, 
more than one hundred spectra including data from previous published works, 
proprietary datasets, and publicly available datasets not yet published.
We have leveraged this dataset to 
investigate the 
puzzling excess of strong \MgII\ absorbers along GRB sightlines
as first noted by \cite{Prochter:2006fr}.
Most importantly, we have performed such analysis on a fully independent
dataset to the original P06 study in order to test their findings.  

This independent sample, our \samI, comprises
\ninde\ GRB lines of sights, yielding a 
redshift path length $\Delta z=44.9$ over the interval $z=0.4-2.2$. 
Along these spectra,
we detect only 8 absorbers, for a total
incidence of strong \ion{Mg}{2} absorbers ($W_r > 1$\AA) of
\gloz$_{,I} = 0.18$. This incidence lies in good
agreement with estimations along QSO lines of sight
taken from the latest work by \cite{zhu:2012aa} ($l(z)_{QSO}=0.26$).
No excess has been identified in the independent sample and, therefore, we do not
confirm the original findings of P06 that an excess of \ion{Mg}{2}
absorbers lie along GRB sightlines.

It is likely that the earlier works on the incidence of \ion{Mg}{2} 
absorption along GRB sightlines were biased by a remarkable,
statistical fluke.   In particular, the presence of a small set of
lines of sight with multiple absorbers appears to have driven the
results \citep[as suggested by][]{Kann:2010fk}. 

Even including the original P06 data  (i.e. our full dataset, \samF), which 
maximizes the redshift path
coverage observed along GRBs sightlines ($\Delta z=55.5$), 
we estimate \gloz$_{,F} = 0.36\pm0.09$, which corresponds to an excess
of strong \MgII\
absorbers by a factor $\sim1.5$ over QSO sightlines (at $90\%$ c.l. for
Poisson distribution).
We tested the significance of this excess using a Monte Carlo analysis
and find that 
$6\%$ of random QSO samples exhibit as many absorbers as the GRB
survey.  This suggests the null hypothesis is ruled out at 
 $\lesssim 2 \sigma$ confidence level.   In conclusion, the data
no longer demand a different incidence of strong \ion{Mg}{2}
absorption along GRB and QSO sightlines.

We wish to emphasize that the P06 analysis was not inherently flawed.
Indeed, if we restrict our analysis to the set of high-resolution
data, which has large overlap with the P06 sample, we find a
significant excess ($\approx 3$ times) at a high statistical
significance ($\approx 4 \sigma$).   At face value, this could suggest
that we have underestimated \gloz\ for the low-resolution sample,
e.g.\ because we mis-estimated our sensitivity to 1\AA\ absorbers.
Our sample of low-resolution data, however, includes a large diversity
of S/N.
In order to investigate the effect of these diversity on our detection rate, we 
degraded the spectra in \samH\ to the lowest S/N and resolution for which 
we are able to estimate the redshift path length (e.g. the ALFOSC spectrum 
of GRB 050802, which has S/N=7 and $R\approx440$): all the strong \MgII\ 
doublets could still be detected at $5\sigma$ level.
Furthermore, we have identified many additional
\ion{Mg}{2} absorbers in these spectra (Table~\ref{tab:interv}) where
the selection criteria are not fully satisfied.
We also established our completeness level and the reliability
of our automatic searching algorithm creating a larger ($\sim$5000) 
set of spectra, derived by the original full sample, where we randomly injected 
mocked doublet profiles with different equivalent widths. 
The automatic identification process recovered $\sim 98\%$ of the 
mocked features.
At this stage, 
we suspect that the few lines of sight observed with high-resolution
spectrographs were simply ``peculiar'' with respect the presence of strong \MgII\
doublets.  Surely a larger collection of such data
(e.g.\ the sample building with X-Shooter) will allow for an
independent test of the high-resolution results.

It is also worth noting, in this context, that other authors have
explored whether the brightness of the GRB afterglow correlates with
the presence of intervening \ion{Mg}{2} absorption, i.e.\ to bias the
observations towards such sightlines.
\cite{Kann:2010fk} have investigated the
optical properties of these GRBs in relation to the presence/absence  of 
\MgII\ absorbers and the possibility that GRB optical afterglows brightness
may be boosted due to gravitational lensing \citep[see][]{Porciani:2007zr,mnt08}.
In particular they compared the absolute mean $B$-band  magnitude
 (estimated at one day post-burst and normalized at $z=1$)
of GRB with strong absorbers and without (which 
usually present weak absorbers). 
For this purpose they used afterglow spectra obtained with echelle 
spectrographs which provide high-$S/N$. 
No appreciable difference was noticed between 
the two samples.
While we defer the reader to \cite{Menard:2005kl} for a quantitative estimate of 
possible gravitational lensing effects, we note that considering only the lines of sight with strong absorbers,
our \samH\ extends the original work of \cite{Kann:2010fk} by only one object, leading
to inconclusive progress on this aspect due small size samples.

Moreover, we compared the equivalent width distribution of the detected absorbers
in our \samF\ and our Monte Carlo analysis: a Kolmogorov-Smirnov test
shows that no significant difference is present between the two samples ($P_{KS}=0.48$).
Similarly, if considering the relative velocity of the two populations of absorbers
as they were, instead of intervening, moving at high velocity towards the observer so as to 
mimic a lower redshift  we also do not find any particular difference ($P_{KS}=0.39$),
further disfavouring an intrinsic nature for the absorbers.

Undoubtedly, the most robust results are
obtained from high S/N, high resolution (Echelle or Echellette) data, of
which we only have a limited sample for GRB afterglows to-date.  For
this reason new samples (such as that being gathered by X-shooter)
obtained at high resolution will provide an important test of our
conclusions.

\acknowledgements
A.Cucchiara thanks, J.X. Prochaska for the fundamental guidances, without which this
work could not be possible. I also thank B. Menard and B.Zhu for the useful comments and 
to have provided the best to date compilation of high signal to noise quasars spectra
as well their \MgII\ search results. SL has been supported by FONDECYT grant number 1100214 and
received partial support from the Center of Excellence in Astrophysics
and Associated Technologies (PFB 06).


\bibliographystyle{apj_8}
\bibliography{bibi090429B,bibthesis,110205A,bib_master,MgII}

\begin{thebibliography}{84}
\expandafter\ifx\csname natexlab\endcsname\relax\def\natexlab#1{#1}\fi

\bibitem[{{Barth} {et~al.}(2003){Barth}, {Sari}, {Cohen}, {Goodrich}, {Price},
  {Fox}, {Bloom}, {Soderberg}, \& {Kulkarni}}]{Barth:2003ve}
{Barth}, A.~J. {et al.}\  2003, \apjl, 584, L47

\bibitem[{{Barthelmy} {et~al.}(1995){Barthelmy}, {Butterworth}, {Cline},
  {Gehrels}, {Fishman}, {Kouveliotou}, \& {Meegan}}]{Barthelmy:1995lr}
{Barthelmy}, S.~D., {Butterworth}, P., {Cline}, T.~L., {Gehrels}, N.,
  {Fishman}, G.~J., {Kouveliotou}, C., \& {Meegan}, C.~A. 1995, \apss, 231, 235

\bibitem[{{Barton} \& {Cooke}(2009)}]{Barton:2009zr}
{Barton}, E.~J. \& {Cooke}, J. 2009, \aj, 138, 1817

\bibitem[{{Bergeron}(1986)}]{Bergeron:1986fk}
{Bergeron}, J. 1986, \aap, 155, L8

\bibitem[{{Bergeron} {et~al.}(2011){Bergeron}, {Boiss{\'e}}, \&
  {M{\'e}nard}}]{Bergeron:2011lr}
{Bergeron}, J., {Boiss{\'e}}, P., \& {M{\'e}nard}, B. 2011, \aap, 525, A51

\bibitem[{{Bordoloi} {et~al.}(2011){Bordoloi}, {Lilly}, {Knobel}, {Bolzonella},
  {Kampczyk}, {Carollo}, {Iovino}, {Zucca}, {Contini}, {Kneib}, {Le Fevre},
  {Mainieri}, {Renzini}, {Scodeggio}, {Zamorani}, {Balestra}, {Bardelli},
  {Bongiorno}, {Caputi}, {Cucciati}, {de la Torre}, {de Ravel}, {Garilli},
  {Kova{\v c}}, {Lamareille}, {Le Borgne}, {Le Brun}, {Maier}, {Mignoli},
  {Pello}, {Peng}, {Perez Montero}, {Presotto}, {Scarlata}, {Silverman},
  {Tanaka}, {Tasca}, {Tresse}, {Vergani}, {Barnes}, {Cappi}, {Cimatti},
  {Coppa}, {Diener}, {Franzetti}, {Koekemoer}, {L{\'o}pez-Sanjuan},
  {McCracken}, {Moresco}, {Nair}, {Oesch}, {Pozzetti}, \&
  {Welikala}}]{Bordoloi:2011ve}
{Bordoloi}, R. {et al.}\  2011, \apj, 743, 10

\bibitem[{{Bowen} {et~al.}(1995){Bowen}, {Blades}, \& {Pettini}}]{Bowen:1995bh}
{Bowen}, D.~V., {Blades}, J.~C., \& {Pettini}, M. 1995, \apj, 448, 634

\bibitem[{{Bowen} \& {Chelouche}(2011)}]{Bowen:2011nx}
{Bowen}, D.~V. \& {Chelouche}, D. 2011, \apj, 727, 47

\bibitem[{{Budzynski} \& {Hewett}(2011)}]{Budzynski:2011uq}
{Budzynski}, J.~M. \& {Hewett}, P.~C. 2011, \mnras, 416, 1871

\bibitem[{{Castro} {et~al.}(2003){Castro}, {Galama}, {Harrison}, {Holtzman},
  {Bloom}, {Djorgovski}, \& {Kulkarni}}]{Castro:2003bh}
{Castro}, S., {Galama}, T.~J., {Harrison}, F.~A., {Holtzman}, J.~A., {Bloom},
  J.~S., {Djorgovski}, S.~G., \& {Kulkarni}, S.~R. 2003, \apj, 586, 128

\bibitem[{{Cenko} {et~al.}(2008){Cenko}, {Fox}, {Penprase}, {Cucchiara},
  {Price}, {Berger}, {Kulkarni}, {Harrison}, {Gal-Yam}, {Ofek}, {Rau},
  {Chandra}, {Frail}, {Kasliwal}, {Schmidt}, {Soderberg}, {Cameron}, \&
  {Roth}}]{Cenko:2008zr}
{Cenko}, S.~B. {et al.}\  2008, \apj, 677, 441

\bibitem[{{Chen}(2012)}]{Chen:2012kx}
{Chen}, H.-W. 2012, \mnras, 419, 3039

\bibitem[{{Chen} {et~al.}(2010){Chen}, {Helsby}, {Gauthier}, {Shectman},
  {Thompson}, \& {Tinker}}]{Chen:2010uq}
{Chen}, H.-W., {Helsby}, J.~E., {Gauthier}, J.-R., {Shectman}, S.~A.,
  {Thompson}, I.~B., \& {Tinker}, J.~L. 2010, \apj, 714, 1521

\bibitem[{{Chen} {et~al.}(2009){Chen}, {Perley}, {Pollack}, {Prochaska},
  {Bloom}, {Dessauges-Zavadsky}, {Pettini}, {Lopez}, {Dall'aglio}, \&
  {Becker}}]{Chen:2009ly}
{Chen}, H.-W. {et al.}\  2009, \apj, 691, 152

\bibitem[{{Chen} \& {Tinker}(2008)}]{Chen:2008dq}
{Chen}, H.-W. \& {Tinker}, J.~L. 2008, \apj, 687, 745

\bibitem[{{Christensen} {et~al.}(2011){Christensen}, {Fynbo}, {Prochaska},
  {Th{\"o}ne}, {de Ugarte Postigo}, \& {Jakobsson}}]{Christensen:2011fk}
{Christensen}, L., {Fynbo}, J.~P.~U., {Prochaska}, J.~X., {Th{\"o}ne}, C.~C.,
  {de Ugarte Postigo}, A., \& {Jakobsson}, P. 2011, \apj, 727, 73

\bibitem[{{Cooksey} {et~al.}(2010){Cooksey}, {Thom}, {Prochaska}, \&
  {Chen}}]{Cooksey:2010uq}
{Cooksey}, K.~L., {Thom}, C., {Prochaska}, J.~X., \& {Chen}, H.-W. 2010, \apj,
  708, 868

\bibitem[{{Cucchiara}(2010)}]{Cucchiara:2010fk}
{Cucchiara}, A. 2010, PhD thesis, The Pennsylvania State University

\bibitem[{{Cucchiara} {et~al.}(2011{\natexlab{a}}){Cucchiara}, {Cenko},
  {Bloom}, {Melandri}, {Morgan}, {Kobayashi}, {Smith}, {Perley}, {Li}, {Hora},
  {da Silva}, {Prochaska}, {Milne}, {Butler}, {Cobb}, {Worseck}, {Mundell},
  {Steele}, {Filippenko}, {Fumagalli}, {Klein}, {Stephens}, {Bluck}, \&
  {Mason}}]{Cucchiara:2011fk}
{Cucchiara}, A. {et al.}\  2011{\natexlab{a}}, \apj, 743, 154

\bibitem[{{Cucchiara} {et~al.}(2009){Cucchiara}, {Jones}, {Charlton}, {Fox},
  {Einsig}, \& {Narayanan}}]{Cucchiara:2009mgii}
{Cucchiara}, A., {Jones}, T., {Charlton}, J.~C., {Fox}, D.~B., {Einsig}, D., \&
  {Narayanan}, A. 2009, \apj, 697, 345

\bibitem[{{Cucchiara} {et~al.}(2011{\natexlab{b}}){Cucchiara}, {Levan}, {Fox},
  {Tanvir}, {Ukwatta}, {Berger}, {Kr{\"u}hler}, {K{\"u}pc{\"u} Yolda{\c s}},
  {Wu}, {Toma}, {Greiner}, {Olivares}, {Rowlinson}, {Amati}, {Sakamoto},
  {Roth}, {Stephens}, {Fritz}, {Fynbo}, {Hjorth}, {Malesani}, {Jakobsson},
  {Wiersema}, {O'Brien}, {Soderberg}, {Foley}, {Fruchter}, {Rhoads},
  {Rutledge}, {Schmidt}, {Dopita}, {Podsiadlowski}, {Willingale}, {Wolf},
  {Kulkarni}, \& {D'Avanzo}}]{Cucchiara:2011lr}
{Cucchiara}, A. {et al.}\  2011{\natexlab{b}}, \apj, 736, 7

\bibitem[{{de Ugarte Postigo} {et~al.}(2012{\natexlab{a}}){de Ugarte Postigo},
  {Fynbo}, {Thoene}, {Christensen}, {Gorosabel}, {Milvang-Jensen}, {Schulze},
  {Jakobsson}, {Wiersema}, {Sanchez-Ramirez}, {Leloudas}, {Zafar}, {Malesani},
  \& {Hjorth}}]{de-Ugarte-Postigo:2012kx}
{de Ugarte Postigo}, A. {et al.}\  2012{\natexlab{a}}, ArXiv e-prints

\bibitem[{{de Ugarte Postigo} {et~al.}(2012{\natexlab{b}}){de Ugarte Postigo},
  {Lundgren}, {Mart{\'{\i}}n}, {Garcia-Appadoo}, {de Gregorio Monsalvo},
  {Peck}, {Micha{\l}owski}, {Th{\"o}ne}, {Campana}, {Gorosabel}, {Tanvir},
  {Wiersema}, {Castro-Tirado}, {Schulze}, {De Breuck}, {Petitpas}, {Hjorth},
  {Jakobsson}, {Covino}, {Fynbo}, {Winters}, {Bremer}, {Levan}, {Llorente},
  {S{\'a}nchez-Ram{\'{\i}}rez}, {Tello}, \&
  {Salvaterra}}]{de-Ugarte-Postigo:2012ys}
--- 2012{\natexlab{b}}, \aap, 538, A44

\bibitem[{{de Ugarte Postigo} {et~al.}(2011){de Ugarte Postigo}, {Th{\"o}ne},
  {Goldoni}, {Fynbo}, \& {X-shooter GRB
  Collaboration}}]{de-Ugarte-Postigo:2011uq}
{de Ugarte Postigo}, A., {Th{\"o}ne}, C.~C., {Goldoni}, P., {Fynbo}, J.~P.~U.,
  \& {X-shooter GRB Collaboration} 2011, Astronomische Nachrichten, 332, 297

\bibitem[{{D'Elia} {et~al.}(2011){D'Elia}, {Campana}, {Covino}, {D'Avanzo},
  {Piranomonte}, \& {Tagliaferri}}]{DElia:2011vn}
{D'Elia}, V., {Campana}, S., {Covino}, S., {D'Avanzo}, P., {Piranomonte}, S.,
  \& {Tagliaferri}, G. 2011, \mnras, 418, 680

\bibitem[{{D'Elia} {et~al.}(2010){D'Elia}, {Fynbo}, {Covino}, {Goldoni},
  {Jakobsson}, {Matteucci}, {Piranomonte}, {Sollerman}, {Th{\"o}ne}, {Vergani},
  {Vreeswijk}, {Watson}, {Wiersema}, {Zafar}, {de Ugarte Postigo}, {Flores},
  {Hjorth}, {Kaper}, {Levan}, {Malesani}, {Milvang-Jensen}, {Pian},
  {Tagliaferri}, \& {Tanvir}}]{DElia:2010kx}
{D'Elia}, V. {et al.}\  2010, \aap, 523, A36

\bibitem[{{Frank} {et~al.}(2007){Frank}, {Bentz}, {Stanek}, {Mathur},
  {Dietrich}, {Peterson}, \& {Atlee}}]{fbs07}
{Frank}, S., {Bentz}, M.~C., {Stanek}, K.~Z., {Mathur}, S., {Dietrich}, M.,
  {Peterson}, B.~M., \& {Atlee}, D.~W. 2007, \apss, 312, 325

\bibitem[{{Fynbo} {et~al.}(2009){Fynbo}, {Jakobsson}, {Prochaska}, {Malesani},
  {Ledoux}, {de Ugarte Postigo}, {Nardini}, {Vreeswijk}, {Wiersema}, {Hjorth},
  {Sollerman}, {Chen}, {Th{\"o}ne}, {Bj{\"o}rnsson}, {Bloom}, {Castro-Tirado},
  {Christensen}, {De Cia}, {Fruchter}, {Gorosabel}, {Graham}, {Jaunsen},
  {Jensen}, {Kann}, {Kouveliotou}, {Levan}, {Maund}, {Masetti},
  {Milvang-Jensen}, {Palazzi}, {Perley}, {Pian}, {Rol}, {Schady}, {Starling},
  {Tanvir}, {Watson}, {Xu}, {Augusteijn}, {Grundahl}, {Telting}, \&
  {Quirion}}]{Fynbo:2009lr}
{Fynbo}, J.~P.~U. {et al.}\  2009, \apjs, 185, 526

\bibitem[{{Gehrels} {et~al.}(2004){Gehrels}, {Chincarini}, {Giommi}, {Mason},
  {Nousek}, {Wells}, {White}, {Barthelmy}, {Burrows}, {Cominsky}, {Hurley},
  {Marshall}, {M{\'e}sz{\'a}ros}, {Roming}, {Angelini}, {Barbier}, {Belloni},
  {Campana}, {Caraveo}, {Chester}, {Citterio}, {Cline}, {Cropper}, {Cummings},
  {Dean}, {Feigelson}, {Fenimore}, {Frail}, {Fruchter}, {Garmire}, {Gendreau},
  {Ghisellini}, {Greiner}, {Hill}, {Hunsberger}, {Krimm}, {Kulkarni}, {Kumar},
  {Lebrun}, {Lloyd-Ronning}, {Markwardt}, {Mattson}, {Mushotzky}, {Norris},
  {Osborne}, {Paczynski}, {Palmer}, {Park}, {Parsons}, {Paul}, {Rees},
  {Reynolds}, {Rhoads}, {Sasseen}, {Schaefer}, {Short}, {Smale}, {Smith},
  {Stella}, {Tagliaferri}, {Takahashi}, {Tashiro}, {Townsley}, {Tueller},
  {Turner}, {Vietri}, {Voges}, {Ward}, {Willingale}, {Zerbi}, \&
  {Zhang}}]{Gehrels:2004fj}
{Gehrels}, N. {et al.}\  2004, \apj, 611, 1005

\bibitem[{{Gehrels} {et~al.}(2009){Gehrels}, {Ramirez-Ruiz}, \&
  {Fox}}]{Gehrels:2009fk}
{Gehrels}, N., {Ramirez-Ruiz}, E., \& {Fox}, D.~B. 2009, \araa, 47, 567

\bibitem[{{Goldoni} {et~al.}(2006){Goldoni}, {Royer}, {Fran{\c c}ois},
  {Horrobin}, {Blanc}, {Vernet}, {Modigliani}, \& {Larsen}}]{Goldoni:2006fk}
{Goldoni}, P., {Royer}, F., {Fran{\c c}ois}, P., {Horrobin}, M., {Blanc}, G.,
  {Vernet}, J., {Modigliani}, A., \& {Larsen}, J. 2006, in Society of
  Photo-Optical Instrumentation Engineers (SPIE) Conference Series, Vol. 6269,
  Society of Photo-Optical Instrumentation Engineers (SPIE) Conference Series

\bibitem[{{Greiner} {et~al.}(2008){Greiner}, {Bornemann}, {Clemens}, {Deuter},
  {Hasinger}, {Honsberg}, {Huber}, {Huber}, {Krauss}, {Kr{\"u}hler},
  {K{\"u}pc{\"u} Yolda{\c s}}, {Mayer-Hasselwander}, {Mican}, {Primak},
  {Schrey}, {Steiner}, {Szokoly}, {Th{\"o}ne}, {Yolda{\c s}}, {Klose}, {Laux},
  \& {Winkler}}]{Greiner:2008fk}
{Greiner}, J. {et al.}\  2008, \pasp, 120, 405

\bibitem[{{Hook} {et~al.}(2004){Hook}, {J{\o}rgensen}, {Allington-Smith},
  {Davies}, {Metcalfe}, {Murowinski}, \& {Crampton}}]{Hook:2004fj}
{Hook}, I.~M., {J{\o}rgensen}, I., {Allington-Smith}, J.~R., {Davies}, R.~L.,
  {Metcalfe}, N., {Murowinski}, R.~G., \& {Crampton}, D. 2004, \pasp, 116, 425

\bibitem[{{Horne}(1986)}]{Horne:1986dq}
{Horne}, K. 1986, \pasp, 98, 609

\bibitem[{{Jakobsson} {et~al.}(2006){Jakobsson}, {Fynbo}, {Ledoux},
  {Vreeswijk}, {Kann}, {Hjorth}, {Priddey}, {Tanvir}, {Reichart}, {Gorosabel},
  {Klose}, {Watson}, {Sollerman}, {Fruchter}, {de Ugarte Postigo}, {Wiersema},
  {Bj{\"o}rnsson}, {Chapman}, {Th{\"o}ne}, {Pedersen}, \&
  {Jensen}}]{Jakobsson:2006lr}
{Jakobsson}, P. {et al.}\  2006, \aap, 460, L13

\bibitem[{{Jakobsson} {et~al.}(2004){Jakobsson}, {Hjorth}, {Fynbo},
  {Weidinger}, {Gorosabel}, {Ledoux}, {Watson}, {Bj{\"o}rnsson}, {Gudmundsson},
  {Wijers}, {M{\"o}ller}, {Pedersen}, {Sollerman}, {Henden}, {Jensen},
  {Gilmore}, {Kilmartin}, {Levan}, {Castro Cer{\'o}n}, {Castro-Tirado},
  {Fruchter}, {Kouveliotou}, {Masetti}, \& {Tanvir}}]{Jakobsson:2004vn}
--- 2004, \aap, 427, 785

\bibitem[{{Kacprzak} {et~al.}(2012){Kacprzak}, {Churchill}, \&
  {Nielsen}}]{Kacprzak:2012kx}
{Kacprzak}, G.~G., {Churchill}, C.~W., \& {Nielsen}, N.~M. 2012, \apjl, 760, L7

\bibitem[{{Kacprzak} {et~al.}(2008){Kacprzak}, {Churchill}, {Steidel}, \&
  {Murphy}}]{Kacprzak:2008qf}
{Kacprzak}, G.~G., {Churchill}, C.~W., {Steidel}, C.~C., \& {Murphy}, M.~T.
  2008, \aj, 135, 922

\bibitem[{{Kann} {et~al.}(2010){Kann}, {Klose}, {Zhang}, {Malesani}, {Nakar},
  {Pozanenko}, {Wilson}, {Butler}, {Jakobsson}, {Schulze}, {Andreev},
  {Antonelli}, {Bikmaev}, {Biryukov}, {B{\"o}ttcher}, {Burenin}, {Castro
  Cer{\'o}n}, {Castro-Tirado}, {Chincarini}, {Cobb}, {Covino}, {D'Avanzo},
  {D'Elia}, {Della Valle}, {de Ugarte Postigo}, {Efimov}, {Ferrero}, {Fugazza},
  {Fynbo}, {G{\aa}lfalk}, {Grundahl}, {Gorosabel}, {Gupta}, {Guziy}, {Hafizov},
  {Hjorth}, {Holhjem}, {Ibrahimov}, {Im}, {Israel}, {Je{\'l}inek}, {Jensen},
  {Karimov}, {Khamitov}, {Kizilo{\v g}lu}, {Klunko}, {Kub{\'a}nek}, {Kutyrev},
  {Laursen}, {Levan}, {Mannucci}, {Martin}, {Mescheryakov}, {Mirabal},
  {Norris}, {Ovaldsen}, {Paraficz}, {Pavlenko}, {Piranomonte}, {Rossi},
  {Rumyantsev}, {Salinas}, {Sergeev}, {Sharapov}, {Sollerman}, {Stecklum},
  {Stella}, {Tagliaferri}, {Tanvir}, {Telting}, {Testa}, {Updike}, {Volnova},
  {Watson}, {Wiersema}, \& {Xu}}]{Kann:2010fk}
{Kann}, D.~A. {et al.}\  2010, \apj, 720, 1513

\bibitem[{{Kawai} {et~al.}(2006){Kawai}, {Kosugi}, {Aoki}, {Yamada}, {Totani},
  {Ohta}, {Iye}, {Hattori}, {Aoki}, {Furusawa}, {Hurley}, {Kawabata},
  {Kobayashi}, {Komiyama}, {Mizumoto}, {Nomoto}, {Noumaru}, {Ogasawara},
  {Sato}, {Sekiguchi}, {Shirasaki}, {Suzuki}, {Takata}, {Tamagawa}, {Terada},
  {Watanabe}, {Yatsu}, \& {Yoshida}}]{Kawai:2006fk}
{Kawai}, N. {et al.}\  2006, \nat, 440, 184

\bibitem[{{Klose} {et~al.}(2004{\natexlab{a}}){Klose}, {Greiner}, {Rau},
  {Henden}, {Hartmann}, {Zeh}, {Ries}, {Masetti}, {Malesani}, {Guenther},
  {Gorosabel}, {Stecklum}, {Antonelli}, {Brinkworth}, {Castro Cer{\'o}n},
  {Castro-Tirado}, {Covino}, {Fruchter}, {Fynbo}, {Ghisellini}, {Hjorth},
  {Hudec}, {Jel{\'{\i}}nek}, {Kaper}, {Kouveliotou}, {Lindsay}, {Maiorano},
  {Mannucci}, {Nysewander}, {Palazzi}, {Pedersen}, {Pian}, {Reichart},
  {Rhoads}, {Rol}, {Smail}, {Tanvir}, {de Ugarte Postigo}, {Vreeswijk},
  {Wijers}, \& {van den Heuvel}}]{Klose:2004qf}
{Klose}, S. {et al.}\  2004{\natexlab{a}}, \aj, 128, 1942

\bibitem[{{Klose} {et~al.}(2004{\natexlab{b}}){Klose}, {Greiner}, {Rau},
  {Henden}, {Hartmann}, {Zeh}, {Ries}, {Masetti}, {Malesani}, {Guenther},
  {Gorosabel}, {Stecklum}, {Antonelli}, {Brinkworth}, {Castro Cer{\'o}n},
  {Castro-Tirado}, {Covino}, {Fruchter}, {Fynbo}, {Ghisellini}, {Hjorth},
  {Hudec}, {Jel{\'{\i}}nek}, {Kaper}, {Kouveliotou}, {Lindsay}, {Maiorano},
  {Mannucci}, {Nysewander}, {Palazzi}, {Pedersen}, {Pian}, {Reichart},
  {Rhoads}, {Rol}, {Smail}, {Tanvir}, {de Ugarte Postigo}, {Vreeswijk},
  {Wijers}, \& {van den Heuvel}}]{Klose:2004fj}
--- 2004{\natexlab{b}}, \aj, 128, 1942

\bibitem[{{Lanzetta} {et~al.}(1987){Lanzetta}, {Turnshek}, \&
  {Wolfe}}]{Lanzetta:1987vn}
{Lanzetta}, K.~M., {Turnshek}, D.~A., \& {Wolfe}, A.~M. 1987, \apj, 322, 739

\bibitem[{{Lawther} {et~al.}(2012){Lawther}, {Paarup}, {Schmidt},
  {Vestergaard}, {Hjorth}, \& {Malesani}}]{Lawther:2012fk}
{Lawther}, D., {Paarup}, T., {Schmidt}, M., {Vestergaard}, M., {Hjorth}, J., \&
  {Malesani}, D. 2012, \aap, 546, A67

\bibitem[{{L{\'o}pez} \& {Chen}(2012)}]{Lopez:2012cr}
{L{\'o}pez}, G. \& {Chen}, H.-W. 2012, \mnras, 419, 3553

\bibitem[{{Matejek} \& {Simcoe}(2012)}]{Matejek:2012bh}
{Matejek}, M.~S. \& {Simcoe}, R.~A. 2012, ArXiv e-prints

\bibitem[{{M{\'e}nard}(2005)}]{Menard:2005kl}
{M{\'e}nard}, B. 2005, \apj, 630, 28

\bibitem[{{M{\'e}nard} {et~al.}(2008){M{\'e}nard}, {Nestor}, {Turnshek},
  {Quider}, {Richards}, {Chelouche}, \& {Rao}}]{mnt08}
{M{\'e}nard}, B., {Nestor}, D., {Turnshek}, D., {Quider}, A., {Richards}, G.,
  {Chelouche}, D., \& {Rao}, S. 2008, \mnras, 385, 1053

\bibitem[{{M{\'e}nard} {et~al.}(2011){M{\'e}nard}, {Wild}, {Nestor}, {Quider},
  {Zibetti}, {Rao}, \& {Turnshek}}]{Menard:2011vn}
{M{\'e}nard}, B., {Wild}, V., {Nestor}, D., {Quider}, A., {Zibetti}, S., {Rao},
  S., \& {Turnshek}, D. 2011, \mnras, 417, 801

\bibitem[{{Metzger} {et~al.}(1997){Metzger}, {Djorgovski}, {Kulkarni},
  {Steidel}, {Adelberger}, {Frail}, {Costa}, \& {Frontera}}]{Metzger:1997lr}
{Metzger}, M.~R., {Djorgovski}, S.~G., {Kulkarni}, S.~R., {Steidel}, C.~C.,
  {Adelberger}, K.~L., {Frail}, D.~A., {Costa}, E., \& {Frontera}, F. 1997,
  \nat, 387, 878

\bibitem[{{Milvang-Jensen} {et~al.}(2012){Milvang-Jensen}, {Fynbo}, {Malesani},
  {Hjorth}, {Jakobsson}, \& {M{\o}ller}}]{Milvang-Jensen:2012ly}
{Milvang-Jensen}, B., {Fynbo}, J.~P.~U., {Malesani}, D., {Hjorth}, J.,
  {Jakobsson}, P., \& {M{\o}ller}, P. 2012, \apj, 756, 25

\bibitem[{{Mirabal} {et~al.}(2002){Mirabal}, {Halpern}, {Kulkarni}, {Castro},
  {Bloom}, {Djorgovski}, {Galama}, {Harrison}, {Frail}, {Price}, {Reichart},
  {Ebeling}, {Bunker}, {Dawson}, {Dey}, {Spinrad}, \& {Stern}}]{Mirabal:2002qf}
{Mirabal}, N. {et al.}\  2002, \apj, 578, 818

\bibitem[{{Nestor} {et~al.}(2011){Nestor}, {Johnson}, {Wild}, {M{\'e}nard},
  {Turnshek}, {Rao}, \& {Pettini}}]{Nestor:2011fk}
{Nestor}, D.~B., {Johnson}, B.~D., {Wild}, V., {M{\'e}nard}, B., {Turnshek},
  D.~A., {Rao}, S., \& {Pettini}, M. 2011, \mnras, 412, 1559

\bibitem[{{Nestor} {et~al.}(2005){Nestor}, {Turnshek}, \&
  {Rao}}]{Nestor:2005fk}
{Nestor}, D.~B., {Turnshek}, D.~A., \& {Rao}, S.~M. 2005, \apj, 628, 637

\bibitem[{{Perley} {et~al.}(2008){Perley}, {Li}, {Chornock}, {Prochaska},
  {Butler}, {Chandra}, {Pollack}, {Bloom}, {Filippenko}, {Swan}, {Yuan},
  {Akerlof}, {Auger}, {Cenko}, {Chen}, {Fassnacht}, {Fox}, {Frail},
  {Johansson}, {McKay}, {Le Mignant}, {Modjaz}, {Rujopakarn}, {Russel},
  {Skinner}, {Smith}, {Smith}, {van Dam}, \& {Yost}}]{Perley:2008zr}
{Perley}, D.~A. {et al.}\  2008, \apj, 688, 470

\bibitem[{{Pollack} {et~al.}(2009){Pollack}, {Chen}, {Prochaska}, \&
  {Bloom}}]{Pollack:2009zr}
{Pollack}, L.~K., {Chen}, H.-W., {Prochaska}, J.~X., \& {Bloom}, J.~S. 2009,
  \apj, 701, 1605

\bibitem[{{Porciani} \& {Madau}(2001)}]{Porciani:2001vn}
{Porciani}, C. \& {Madau}, P. 2001, \apj, 548, 522

\bibitem[{{Porciani} {et~al.}(2007{\natexlab{a}}){Porciani}, {Viel}, \&
  {Lilly}}]{pvs07}
{Porciani}, C., {Viel}, M., \& {Lilly}, S.~J. 2007{\natexlab{a}}, \apj, 659,
  218

\bibitem[{{Porciani} {et~al.}(2007{\natexlab{b}}){Porciani}, {Viel}, \&
  {Lilly}}]{Porciani:2007zr}
--- 2007{\natexlab{b}}, \apj, 659, 218

\bibitem[{{Prochaska} {et~al.}(2006){Prochaska}, {Chen}, \&
  {Bloom}}]{Prochaska:2006vn}
{Prochaska}, J.~X., {Chen}, H.-W., \& {Bloom}, J.~S. 2006, \apj, 648, 95

\bibitem[{{Prochaska} {et~al.}(2007){Prochaska}, {Chen}, {Bloom},
  {Dessauges-Zavadsky}, {O'Meara}, {Foley}, {Bernstein}, {Burles}, {Dupree},
  {Falco}, \& {Thompson}}]{Prochaska:2007ly}
{Prochaska}, J.~X. {et al.}\  2007, \apjs, 168, 231

\bibitem[{{Prochter} {et~al.}(2006{\natexlab{a}}){Prochter}, {Prochaska}, \&
  {Burles}}]{Prochter:2006kx}
{Prochter}, G.~E., {Prochaska}, J.~X., \& {Burles}, S.~M. 2006{\natexlab{a}},
  \apj, 639, 766

\bibitem[{{Prochter} {et~al.}(2006{\natexlab{b}}){Prochter}, {Prochaska},
  {Chen}, {Bloom}, {Dessauges-Zavadsky}, {Foley}, {Lopez}, {Pettini}, {Dupree},
  \& {Guhathakurta}}]{Prochter:2006fr}
{Prochter}, G.~E. {et al.}\  2006{\natexlab{b}}, \apjl, 648, L93

\bibitem[{{Quider} {et~al.}(2011){Quider}, {Nestor}, {Turnshek}, {Rao},
  {Monier}, {Weyant}, \& {Busche}}]{Quider:2011fk}
{Quider}, A.~M., {Nestor}, D.~B., {Turnshek}, D.~A., {Rao}, S.~M., {Monier},
  E.~M., {Weyant}, A.~N., \& {Busche}, J.~R. 2011, \aj, 141, 137

\bibitem[{{Rapoport} {et~al.}(2011){Rapoport}, {Onken}, {Schmidt}, {Wyithe},
  {Tucker}, \& {Levan}}]{Rapoport:2011kx}
{Rapoport}, S., {Onken}, C.~A., {Schmidt}, B.~P., {Wyithe}, J.~S.~B., {Tucker},
  B.~E., \& {Levan}, A.~J. 2011, ArXiv e-prints

\bibitem[{{Rodr{\'{\i}}guez Hidalgo} {et~al.}(2012){Rodr{\'{\i}}guez Hidalgo},
  {Wessels}, {Charlton}, {Narayanan}, {Mshar}, {Cucchiara}, \&
  {Jones}}]{Rodriguez-Hidalgo:2012oq}
{Rodr{\'{\i}}guez Hidalgo}, P., {Wessels}, K., {Charlton}, J., {Narayanan}, A.,
  {Mshar}, A., {Cucchiara}, A., \& {Jones}, T. 2012, ArXiv e-prints

\bibitem[{{Rubin} {et~al.}(2011){Rubin}, {Prochaska}, {M{\'e}nard}, {Murray},
  {Kasen}, {Koo}, \& {Phillips}}]{Rubin:2011ly}
{Rubin}, K.~H.~R., {Prochaska}, J.~X., {M{\'e}nard}, B., {Murray}, N., {Kasen},
  D., {Koo}, D.~C., \& {Phillips}, A.~C. 2011, \apj, 728, 55

\bibitem[{{Salvaterra} {et~al.}(2009){Salvaterra}, {Della Valle}, {Campana},
  {Chincarini}, {Covino}, {D'Avanzo}, {Fern{\'a}ndez-Soto}, {Guidorzi},
  {Mannucci}, {Margutti}, {Th{\"o}ne}, {Antonelli}, {Barthelmy}, {de Pasquale},
  {D'Elia}, {Fiore}, {Fugazza}, {Hunt}, {Maiorano}, {Marinoni}, {Marshall},
  {Molinari}, {Nousek}, {Pian}, {Racusin}, {Stella}, {Amati}, {Andreuzzi},
  {Cusumano}, {Fenimore}, {Ferrero}, {Giommi}, {Guetta}, {Holland}, {Hurley},
  {Israel}, {Mao}, {Markwardt}, {Masetti}, {Pagani}, {Palazzi}, {Palmer},
  {Piranomonte}, {Tagliaferri}, \& {Testa}}]{sdc09}
{Salvaterra}, R. {et al.}\  2009, \nat, 461, 1258

\bibitem[{{Schulze} {et~al.}(2012){Schulze}, {Fynbo}, {Milvang-Jensen},
  {Rossi}, {Jakobsson}, {Ledoux}, {De Cia}, {Kruehler}, {Mehner}, {Bjoernsson},
  {Chen}, {Vreeswijk}, {Perley}, {Hjorth}, {Levan}, {Tanvir}, {Ellison},
  {Moller}, {Worseck}, {Chapman}, {Dall'Aglio}, \& {Letawe}}]{Schulze:2012ys}
{Schulze}, S. {et al.}\  2012, ArXiv e-prints

\bibitem[{{Simcoe} {et~al.}(2011){Simcoe}, {Cooksey}, {Matejek}, {Burgasser},
  {Bochanski}, {Lovegrove}, {Bernstein}, {Pipher}, {Forrest}, {McMurtry},
  {Fan}, \& {O'Meara}}]{Simcoe:2011uq}
{Simcoe}, R.~A. {et al.}\  2011, \apj, 743, 21

\bibitem[{{Steidel}(1993)}]{Steidel:1993fk}
{Steidel}, C.~C. 1993, in Astronomical Society of the Pacific Conference
  Series, Vol.~49, Galaxy Evolution. The Milky Way Perspective, ed. S.~R.
  {Majewski}, 227

\bibitem[{{Steidel} \& {Sargent}(1992)}]{Steidel:1992kx}
{Steidel}, C.~C. \& {Sargent}, W.~L.~W. 1992, \apjs, 80, 1

\bibitem[{{Tanvir} {et~al.}(2009){Tanvir}, {Fox}, {Levan}, {Berger},
  {Wiersema}, {Fynbo}, {Cucchiara}, {Kr{\"u}hler}, {Gehrels}, {Bloom},
  {Greiner}, {Evans}, {Rol}, {Olivares}, {Hjorth}, {Jakobsson}, {Farihi},
  {Willingale}, {Starling}, {Cenko}, {Perley}, {Maund}, {Duke}, {Wijers},
  {Adamson}, {Allan}, {Bremer}, {Burrows}, {Castro-Tirado}, {Cavanagh}, {de
  Ugarte Postigo}, {Dopita}, {Fatkhullin}, {Fruchter}, {Foley}, {Gorosabel},
  {Kennea}, {Kerr}, {Klose}, {Krimm}, {Komarova}, {Kulkarni}, {Moskvitin},
  {Mundell}, {Naylor}, {Page}, {Penprase}, {Perri}, {Podsiadlowski}, {Roth},
  {Rutledge}, {Sakamoto}, {Schady}, {Schmidt}, {Soderberg}, {Sollerman},
  {Stephens}, {Stratta}, {Ukwatta}, {Watson}, {Westra}, {Wold}, \&
  {Wolf}}]{tfl09}
{Tanvir}, N.~R. {et al.}\  2009, \nat, 461, 1254

\bibitem[{{Thoene} {et~al.}(2008){Thoene}, {de Ugarte Postigo}, {Vreeswijk},
  {Malesani}, \& {Jakobsson}}]{Thoene:2008ys}
{Thoene}, C.~C., {de Ugarte Postigo}, A., {Vreeswijk}, P.~M., {Malesani}, D.,
  \& {Jakobsson}, P. 2008, GRB Coordinates Network, 8058, 1

\bibitem[{{van Dokkum}(2001)}]{van-Dokkum:2001fk}
{van Dokkum}, P.~G. 2001, \pasp, 113, 1420

\bibitem[{{Vergani} {et~al.}(2009){Vergani}, {Petitjean}, {Ledoux},
  {Vreeswijk}, {Smette}, \& {Meurs}}]{Vergani:2009fj}
{Vergani}, S.~D., {Petitjean}, P., {Ledoux}, C., {Vreeswijk}, P., {Smette}, A.,
  \& {Meurs}, E.~J.~A. 2009, \aap, 503, 771

\bibitem[{{Vestrand} {et~al.}(2002){Vestrand}, {Borozdin}, {Brumby},
  {Casperson}, {Fenimore}, {Galassi}, {McGowan}, {Perkins}, {Priedhorsky},
  {Starr}, {White}, {Wozniak}, \& {Wren}}]{Vestrand:2002uq}
{Vestrand}, W.~T. {et al.}\  2002, in Society of Photo-Optical Instrumentation
  Engineers (SPIE) Conference Series, Vol. 4845, Society of Photo-Optical
  Instrumentation Engineers (SPIE) Conference Series, ed. R.~I. {Kibrick},
  126--136

\bibitem[{{Vreeswijk} {et~al.}(2004){Vreeswijk}, {Ellison}, {Ledoux}, {Wijers},
  {Fynbo}, {M{\o}ller}, {Henden}, {Hjorth}, {Masi}, {Rol}, {Jensen}, {Tanvir},
  {Levan}, {Castro Cer{\'o}n}, {Gorosabel}, {Castro-Tirado}, {Fruchter},
  {Kouveliotou}, {Burud}, {Rhoads}, {Masetti}, {Palazzi}, {Pian}, {Pedersen},
  {Kaper}, {Gilmore}, {Kilmartin}, {Buckle}, {Seigar}, {Hartmann}, {Lindsay},
  \& {van den Heuvel}}]{Vreeswijk:2004ve}
{Vreeswijk}, P.~M. {et al.}\  2004, \aap, 419, 927

\bibitem[{{Vreeswijk} {et~al.}(2007){Vreeswijk}, {Ledoux}, {Smette}, {Ellison},
  {Jaunsen}, {Andersen}, {Fruchter}, {Fynbo}, {Hjorth}, {Kaufer}, {M{\o}ller},
  {Petitjean}, {Savaglio}, \& {Wijers}}]{Vreeswijk:2007rt}
--- 2007, \aap, 468, 83

\bibitem[{{Vreeswijk} {et~al.}(2003){Vreeswijk}, {M{\o}ller}, \&
  {Fynbo}}]{Vreeswijk:2003uq}
{Vreeswijk}, P.~M., {M{\o}ller}, P., \& {Fynbo}, J.~P.~U. 2003, \aap, 409, L5

\bibitem[{{Werk} {et~al.}(2012){Werk}, {Prochaska}, {Thom}, {Tumlinson},
  {Tripp}, {O'Meara}, \& {Meiring}}]{Werk:2012vn}
{Werk}, J.~K., {Prochaska}, J.~X., {Thom}, C., {Tumlinson}, J., {Tripp}, T.~M.,
  {O'Meara}, J.~M., \& {Meiring}, J.~D. 2012, \apjs, 198, 3

\bibitem[{{York} {et~al.}(2000){York}, {Adelman}, {Anderson}, {Anderson},
  {Annis}, {Bahcall}, {Bakken}, {Barkhouser}, {Bastian}, {Berman}, {Boroski},
  {Bracker}, {Briegel}, {Briggs}, {Brinkmann}, {Brunner}, {Burles}, {Carey},
  {Carr}, {Castander}, {Chen}, {Colestock}, {Connolly}, {Crocker}, {Csabai},
  {Czarapata}, {Davis}, {Doi}, {Dombeck}, {Eisenstein}, {Ellman}, {Elms},
  {Evans}, {Fan}, {Federwitz}, {Fiscelli}, {Friedman}, {Frieman}, {Fukugita},
  {Gillespie}, {Gunn}, {Gurbani}, {de Haas}, {Haldeman}, {Harris}, {Hayes},
  {Heckman}, {Hennessy}, {Hindsley}, {Holm}, {Holmgren}, {Huang}, {Hull},
  {Husby}, {Ichikawa}, {Ichikawa}, {Ivezi{\'c}}, {Kent}, {Kim}, {Kinney},
  {Klaene}, {Kleinman}, {Kleinman}, {Knapp}, {Korienek}, {Kron}, {Kunszt},
  {Lamb}, {Lee}, {Leger}, {Limmongkol}, {Lindenmeyer}, {Long}, {Loomis},
  {Loveday}, {Lucinio}, {Lupton}, {MacKinnon}, {Mannery}, {Mantsch}, {Margon},
  {McGehee}, {McKay}, {Meiksin}, {Merelli}, {Monet}, {Munn}, {Narayanan},
  {Nash}, {Neilsen}, {Neswold}, {Newberg}, {Nichol}, {Nicinski}, {Nonino},
  {Okada}, {Okamura}, {Ostriker}, {Owen}, {Pauls}, {Peoples}, {Peterson},
  {Petravick}, {Pier}, {Pope}, {Pordes}, {Prosapio}, {Rechenmacher}, {Quinn},
  {Richards}, {Richmond}, {Rivetta}, {Rockosi}, {Ruthmansdorfer}, {Sandford},
  {Schlegel}, {Schneider}, {Sekiguchi}, {Sergey}, {Shimasaku}, {Siegmund},
  {Smee}, {Smith}, {Snedden}, {Stone}, {Stoughton}, {Strauss}, {Stubbs},
  {SubbaRao}, {Szalay}, {Szapudi}, {Szokoly}, {Thakar}, {Tremonti}, {Tucker},
  {Uomoto}, {Vanden Berk}, {Vogeley}, {Waddell}, {Wang}, {Watanabe},
  {Weinberg}, {Yanny}, {Yasuda}, \& {SDSS Collaboration}}]{York:2000uq}
{York}, D.~G. {et al.}\  2000, \aj, 120, 1579

\bibitem[{{Zhu} \& {Menard}(2012)}]{zhu:2012aa}
{Zhu}, G.~B. \& {Menard}, B. 2012, arXiv:1211.6215

\bibitem[{{Zibetti} {et~al.}(2007){Zibetti}, {M{\'e}nard}, {Nestor}, {Quider},
  {Rao}, \& {Turnshek}}]{Zibetti:2007ys}
{Zibetti}, S., {M{\'e}nard}, B., {Nestor}, D.~B., {Quider}, A.~M., {Rao},
  S.~M., \& {Turnshek}, D.~A. 2007, \apj, 658, 161

\end{thebibliography}


\begin{deluxetable}{lcccccc}
\tablewidth{0in}
\tablecaption{List of objects considered for the \MgII\ analysis  \label{tab:info}}
\tabletypesize{\footnotesize} 
\tablehead{\colhead{GRB} & \colhead{$z_{GRB}$} &  \colhead{Telescope} & \colhead{Instrument} & \colhead{Resolution}& \colhead{$S/N^a$} &\colhead{Reference}   \\
&&&&(\AA)&&
}
\startdata
111229A & 1.380 &Gemini&GMOS& $5.8$&$6.7$&this work\\ 
111107A & 2.893 &Gemini&GMOS& $5.8$&$3.5$&this work\\ 
111008A & 4.989 &Gemini&GMOS& $5.8$&$3.8$&this work\\ 
110918A&0.982 & Gemini&GMOS& $5.8$&$20$&this work\\
110731A & 2.830& Gemini&GMOS& $5.8$&$26$&this work\\ 
110726A &1.036 & Gemini&GMOS &$5.8$&$9$&this work\\
110213B & 1.083 & Gemini&GMOS &$3.4$&$5$&this work\\
110213A&1.460&Bok&FAST & $6$&$20$&(4)\\
110205A & 2.214 &Lick& KAST& $11$&14&(4)\\
100906A &1.727 & Gemini&GMOS& $5.8$&$21$&this work\\
100901A& 1.408 & Gemini&GMOS& $3.4$&$6$&this work\\
100814A&1.438&MAGELLAN & MagE&$1.8$&10&this work\\
100513A &4.798&Gemini&GMOS& $5.8$&$17$& this work\\
100418A&0.624 & VLT&X-Shooter&0.86/0.72/2$^*$&$12-38$& (19)\\
100414A& 1.368 & Gemini&GMOS& $5.8$&$11$&this work\\
100302A& 4.813 &Gemini&GMOS&$5.8$ &$3$& this work\\
100219A & 4.667 &Gemini& GMOS  & $1.6$&$1.2$&this work\\
091208B & 1.063 & Gemini &GMOS& $5.8$&$26$&(2)\\
091109A&3.076&VLT&FORS2&$13$&$3$&this work\\ 
091029 & 2.752 & Gemini&GMOS&$5.8$& $39$&this work\\ 
091024 & 1.092 &Gemini& GMOS& $5.8$&$55$&(2)\\  
091020A&1.713&NOT&ALFOSC&$13$&$7$&this work\\
090926A & 2.106 & VLT&X-Shooter& $1.0$&$15-30$&(6)\\ 
090902B& 1.822 & Gemini&GMOS&$4$& $14$&this work\\
090812A&2.454&VLT&FORS2&$13$&$15$&(18)\\
090529A&2.625&VLT&FORS2 &$13$&$4$&(18)\\
090519A&3.851&VLT&FORS2&$13$&$3$&(18)\\
090516A&4.109&VLT&FORS2&$13$&$24$&(18)\\
090426 & 2.609 & Keck&LRIS& $5.5$&8& this work\\ 
090424 & 0.544 & Gemini&GMOS&$5.8$& $22$&this work\\ 
090323 & 3.567 &Gemini& GMOS& $5.8$&$17$&this work\\ 
090313& 3.375 & Gemini& GMOS&$5.8$&$11$&this work\\
081222 & 2.771 & Gemini&GMOS& $5.8$&$21$&this work\\ 
081029 & 3.847 &Gemini&GMOS&$5.8$& $42$ & (2)\\ 
081008 & 1.967 & Gemini&GMOS&$3.4$& $35$&(2)\\ 
081007 & 0.529 & Gemini&GMOS& $5.8$&$31$&(2)\\ 
080928 & 1.690 & Gemini/VLT&GMOS/FORS2&$5.8/13$ &$8/25$& (2)/(3)\\ 
080916A&0.689&VLT&FORS1 &$13$&$5$&(18)\\
080913A&6.700&VLT&FORS2&$13$&$2.5$&(13)\\
080905B&2.374&VLT&FORS1&13&$13$&(18)\\
080810 & 3.350 & Keck&HIRES& $0.18$&16&(18)\\ 
080805& 1.505 & VLT&FORS2& $13$&$3$&(3)\\
080804& 2.205 & Gemini&GMOS& $5.8$&$17$&(2)\\
080721&2.608&TNG&Dolores&$8.1$&$9$&(18)\\
080710 & 0.845 & Gemini&GMOS& $3.4$&$35$&(2)\\ 
080707& 1.234 & VLT&FORS1& $13$&$5$&(3)\\
080607 & 3.036 & Keck&LRIS& $4$&11&(3)\\ 
080605A& 1.639 & VLT&FORS2& $13$&$30$&(3)\\
080604 & 1.416 & Gemini&GMOS& $5.8$&$4$&(2)\\ 
080603B& 2.686 & NOT&ALFOSC& $13$&$41$&(3)\\
080603A & 1.688 & Gemini&GMOS&$5.8$& $38$&(2)\\ 
080520& 1.545 & VLT&FORS2& $13$&$5$&(3)\\
080413B & 1.100 & Gemini&GMOS&$3.4$& $2$&(2)\\ 
080413A & 2.433 &Gemini &GMOS & $4$&$14$&(2)/(9)\\ 
080411&1.030 & VLT&FORS1 &$13$&$60$&(18)\\
080330& 1.513 &NOT& ALFOSC& $13$&$18$&(3)\\
080319C &1.949  & Gemini&GMOS& $3.4$&$4$&(2)\\ 
080319B & 0.937 & Gemini/VLT&GMOS/UVES& $5.8/0.13$&$45/70$&(2)/(9)\\ 
080310A& 2.4272 & VLT&UVES& $0.13$&15&(9)\\
080210 & 2.6419 & VLT&FORS2& $13$&$33$&(3)\\ 
071122 & 1.141 & Gemini&GMOS& $5.8$&$12$&(2)\\ 
071117& 1.334 &VLT& FORS1& $13$&$4$&(3)\\
071112C & 0.823 & Gemini&GMOS& $4$&$3$&(2)\\ 
071031 & 2.692 & VLT&UVES/FORS2& $0.13/13$&$70/40$&(3)\\ 
071020& 2.145 & VLT&FORS2& $13$&$6$&(3)\\
071010B & 0.947 & Gemini&GMOS&$5.8$& $15$&(2)\\ 
071003&1.6044&Keck&LRIS & $5$&34&(20)\\
070810A& 2.170& Keck&LRIS& $5$&6&(21)\\
070802& 2.453 & VLT&FORS2& $13$&$8$&(3)\\
070721B&3.626&VLT&FORS2&$13$&$6$&(18)\\
070611& 2.039 & VLT&FORS2& $13$&$15$&(3)\\
070529 & 2.498 & Gemini&GMOS& $3.4$&$12$&(2)\\ 
070506& 2.306 & VLT&FORS1& $13$&$3$&(3)\\
070411& 2.954 & VLT&FORS1& $13$&$6.5$&(3)\\
070318 & 0.836 & Gemini&GMOS&$4$& $8$&(2)\\ 
070306& 1.496 & VLT&FORS2& $13$&$4$&(3)\\
070125 & 1.547 & Gemini/VLT&GMOS/FORS1&$5.8/$& $6/15$&(10)/(3)\\ 
070110& 2.352 & VLT&FORS2& $13$&$30$&(3)\\
061121&1.314&Keck&LRIS & $13$&28&(3)\\
061110B& 3.434 & VLT&FORS1& $13$&$11$& (3)\\
061110A &0.758 &VLT&FORS1&$13$&$6$&(3)\\
061007& 1.261 & VLT&FORS1& $13$&$6$&(3)\\
060927 & 5.468 &VLT&FORS1& $8$&$2.5$& (1)\\  
060926 & 3.205 & VLT&FORS1& $13$&$8$&(1)\\
060908 &1.884 & Gemini&GMOS& $5.8$&$8$&this work\\
060906 & 3.686 &VLT& FORS1&13& $5.8$&(1)\\ 
060904B&0.703&VLT&FORS1 &$13$&$12$&(18)\\
060729 & 0.543 & Gemini&GMOS& $3.4$&$26$&(2)\\ 
060714& 2.711 & VLT&FORS1& $13$&$50$&(3)\\
060708 & 1.923&VLT&FORS2& $13$&$5$&(3)\\
060707 & 3.425 & VLT&FORS2& $13$&$7$&(3)\\ 
060607A& 3.047 & VLT&UVES& $0.13$&43&(9)\\
060526 & 3.221 & VLT&FORS1& $13$&$38$&(1)\\ 
060522& 5.111 &Keck&LRIS&$5$ &2.3&(1)\\    
060512&2.092&VLT&FORS1 &$13$&3&(3)\\
060510B & 4.922 &Gemini&GMOS& $5.8$&$8.4$&(2)\\ 
060502A & 1.515 &Gemini& GMOS& $5.8$&$8$&(2)\\ 
060418 & 1.489 & Gemini/VLT&GMOS/UVES& $5.8/0.13$&$86/60$&(2)/(11)\\ 
060210 & 3.912 &Gemini& GMOS&$5.7$ &$26$&(2)\\ 
060206 & 4.046 &NOT&ALFOSC&$13$& $40$& (3)\\
060124& 2.296 & Keck&ESI& $13$&8&(3)\\
060115 & 3.5328 & VLT&FORS1& $13$&$10$&(3)\\ 
051111& 1.5489 &Keck& HIRES& $0.18$&20&(12)\\
050922C& 2.1996 & VLT&UVES& $0.13$&12& (9)\\
050908 & 3.339& Gemini/Keck&GMOS/Deimos& $4/1.6$&$9/12$&(2)/this work\\ 
050820 & 2.614 & VLT&UVES& $0.13$&23&(9)\\ 
050802& 1.711 &NOT& ALFOSC& $13$&$7$&(3)\\
050801& 1.559 &Keck& LRIS& $5$&5&(3)\\
050730 & 3.9687 &VLT& UVES& $0.13$&40& (9)\\ 
050401& 2.896 &VLT& FORS2& $13$&$23$&(3)\\
050319& 3.240 &NOT& ALFOSC& $13$&$6$&(3)\\
030429 & 2.655 & VLT& FORS1& $13$&$7$&(1)\\ 
030323& 3.372 & VLT&FORS1& $13$&$8$&(13)\\
030226&1.986&VLT&FORS1 & $13$&$30$&(14)\\
021004& 2.323 & VLT&UVES &$0.13$&40&(9)\\
020813&1.255&VLT&UVES & $0.13$&$60$&(15)\\
010222&1.477&Keck&ESI & $0.6$&4&(16)\\
000926&2.038& Keck& ESI& $0.6$& 12&(17)\\
\enddata
\tablenotetext{a}{Signal-to-noise ratio is estimated as the median at the continuum level over a wavelength range clean of telluric lines}
\tablerefs{(1) \cite{Jakobsson:2006lr}; (2) \cite{Cucchiara:2010fk}; (3) \cite{Fynbo:2009lr} and reference therein;
(4) \cite{Cucchiara:2011fk}; (5) \cite{de-Ugarte-Postigo:2011uq}; (6) \cite{DElia:2010kx}; (7)\cite{DElia:2011vn};
(8) \cite{Thoene:2008ys}; (9) \cite{Vergani:2009fj}; (10) \cite{Cenko:2008zr}; (11) \cite{Vreeswijk:2007rt}; (12) \cite{Prochaska:2007ly}; (13) \cite{Vreeswijk:2004ve}; (14) \cite{Klose:2004qf}; (15) \cite{Barth:2003ve}; (16) \cite{Mirabal:2002qf}; (17) \cite{Castro:2003bh}; (18) \cite{de-Ugarte-Postigo:2012kx}; (19) \cite{de-Ugarte-Postigo:2012ys}; (20) \cite{Perley:2008zr};(21) \cite{Milvang-Jensen:2012ly} }\label{tab:sample}
\end{deluxetable}

\begin{deluxetable}{lccccc}
\tablewidth{0in}
\tablecaption{Studied Sample}
\tabletypesize{\footnotesize} 
\tablehead{\colhead{} & \colhead{Number}$^a$ &   \colhead{$\Delta z_{1.0{\rm \AA}}$} &  \colhead{$N_{1.0{\rm \AA}}$}&  $\ell_{GRB}(z)$& $\ell_{QSO}(z)$\\
&of GRBs&& &
}
\startdata
\samI &  83     & 44.9   &8&$0.18\pm0.06$ &0.26\\
\samF &  95      & 55.5  &20&$0.36\pm0.09$ &0.24\\
\samH &    18    & 20.3  &13&$0.64\pm0.26$&0.25\\
\samL &      79  & 35.3  &7&$0.19\pm0.08$&0.12\\
\enddata
\tablenotetext{a}{The total number of GRB lines of sight in each sample corresponds to the sum of all those GRBs 
where $\ell(z)\neq0$.}
\tablecomments{Summary of our \MgII\ search: the sample name and the number of lines of sight included are 
listed in the first two columns; the total redshift path density explored and the number of absorbers identified are listed in 
the third and forth column. Based on these we could determine the incidence of the absorbers in each sample and compare 
it with the expected incidence along our QSOs sample (last two columns).}
\label{tab:subsam}

\end{deluxetable}  
\begin{deluxetable}{llll}
\tablewidth{0in}
\tablecaption{List of excluded lines and regions in the redshift pathlength estimate }
\tabletypesize{\footnotesize} 
\tablehead{\colhead{Description}&\colhead{$\lambda_{rest}$(\AA)} &\colhead{Description}&\colhead{$\lambda_{rest}$(\AA)}  
}
\startdata
\NV & 1238,1242 &\CI&1560,\\
\SII  & 1250,1253,1259 &\FeII&1608,1611,2249,2260,2344,2374,2382,2586,2600\\
\SiII & 1260,1304,1526,1808 &\AlII&1670\\
\SiII*& 1264,1309,1533,1816 &\AlIII&1854,862\\
\OI  & 1302 &\CrII&2017,2026,2056,2066\\
\NiII & 1317,1370,1454,1703,1709,1741,1751&\ZnII&2026,2062\\
\CII & 1334 &\NiII*&2217\\
\CII* & 1335&\MnII&2576,2594,2606\\
\SiIV & 1393,1402 &Band B$^a$&$6860-7000$\\
\CIV & 1548,1550 &Band A$^a$&$7600-7704$\\
Atm. Band$^a$ & $8130-8323$ & Atm. Band$^a$ & $8930-9020$\\
\enddata
\tablenotetext{a}{Atmospheric absorption bands from the HIRES telluric
  line list: {\tt
    http://www2.keck.hawaii.edu/inst\/hires/makeewww/Atmosphere/atmabs.txt}. The
  indicated wavelengths are (obviously) independent of the GRB redshift.}
\label{tab:exclreg}
\end{deluxetable}

\begin{deluxetable}{lccc}
\tablewidth{0in}
\tablecaption{List of Redshifts Intervals where $g(z)=1$ for 1\AA\ \MgII\ survey  \label{tab:path}}
\tabletypesize{\footnotesize} 
\tablehead{\colhead{GRB} & \colhead{$z_{GRB}$} &  \colhead{$z_{start}$} & \colhead{$z_{end}$}
}
\startdata
 030226 & 1.986  &        1.30132 &      1.31706   \\
  &   &        1.32636 &      1.32922   \\
  &   &        1.35426 &      1.38430  \\
  &   &        1.41005 &      1.41935   \\
  &   &        1.42936 &      1.45655   \\
  &   &        1.46656 &      1.48158   \\
  &   &        1.50161 &      1.54667   \\
  &   &        1.55669 &      1.62393   \\
  &   &        1.64038 &      1.64682   \\
  &   &        1.66971 &      1.71119   \\
  &   &        1.72407 &      1.76127   \\
  &   &        1.77200 &      1.77700   \\
  &   &        1.78773 &      1.81134  \\
  &   &        1.82922 &      1.85211  \\
  &   &        1.86284 &      1.86356  \\
  &   &        1.87428 &      1.92364   \\
  &   &        1.94367 &      1.97300   \\
  &   &        1.99231 &      2.14664   \\
  &   &        2.16738 &      2.18741  \\
  &   &        2.20959 &      2.35908  \\
  &   &        2.37124 &      2.39413   \\
  &   &        2.41774 &      2.44850  \\
  &   &        2.48927 &      2.48999   \\
  &   &        2.50644 &      2.52718  \\
  &   &        2.54793 &      2.70959   \\
  &   &        2.75894 &      2.75966   \\
  &   &        2.78040 &      2.78112   \\
  &   &        2.78684 &      2.90057   \\

\enddata
\end{deluxetable}

\begin{deluxetable}{lllllcc}
\tablewidth{0in}
\tablecaption{Intervening systems  \label{tab:interv}}
\tabletypesize{\footnotesize} 
\tablehead{\colhead{GRB} & \colhead{$z_{GRB}$} &  \colhead{$z_{abs}$} & \colhead{$W_r(2796)$} & \colhead{$W_r(2803)$}& \colhead{Statistical} &Other  \\
&&&(\AA)&(\AA)&sample&transition
}
\startdata
010222&$1.477$&$1.156$&$2.22 (0.14)$&$1.69 (0.11) $& F,H &\FeII\\
020813&$1.255$&$1.224$&$1.58 (0.03)$&$1.43 (0.03)$ & F,H &\MgI,\FeII\\
021004$^c$&$2.3295$&$0.555$&$0.66 (0.045)$&$0.36 (0.034)$ & N &\MgI,\FeII\\
		&		&$1.380$&$1.637 (0.020)$&$1.574 (0.043)$&F,H&\MgI,\MnII,\FeII\\
		&		&$1.6026$&$1.407 (0.024)$&$1.02 (0.013)$&F,H&\MgI, \FeII,\MnII\\
030226$^d$&$1.986$&$1.043$&$0.68 (0.25)$&$0.41(0.25)$&N&\AlII\\
		&		&$1.963$&$2.22 (0.10)$&$2.47 (0.10)$&N&\MgI, \CIV,\SiII\\
050730$^c$&$3.9687$&$1.7732$&$0.927 (0.030)$&$0.718 (0.016)$ &N &\MgI,\FeII\\
		&		&$2.2531$&$0.783 (0.650)^b$&$0.677 (0.017)$&N&\SiII,\AlII,\FeII,\MgI\\
050820$^c$&$2.6147$&$0.6915$&$2.723 (0.007)$&$1.576 (0.031)$&F,H&\MgI\\
		&		&$1.4288$&$1.203 (0.023)$&$1.265 (0.026)$&F,H&\MgI,\FeII,\AlIII\\
		&		&$1.6204$&$0.277 (0.024)$&$0.214 (0.008)$&N&\MgI,\FeII,\ZnII,\SiII\\
		&		&$2.3598$&$0.424 (0.306)^b$&$0.517 (0.024)$&N&\FeII,\SiII,\ZnII,\CIV\\
050908	&$3.339$	&$1.548$&$1.21 (0.02)$&$0.92 (0.02)$&F,H&\FeII\\
050922C$^c$&$2.1996$&$0.6369$&$0.187 (0.018)$&$0.121 (0.011)$&N&\MgI,\FeII\\
		&		&$1.1076$&$0.476 (0.029)$&$0.422 (0.19)$&N&\MgI,\FeII\\
		&		&$1.5670$&$0.121 (0.080)^b$&$0.088 (0.007)$&N&\CIV,\FeII\\
051111	&$1.55$&$1.190$&$1.56 (0.02)$&$1.92 (0.01)$&F,H&\MgI,\FeII\\
		&		&$0.827$ & $0.39 (0.02)$ &   $0.29 (0.01)$      &N  &\MgI\\ 
060418$^b$&$1.489$	&$1.107$&$1.84 (0.2)$&$1.58 (0.1)$&F,H&\MgI,\FeII,\ZnII,\AlIII,\AlII\\
		&		&$0.6559$&$1.52 (0.3)$&$2.15 (0.4)$&F,H&\FeII\\
		&		&$0.603$&$1.49 (0.2)$&$1.47 (0.1)$&F,H&\FeII\\
060502A	&$1.515$	&$1.147$&$2.39 (0.12)$&$2.87 (0.12)$&F,L,I&\MgI\\
		&		&$1.078$&$0.61 (0.12)$&$0.49 (0.12)$&N&\\
		&		&$1.044$&$1.90 (0.15)$&$1.92 (0.16)$&F,L,I&\FeI,\MnII,\MgI\\
060607A$^c$&$3.0748$&$1.5103$&$0.124 (0.011)$&$0.144 (0.007)$&N&\FeII\\
		&		&$1.8033$&$1.916 (0.006)$&$1.600 (0.015)$&F,H,I&\MgI,\FeII,\AlIII\\
		&		&$2.2783$&$0.210 (0.058)$&$0.298 (0.013)$&N&\FeII,\AlIII,\AlII,\CIV,\SiII,\SiIV\\
060906	&$3.685$	&$1.2659$&$1.63 (0.28)^a$&$1.63 (0.28)^a$&N&\MgI\\
060926	&$3.2$	&$0.924$&$2.49 (0.62)^a$&$2.49 (0.62)^a$&N&\MgI,\FeI\\
		&		&$1.7954$&$3.27 (0.69)$&$3.71 (0.87)$&N&\MgI,\FeII,\MnII\\
		&		&$1.8289$&$1.27 (0.11)$&$0.72 (0.07)$&N&\MgI\\
061007	&$1.261$	&$1.065$&$3.14 (0.53)^b$&$4.48 (0.65)^b$&N&\MgI,\FeII,\MnII\\
070529	&$2.498$	&$1.414$&$0.20 (0.02)$&$0.09 (0.02)$&N&\\ 
070506	&$2.306$	&$1.600$&$1.92 (0.04)$&$1.65 (0.05)$&N& \AlIII\\
070611	&$2.039$	&$1.297$&$2.65 (0.27)$&$1.99 (0.23)$&N&\MgI,\FeII\\
070802	&$2.45$   	&$2.0785$&$0.82 (0.12)$&$0.82 (0.12)$&N&\AlII,\NiII,\MgI,\FeII\\
		&		&$2.2921$&$0.55 (0.15) $&$0.55 (0.22)$&N&\NiII,\AlIII,\CrII,\FeII\\
071003	&$1.604$&$0.372$&$2.28 (0.19) $&$1.91(0.19)$&F,L,I&\MgI\\
		&		&$0.943$&$0.61 (0.05)^b $&$0.36 (0.05)$&N&\MgI\\
		&		&$1.101$&$0.80 (0.06) $&$0.64 (0.05)$&N&\MgI\\
071031$^c$&$2.6922$&$1.0743$&$0.330 (0.016)$&0.206 (0.008)&N&\FeII\\
		&		&$1.6419$&$0.806 (0.014)$&0.586 (0.052)&N&\FeII,\AlIII,\CIV\\
		&		&$1.9520$&$0.743 (0.016)$&0.612 (0.016)&N&\MgI,\FeII\\
080310$^c$&$2.4272$&$1.6711$&$0.421 (0.012)$&0.366 (0.016)&N&\MgI,\FeII,\AlII,\SiII,\CIV\\
080319B$^c$&$0.9378$&$0.5308$&$0.614 (0.001)$&0.350 (0.002)&N&\MgI,\FeII\\
		&		&$0.5662$&$0.083 (0.003)$&0.029 (0.001)&N&\MgI,\FeII\\
		&		&$0.7154$&$1.482 (0.001)$&0.736 (0.003)&F,H,I&\MgI,\FeII\\
		&		&$0.7608$&$0.108 (0.002)$&0.039 (0.002)&N&\FeII\\
080319C	&$1.95$	&$0.8104$&$2.04(0.52)$&$1.64(0.42)$&N&\FeII,\MnII\\  
080603A	&$1.688$	&$1.271$&$3.11(0.11)$&$3.17 (0.13)^b$&F,L,I&\MgI,\FeII\\
		&		&$1.563$&$0.77 (0.01)$&$0.92 (0.01)$&N&\FeI\\
080605	&$1.64$	&$1.2987$&$1.08 (0.11)$&$0.77 (0.10)$&F,L,I&\FeII\\
080607A	&$3.036$	&$1.341$&$3.0 (0.08)$&$1.26 (0.05)$&F,L,I&\MgI\\
080805A	&$1.505$	&$1.197$&$8.2 (0.92)^a$&$8.2 (0.92)^a$&N&\MnII,\FeII\\
080905B	&$2.374$	&$0.618$&$6.65 (0.2)^a$&$6.65 (0.1)^a$&N&\MgI\\ 
080928	&$1.691$	&$0.736$&$9.54 (0.25)^a$&$9.54 (0.25)^a$&N&\MgI,\FeII\\
081222	&$2.77$	&$0.8168$&$0.52 (0.01)$&$0.28 (0.11)$&N&\MgI,\FeII\\
		&		&$1.0708$&$1.46 (0.23)$&$0.61 (0.21)$&F,L,I&\FeII\\	
091208B &$1.063$	&$0.784$&$0.65 (0.43) $&$1.03 (0.43)$&N&\MgI\\
100814A &$1.44$  & $1.1574$&$0.426 (0.04)$&$0.379 (0.04)$&N&\MgI\\
100901A	&$1.408$	&$1.314$&$1.74 (0.17)^b$&$1.53 (0.16)^b$&N&\FeII,\MgI\\
100906A	&$1.64$	&$0.994$&$0.87 (0.1)$&$1.19 (0.1)^b$&N&\\
110918A	&$0.982$	&$0.877$&$2.65 (0.20)$&$2.82 (0.20)$&N&\MgI,\FeII\\
\enddata
\tablenotetext{a}{Equivalent Width measurement is largely effected by blending. For these lines we report the total EW for the doublet.}
\tablenotetext{b}{Equivalent Width measurement is lightly effected by blending. EW values are derived via deblending procedure using gaussian fit of the two lines (either the other member of the doublet or other lines) via the {\tt IRAF splot} tool.}
\tablenotetext{c}{UVES}
\tablenotetext{d}{Also see \cite{Klose:2004fj}}

\end{deluxetable}  


%

%
%

\clearpage

\clearpage

\clearpage

\clearpage

\clearpage

\clearpage

\clearpage

\clearpage

\clearpage

\clearpage

\clearpage

\clearpage

\clearpage

\end{document}